\documentclass[a4paper,10pt]{book}
\usepackage{graphicx,subfigure,psfrag}
\usepackage{amsfonts,amsmath}
\usepackage{color}
\usepackage{tabularx}
\newcommand{\pc}[1]{\parbox{0pt}{#1}}
\usepackage[colorlinks=false]{hyperref} 
\usepackage{dsfont}
\usepackage{times}
\usepackage{amsfonts,amsmath}

\usepackage{color}
\usepackage{tabularx}
\usepackage{dsfont}
\usepackage{times}
\usepackage{ifthen}
\usepackage{enumerate}
\usepackage{amsmath}
\usepackage{amssymb}
\usepackage[mathscr]{eucal}
\usepackage[errorshow]{tracefnt}
\usepackage{amsmath}
\usepackage{slashed}
\usepackage{amssymb}
\usepackage{array}
\usepackage{amsthm}
\usepackage{eucal}
\usepackage{epsf}
\usepackage{epsfig}
\usepackage{psfrag}
\usepackage{multirow}

\usepackage[apple mac]{inputenc}

\newcommand{\complex}{{\mathds C}} 

\def\+{{+\!\!\!+}}

\def\P{\Phi}

\def\h{\chi}

\newcommand{\C}{{\mathds C}}

\def\F{{\cal F}}
\def\T{{\rm T}}

\def\Tr{\rm Tr}
\def\log{\rm log}

\def\P{{\cal P}}

\def\pmb#1{\setbox0=\hbox{#1}%
\kern.0em\copy0\kern-\wd0 
\kern-.04em\copy0\kern-\wd0 
\kern.08em\copy0\kern-\wd0 
\kern-.04em\raise.0433em\box0 }         
 
\def\diag{\textstyle{\rm{diag}}} 
\newcommand{\nc}{\newcommand} 
\nc{\beq}{\begin{equation}} 
\nc{\eeq}[1]{\label{#1}\end{equation}} 
\nc{\ber}{\begin{eqnarray}} 
\nc{\eer}[1]{\label{#1}\end{eqnarray}} 
\nc{\pek}[1]{\cite{#1}} 
\nc{\enr}[1]{(\ref{#1})} 
\nc{\kal}[1]{{\cal{#1}}} 
\nc{\dott}{\;\cdot\;}

\def\0 {\nonumber}

\def\+{{+\!\!\!+}}

\def\P{\Phi}

\def\h{\chi}

\def\F{{\cal F}}
\def\T{{\rm T}}

\def\Tr{\rm Tr}
\def\log{\rm log}

\def\P{{\cal P}}

\def\calb         {{\cal B}}
\def\calc         {{\cal C}}
\def\cald         {{\cal D}}

\def\calF         {{\cal F}}

\def\call         {{\cal L}}

\def\caln         {{\cal N}}

\def\complex      {{\mathbb C}}

\def\pmb#1{\setbox0=\hbox{#1}%
\kern.0em\copy0\kern-\wd0 
\kern-.04em\copy0\kern-\wd0 
\kern.08em\copy0\kern-\wd0 
\kern-.04em\raise.0433em\box0 }         
 
\def\diag{\textstyle{\rm{diag}}} 
\def\0 {\nonumber}
 
\newcommand{\be}{\begin{equation}} 
\newcommand{\ee}{\end{equation}} 
\newcommand{\bea}{\begin{eqnarray}} 
\newcommand{\eea}{\end{eqnarray}}

\begin{document}

\title{}
\author{}
\date{}

\frontmatter

  \pagestyle{plain}
    \begin{titlepage}


\begin{center}
\large{Scuola Internazionale Superiore di Studi Avanzati}

\begin{figure}[htd]
\begin{center}
\includegraphics[width=0.5\textwidth]{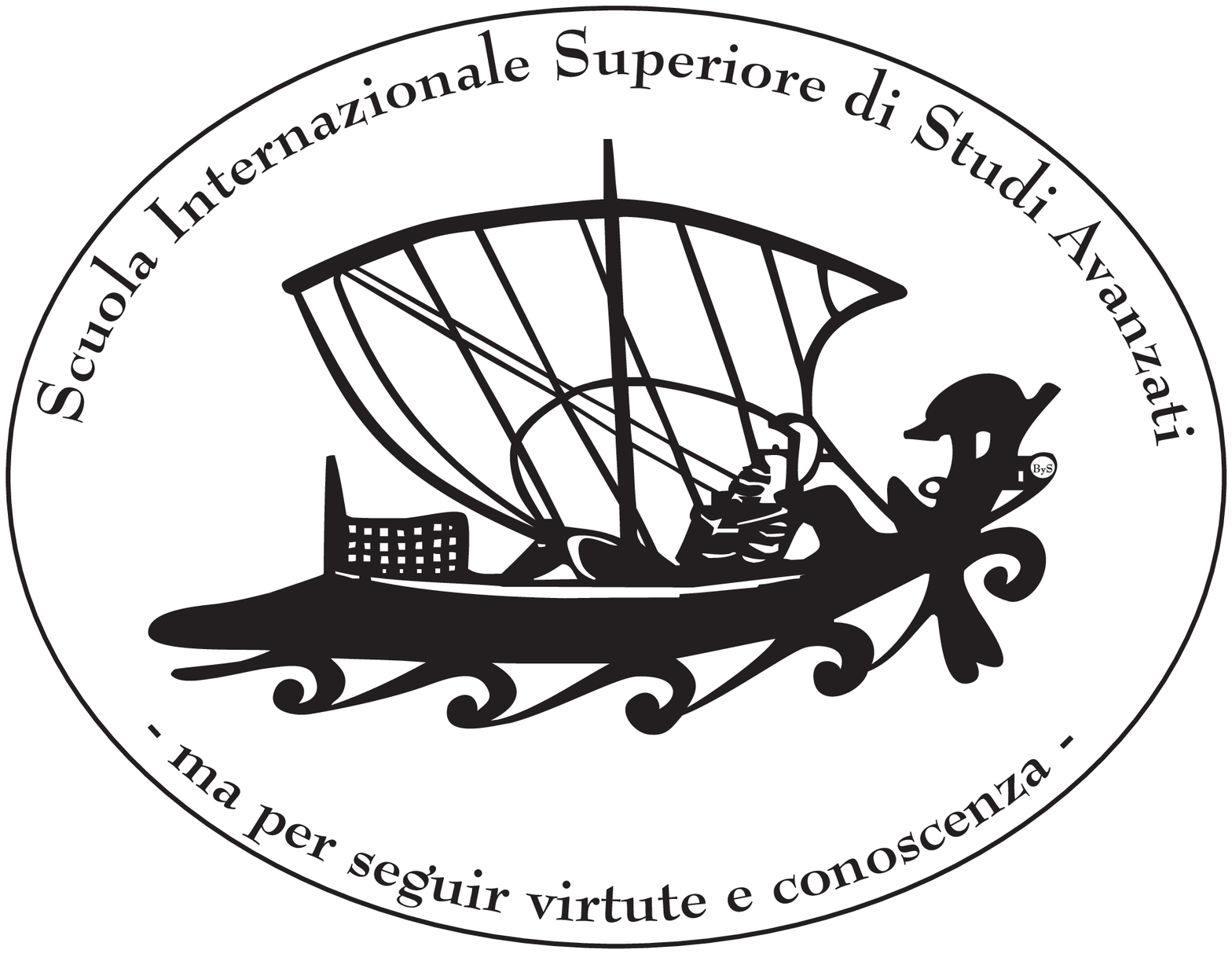}
\end{center}
\end{figure}

\vspace{1cm}
\huge{{\bf Topics in Open Topological Strings }}
\end{center}

\vspace{2cm}
\begin{center}
\large{Thesis submitted for the degree of \\ Doctor Philosophiae }
\end{center}

\vspace{3cm}
\hbox{\hsize=3in \raggedright\parindent=0pt
\vtop {Supervisors: \\Dr. Giulio Bonelli \\Dr. Alessandro Tanzini}
\hspace{3cm}
\vtop {Candidate:  \\Andrea Prudenziati }}

\vspace{2cm}
\begin{center}
{\underline{Trieste, September 2010}}
\end{center}

  
    \end{titlepage}
    \large
    \newpage

  \pagenumbering{roman}
    
    \chapter*{Abstract}
\addcontentsline{toc}{chapter}{Abstract}
\label{sec:abs}
\noindent

This thesis is based on some selected topics in open topological string theory which I have worked on during my Ph.D. It comprises an introductory part where I have focused on the points most needed for the later chapters, trading completeness for conciseness and clarity. Then, following \cite{nostro}, we discuss tadpole cancellation for topological strings where we mainly show how its implementation is needed for ensuring the same "odd" moduli decoupling encountered in the closed theory. Next we move to analyse how the open and closed effective field theories for the B model interact writing the complete Lagrangian. We first check it deriving some already known tree level amplitudes in term of target space quantities, and then we extend the recipe to new results; later we implement  open closed duality from a target field theory perspective. This last subject is also analysed from a worldsheet point of view extending the analysis of \cite{Bonelli:2010cu}. Some ideas for future research are briefly reported.  
    
    \tableofcontents
    \cleardoublepage

    \mainmatter
  
    \chapter*{Introduction}
\addcontentsline{toc}{chapter}{Introduction}
\label{sec:intro}
\noindent

String theory was born in the late sixties as a non so well working method for explaining string-like phenomena observed in processes involving strong interactions. Unfortunately ( or maybe not ) for its creators, QCD came as a much better explanation but the theory, because of some interesting features, was not thrown away and was still studied for a while. In particular it was noted the presence of a massless spin two particle which turned out to be a perfect candidate for the graviton, the particle in charge for transmitting the gravitational force at a quantum level. Having for a long time unsuccessfully looked for the quantum version of general relativity, the discovery gave a tremendous boost to the work around the theory. Indeed the initial enthusiasm  was slowly partially replaced by a more skeptical vision, as the intrinsic complications of the theory came out; since the seventies the string research history has oscillated between periods of low results, where the difficulties appeared too much for our limited ability, and fast leaps forward when it looked as if the quest was practically over, with just a few minor points still to be settled. Nowadays what we are left with is a long developed theory, intrinsically complicated, with a huge amount of theoretical results and predictions; unfortunately none of them is really independent of the specific details of the model we need to embed in the theory for being able to get in touch with our world. It is for example the case of supersymmetry, a necessary condition for making the theory consistent in itself and with what we observe experimentally, or extradimensions, six in the various perturbative descriptions we have. In both cases we know what we have to generically expect but we miss the details, what is the amount of supersymmetry, how it is practically realized, what is the shape of the internal manifold and so on. The time when the hope of uniqueness for string theory was seen as plausible are far away in the past and we are forced to deal with some general constructions whose explicit realization for describing our world is far from being known. Clearly this situation has generated, during the years, some skepticism among the physicists community, having to deal with something which in practice was not falsifiable or, at least, not for the limited knowledge we have. So many criticisms have appeared claiming for strings not to be useful that is, able to predict and contradict whatever experimental result may be found, relying on the numerous obscure patches whose mysterious details could be properly adjusted in order to fit practically everything. 

Present research is essentially divided into four subsectors: there are people trying to develop `` brute force'' phenomenological models in the string framework, looking for proper compactifications, fluxes and brane arrangements, with the explicit goal of being able to obtain, at least, something not in contradiction with the Standard Model or, better, to generate some plausible prediction. Then there are people using string theory as a tool for more sophisticated and indirect checks as, for example, the Ads/CFT correspondence for describing superconductivity, plasmas and so forth. Then we have people dealing with the issue of developing the theory in itself, understanding more deeply what we already know and clarifying the points still obscure. And finally there are people trying to better understand the theory but through the analysis of easier toy models whose features can resemble, at least partially, the ones of the ``true`` theory but much easier to treat. This last one is also the case which better fits the idea of studying topological string models on which this thesis will be focused. 

Topological strings were invented by Witten at the end of the eighties, \cite{Wittentop1,Witten,Wittentop2}, and they grew rapidly as it became clear that: first they were in principle ( and in some cases also in practice ) completely solvable; second they carried both some peculiar and interesting physical properties worth to be studied on their own, and at the same time others shared with physical strings but here in some way more manifest and transparent. Third that they were, even if a different theory, still in close contact with superstrings, explicitly through their ability in computing coefficient for some  F-terms of its space time effective theory \cite{AGNT,ant}. In addition, because of their topological properties, every physical sensible quantity corresponds to some topological invariant. This provides not only many interesting physical applications for pure mathematics and, in general, a playground for geometry, but it also represents a perfect example of how a close interaction between mathematics and physics can be fruitful to both of them. A typical case is given by the link between Gromow-Witten invariants and A-model amplitudes which was noticed long ago; it furnishes not only an interesting application but also a computing procedure and possible hints for their generalization to the open case. 

This thesis will be based on the attempt to develop some issues for open topological strings, whose closed counterpart is generally already known but not easy to extend, having to deal with additional physics and degrees of freedom. It is also true that open strings are in some way much different from the closed ones but nevertheless strictly bound to them; during all the work I have tried to focus on the conceptual and physical differences arising when you treat an open theory and compare it with the closed case: what can be generalized, how it can be done, what needs new inputs and what they are. Few answers have been given, many questions have arisen; and probably this was the best result. 

The thesis is divided in four chapters: chapter one contains a brief review of the topics we will mostly need later on; it is not supposed to be  complete, instead I focused on specific points and looked for a concise and simple treatment. 
Chapter two deals with the issue of orientability for open strings; in particular we have applied the well known mechanism of tadpole cancellation to the topological case finding a clear reason for its implementation, peculiar of the topological theory. Starting from one loop computations, first on a torus and then generalizing, both explicitly and through diagrams, we explain how tadpole cancellation deletes the dependence from ''wrong`` target space moduli and how the discussion can be implemented as well for higher genus amplitudes. Hints for this mechanism had been already present in the literature but clear computations were missing, so we tried to fill the gap.
The necessity of considering also closed amplitudes when dealing with an open computation is clear already from simple topological reasons; tadpole cancellation itself requires the closed sector to be present in an open theory ( which should be also unorientable ) in order to make sense, but in the topological case there are reasons for this open-closed coupling to be weaker then in usual superstring theories. A clear understanding of how the open and closed sector interact with each other was necessary and we decided to tackle the problem from the target space point of view in the third chapter. The natural continuation of this is a treatment of open closed duality as the question of what happens when we integrate out the purely open degrees of freedom in an open-closed theory. We answered in chapter four both using the effective field theories and with worldsheet arguments explaining how a shift in the background moduli of the closed theory can mimic the open part. 
Conclusions and open issues follows.     
    \chapter{An introduction to topological strings}
\label{sec:cap0}
\noindent

\section{Generic construction of topological quantum field theories}\label{01}
The standard construction for a topological quantum field theory starts from a non linear sigma model with $N=(2,2)$ supersymmetry and a $U(1)$ left and right R-symmetry current. Standard arguments force the target space to be K\"ahler having in this way a splitting into holomorphic and antiholomorphic indexes. In addition, if we want a superconformal symmetry without anomalies, we have to require the vanishing of the first Chern class, that is in that case the target space should be a Calabi Yau. The action we will consider is

\begin{equation}\label{0a}
S = t\int_{\Sigma}d^{2}z\left(\frac{1}{2}g_{IJ}\partial X^{I}\bar{\partial} X^{J} + i\psi^{\bar{i}}_{-}D_{z}\psi^{i}_{-}g_{i\bar{i}} + i\psi^{\bar{i}}_{+}D_{\bar{z}}\psi^{i}_{+}g_{i\bar{i}} + R_{i\bar{i}j\bar{j}}\psi^{i}_{+}\psi^{\bar{i}}_{+}\psi^{j}_{-}\psi^{\bar{j}}_{-}\right) 
\end{equation}

where $t$ is a coupling constant, $g$ is the target space metric, $X^{I}$ is an embedding map from the Riemann surface $\Sigma$ to the target space $X$; $I,J$ run over both holomorphic and antiholomorphic indexes and $\psi_{+/-}$ are worldsheet fermions respectively left ($+$) and right ($-$) moving. In addition $D_{z/\bar{z}}$ is the covariant $\partial/\bar{\partial}$ derivative with, as a connection for the target space indexes, the pullback to $\Sigma$ of the Levi Civita connection on the ( complexified tangent of the ) target space itself, and finally $R$ is the Riemann tensor. Every field transforms appropriately under the four supercharges living on the worldsheet, $Q^{-},\;\bar{Q}^{-}$ and $Q^{+},\;\bar{Q}^{+}$, with corresponding fermionic coefficients, $\epsilon^{-}_{+/-}$ and $\epsilon^{+}_{+/-}$ which are ( inverse ) spin $1/2$ holomorphic and antiholomorphic sections on $\Sigma$; the upper index $\pm$ stands for the R-charge ( scalars $X$ has R charge zero while fermions have respectively $+$ or $-$ for $\bar{i}$ and $i$ indexes ). For completeness let us give the explicit form for the transformations

\begin{eqnarray}
\delta X^{i} &=& i\epsilon^{+}_{-}\psi^{i}_{+} + i\epsilon^{+}_{+}\psi^{i}_{-} \nonumber \\
\delta X^{\bar{i}} &=& i\epsilon^{-}_{-}\psi^{\bar{i}}_{+} + i\epsilon^{-}_{+}\psi^{\bar{i}}_{-} \nonumber \\
\delta \psi^{i}_{+} &=& -\epsilon^{-}_{-}\partial X^{i} - i\epsilon^{+}_{+}\psi^{j}_{-}\Gamma^{i}_{jm}\psi^{m}_{+} \nonumber \\
\delta \psi^{\bar{i}}_{+} &=& -\epsilon^{+}_{-}\partial X^{\bar{i}} - i\epsilon^{-}_{+}\psi^{\bar{j}}_{-}\Gamma^{\bar{i}}_{\bar{j}\bar{m}}\psi^{\bar{m}}_{+} \\
\delta \psi^{i}_{-} &=& -\epsilon^{-}_{+}\bar{\partial} X^{i} - i\epsilon^{+}_{-}\psi^{j}_{+}\Gamma^{i}_{jm}\psi^{m}_{-} \nonumber \\
\delta \psi^{\bar{i}}_{-} &=& -\epsilon^{+}_{+}\bar{\partial} X^{\bar{i}} - i\epsilon^{-}_{-}\psi^{\bar{j}}_{+}\Gamma^{\bar{i}}_{\bar{j}\bar{m}}\psi^{\bar{m}}_{-} \nonumber 
\end{eqnarray}

The point now is to introduce the topological twisting of the theory; this procedure, that can be justified in some different ways we will see later, consists in a redefinition of the worldsheet energy momentum tensor involving the derivative of the R-symmetry current $J$. Modulo a switch in the complex structure of the target space, there are two nonequivalent twists that lead respectively to what are called the A and B model.

\begin{eqnarray}\label{0b}
A-model: && T(z) \rightarrow T(z) + \frac{1}{2}\partial J(z) \;\;\;\bar{T}(\bar{z}) \rightarrow \bar{T}(\bar{z}) - \frac{1}{2}\bar{\partial} J(\bar{z}) \nonumber \\
B-model: && T(z) \rightarrow T(z) + \frac{1}{2}\partial J(z) \;\;\;\bar{T}(\bar{z}) \rightarrow \bar{T}(\bar{z}) + \frac{1}{2}\bar{\partial} J(\bar{z})  
\end{eqnarray}

The easiest way to look for the consequences of the twist is to write down the divergent terms in the OPEs before and after the twist. The $N=(2,2)$ algebra is represented by \cite{Polchinski_II} ( $G^{\pm}$ and $\bar{G}^{\pm}$ are the currents associated to the four supercharges):

\begin{eqnarray}
T^{old}(z)T^{old}(0) &\sim& \frac{c}{2z^{4}} + \frac{2}{z^2}T^{old}(0)+\frac{1}{z}\partial T^{old}(0) \nonumber \\
T^{old}(z)G^{\pm}(0) &\sim& \frac{3}{2z^{2}}G^{\pm}(0) + \frac{1}{z}\partial G^{\pm}(0) \nonumber \\
T^{old}(z)J(0) &\sim& \frac{1}{z^2}J(0)+\frac{1}{z}\partial J(0) \nonumber \\
G^+(z)G^-(0) &\sim& \frac{2c}{3z^{3}} + \frac{2}{z^2}J(0)+\frac{2}{z}T^{old}(0) + \frac{1}{z}\partial J(0)  \\
G^{+}(z)G^{+}(0) &\sim& G^{-}(z)G^{-}(0) \sim 0 \nonumber \\
J(z)G^{\pm}(0) &\sim& \pm\frac{1}{z}G^{\pm}(0) \nonumber \\
J(z)J(0) &\sim& \frac{c}{3z^{2}} \nonumber 
\end{eqnarray}

and similarly for the $\bar{z}$ part. They become

\begin{eqnarray}
T(z)T(0) &\sim&  \frac{2}{z^2}T(0)+\frac{1}{z}\partial T(0) \nonumber \\
T(z)G^{\pm}(0) &\sim& \frac{3\mp 1}{2z^{2}}G^{\pm}(0) + \frac{1}{z}\partial G^{\pm}(0) \nonumber \\
T(z)J(0) &\sim& \frac{1}{z^2}J(0)+\frac{1}{z}\partial J(0) - \frac{c}{3z^{3}} \\
G^+(z)G^-(0) &\sim& \frac{2c}{3z^{3}} + \frac{2}{z^2}J(0)+\frac{2}{z}T(0)   \nonumber
\end{eqnarray}

and for the A model

\begin{eqnarray}
\bar{T}(\bar{z})\bar{T}(0) &\sim&  \frac{2}{\bar{z}^2}\bar{T}(0)+\frac{1}{\bar{z}}\bar{\partial} \bar{T}(0) \nonumber \\
\bar{T}(\bar{z})\bar{G}^{\pm}(0) &\sim& \frac{3\pm 1}{2\bar{z}^{2}}\bar{G}^{\pm}(0) + \frac{1}{\bar{z}}\bar{\partial} \bar{G}^{\pm}(0)   \nonumber \\
\bar{T}(\bar{z})\bar{J}(0) &\sim&  \frac{1}{\bar{z}^2}\bar{J}(0)+\frac{1}{\bar{z}}\bar{\partial} \bar{J}(0) + \frac{c}{3\bar{z}^{3}} \\ 
\bar{G}^+(\bar{z})\bar{G}^-(0) &\sim& \frac{2c}{3\bar{z}^{3}} + \frac{2}{\bar{z}^2}\bar{J}(0)+\frac{2}{\bar{z}}\bar{T}(0) + \frac{2}{\bar{z}}\bar{\partial} \bar{J}(0)  \nonumber
\end{eqnarray}

while for the B model

\begin{eqnarray}
\bar{T}(\bar{z})\bar{T}(0) &\sim&  \frac{2}{\bar{z}^2}\bar{T}(0)+\frac{1}{\bar{z}}\bar{\partial} \bar{T}(0) \nonumber \\
\bar{T}(\bar{z})\bar{G}^{\pm}(0) &\sim& \frac{3\mp  1}{2\bar{z}^{2}}\bar{G}^{\pm}(0) + \frac{1}{\bar{z}}\bar{\partial} \bar{G}^{\pm}(0)  \nonumber \\
\bar{T}(\bar{z})\bar{J}(0) &\sim&  \frac{1}{\bar{z}^2}\bar{J}(0)+\frac{1}{\bar{z}}\bar{\partial} \bar{J}(0) - \frac{c}{3\bar{z}^{3}}  \nonumber \\ 
\bar{G}^+(\bar{z})\bar{G}^-(0) &\sim& \frac{2c}{3\bar{z}^{3}} + \frac{2}{\bar{z}^2}\bar{J}(0)+\frac{2}{\bar{z}}\bar{T}(0) +   \nonumber
\end{eqnarray}

Three comments are in order:
\begin{itemize}
\item First we can easily see that the central charge has disappeared from the OPE of the energy-momentum tensor with itself. So we will not have to add ghost in order to delete it and, a priori, it makes sense to consider theories in every target space dimension. Also the restriction that would have come in usual superstring theory from an $N=(2,2)$ model to be with $2+2$ dimensions, drops out.
\item Second from the $\frac{1}{z^{2}}$ coefficient in the second OPE you can see that the currents $G^{\pm}$ and $\bar{G}^{\pm}$ have become either spin one or spin two objects so that the corresponding charges are either scalars or spin one ( similarly among the four fermions in the Lagrangian two will become scalars and two one forms ).
\item Third the $T-J$ OPE, when $c=\bar{c}=9$, replies exactly the one of the ghost number current with the energy momentum tensor in bosonic string. 
\end{itemize}
Related to the third point above let us note that for an $N=(2,2)$ supersymmetric model $c=\bar{c}=9$ means three complex dimensions in the target space. If this theory ( the original untwisted one ) had to be interpreted as the internal six dimensional theory of a ten dimensional one whose 4 dimensional part is a usual superstring theory, then $c=\bar{c}=9$ is exactly what you need in order to delete the complete central charge. Later we will see a third, much more clear reason, for restricting to the three complex dimensional case. 

Still the main point to analyse is the observation of the presence of two scalar charges out of the four original spin one-half supercharges. In particular this allows us to define a globally well defined charge over every Riemann surface defined as the sum of them. So for the two models we have

\begin{eqnarray}
Q_{A} &=& Q^{+} + \bar{Q}^{-} \nonumber \\
Q_{B} &=& Q^{+} + \bar{Q}^{+} 
\end{eqnarray}
   
What we do is to consider $Q_{A/B}$ as a BRST-like operator defining as physical states only the one in its cohomology, and asking to the vacuum to be $Q$-closed. It is worth to mention, for future use, a completely equivalent way for defining the topological twist; in spite of considering the same action with a twisted energy momentum tensor we consider the same energy momentum tensor with a modified action. What we want to do is to change the spin of the fields/operators depending on their R-charge, so we add a background gauge field $A$, coupled to the R-symmetry current, with the requirement for $A$ to be equal to $\frac{1}{2}$ the spin connection. Signs has to be set depending on the twist:

\begin{equation}\label{0t}
S \rightarrow S + \int_{\Sigma}\pm J\Bar{A} \pm \bar{J}A  
\end{equation}
  
In this way we can delete the spin connection for two fermions and double it for the other two so to obtain two scalars and two one forms.

The last point to analyse is the similarity in structure between topological string and bosonic string. We look at the holomorphic side and consider the $+$ twist in (\ref{0b}) ( for the $-$ twist it is enough to redefine the corresponding R-symmetry current as minus itself ). If you made the correspondence between the following quantities 

\begin{equation}\label{0g}
G^{+} \leftrightarrow J_{BRST}, \;\; J \leftrightarrow J_{ghost}, \;\; T \leftrightarrow T_{ghost+matter}, \;\; G^{-}\leftrightarrow b
\end{equation}

then the commutation relations among them are the same on both sides. Still there are two important differences. One we have already seen, that is the vanishing of the central charge for topological theories in every dimension, while for the bosonic string you have to work in 26. The second is a little more subtle but maybe even more important: while the cohomology for the ghost $b$ in bosonic string is vanishing, that is every closed state with respect to $b$ is also exact\footnote{ and this can be easily inferred from the existence of the ghost $c$ whose commutation relations with $b$ are  $\{b_{0},c_{0}\} = 1$}, in topological string the corresponding quantity, the supercurrent of spin two, has a charge whose cohomology is nontrivial. It is in fact the CPT conjugate of the spin zero supercharge cohmology\footnote{and thus the quantity corresponding to the ghost $c$ does not exist }. This will be fundamental in the future when we will study the holomorphic anomaly equation. 
Having seen the twist we pass now to analyse the most important properties for $A$ and $B$ model.

\subsection{A-Model}

The topological charge is 

\begin{equation}
Q_{A} = Q^{+} + \bar{Q}^{-}
\end{equation}

and out of the four fermions in the Lagrangian two become scalars as well, $\psi^{i}_{+} \rightarrow \chi^{i},\;\psi^{\bar{i}}_{-} \rightarrow \chi^{\bar{i}}$, and two one forms, $\psi^{i}_{-} \rightarrow \rho_{\bar{z}}^{i},\; \psi^{\bar{i}}_{+} \rightarrow \rho_{z}^{\bar{i}}$. The BRST transformations are easily obtained keeping only those $\epsilon$-parameters associated to the two scalar supercharges forming our BRST operator, renaming them simply $\epsilon$, and fixing them to be a constant instead of a generic function. The other two will be put to zero. So we have

\begin{eqnarray}\label{0f}
\delta X^{i} &=& i\epsilon\chi^{i} \nonumber \\
\delta X^{\bar{i}} &=&  i\epsilon\chi^{\bar{i}} \nonumber \\
\delta \chi^{i} &=& 0 \nonumber \\
\delta \rho^{\bar{i}}_{z} &=& -\epsilon\partial X^{\bar{i}} - i\epsilon\chi^{\bar{j}}\Gamma^{\bar{i}}_{\bar{j}\bar{m}}\rho^{\bar{m}}_{z} \\
\delta \rho^{i}_{\bar{z}} &=& -\epsilon\bar{\partial} X^{i} - i\epsilon\chi^{j}\Gamma^{i}_{jm}\rho^{m}_{\bar{z}} \nonumber \\
\delta \chi^{\bar{i}} &=&  0 \nonumber 
\end{eqnarray}

It is straightforward to see that these transformations square to zero except for terms proportional to the equations of motion for $\rho$. It is nevertheless possible to add auxiliary fields to make it nihilpotent also off shell. Also, modulo terms again proportional to the $\rho$-equations of motion, we can see that the action (\ref{0a}), opportunely modified by the field changing due to the twist, can be put in the form  

\begin{eqnarray}\label{0c}
S_{A} &=&  t\int_{\Sigma}d^{2}z\left(\frac{1}{2}g_{IJ}\partial X^{I}\bar{\partial} X^{J} + i\chi^{\bar{i}}D_{z}\rho^{i}_{\bar{z}}g_{i\bar{i}} + i\rho^{\bar{i}}_{z}D_{\bar{z}}\chi^{i}g_{i\bar{i}} + R_{i\bar{i}j\bar{j}}\chi^{i}\rho^{\bar{i}}_{z}\rho^{j}_{\bar{z}}\chi^{\bar{j}}\right) = \nonumber \\
&=& \frac{it}{2}\int_{\Sigma}\{ Q_{A},V\} + \frac{t}{2}\int_{\Sigma}X^{*}(K) + (eq.\; of\; mot.\; for \;\rho) 
\end{eqnarray}

for

\[
V = g_{\bar{i},j}\left(\rho^{\bar{i}}_{z}\bar{\partial}X^{j} + \partial X^{\bar{i}}\rho^{j}_{\bar{z}}\right) 
\]

and $X^{*}(K)$ the pullback of the  K\"ahler form $K = -ig_{i\bar{j}}dz^{i}dz^{\bar{j}}$ through the map $X^{I}$. The important consequence from this expression is that the path integral depends on the complex structure of the target space only through terms $Q_{A}$-exact! That is, being the eventual operator insertions $Q_{A}$-closed as well as the vacuum, it is independent by the target space complex structure. It is in this sense that it is called a topological theory. Of course it mantains the dependence trough the K\"ahler moduli so in fact it would be better to call it semi-topological. The only issue can come from the terms proportional to the equations of motion for $\rho$ but we can modify the topological BRST transformations in such a way to make (\ref{0c}) without those terms; and not being $\rho$ present, as we will soon see, in the local physical operators of the theory, this does not effect the computations.
We pass now to briefly describe the physical local operators in the theory, that is those in the cohomology of $Q_{A}$. They can be formed only by the scalars $X$ and $\chi$. The most generic form, obeying $\chi$ the statistics of a fermion, is 

\begin{equation}
O_{W} = W_{I_{1},\dots,I_{n}} \chi^{I_{1}}\dots\chi^{I_{n}}
\end{equation}

and, to be in the BRST cohomology, it is immediate to check that $W$ should be in the de Rham cohomology for the target space and vice versa. That is there is an isomorphism between the $d$ operator in the target space and $Q_{A}$. 

The last point I want to focus on for the moment is about zero modes in the path integral. Let us define $p$ as the number of the $\chi^{i}$ and $\chi^{\bar{i}}$ zero modes and $q$ the corresponding one for the $\rho^{i}_{\bar{z}}$ and $\rho^{\bar{i}}_{z}$ ones. The Riemann Roch theorem tells us that $p-q=d(2-2g)$ where $d$ is the complex dimension of the target space and $g$ the genus of the worldsheet ( presently we will consider only closed surfaces ). For having non zero amplitudes we should soak up these zero modes with appropriate insertions of physical operators and, being usually $p-q>0$, the amplitude will be nonzero only if there are enough $\chi$'s in the path integral insertions. 

\subsection{B-Model}\label{0sezcit}

The topological charge is 

\begin{equation}
Q_{B} = Q^{+} + \bar{Q}^{+}
\end{equation}

and the fermions split into scalars, $\psi^{\bar{i}}_{+}$ and $ \psi^{\bar{i}}_{-}$ renamed for convenience $\eta^{\bar{i}} \equiv \psi^{\bar{i}}+\psi^{\bar{i}}$ and $\theta_{i}\equiv g_{i\bar{i}}\left(\psi^{\bar{i}}-\psi^{\bar{i}}\right)$ , and one forms, $\psi^{i}_{+} \rightarrow \rho_{z}^{i},\; \psi^{i}_{-} \rightarrow \rho_{\bar{z}}^{i}$. The BRST transformations will be 

\begin{eqnarray}\label{0d}
\delta X^{i} &=& 0\nonumber \\
\delta X^{\bar{i}} &=&  i\epsilon\eta^{\bar{i}} \nonumber \\
\delta \theta_{i} &=& 0 \nonumber \\
\delta \eta^{\bar{i}} &=& 0 \\
\delta \rho^{i}_{\bar{z}} &=& -\epsilon\bar{\partial} X^{i}  \nonumber \\
\delta \rho^{i}_{z} &=&  -\epsilon\partial X^{i} \nonumber 
\end{eqnarray}

Also in this case these transformations square to zero on shell ( but it is possible an off shell formulation ). And also in this case we can rewrite the action in a more useful form

\[
S_{B} =  \frac{t}{2}\int_{\Sigma}d^{2}z \frac{1}{2}g_{IJ}\partial X^{I}\bar{\partial} X^{J} + i\eta^{\bar{i}}\left(D_{z}\rho^{i}_{\bar{z}} +D_{\bar{z}}\rho^{i}_{z} \right)g_{i\bar{i}} + i\theta_{i}\left(D_{\bar{z}}\rho^{i}_{z} - D_{z}\rho^{i}_{\bar{z}} \right) +  
\]
\begin{equation}\label{0e}
+ R_{i\bar{i}j\bar{j}}\rho^{i}_{z}\eta^{\bar{i}}\rho^{j}_{\bar{z}}\theta_{m}g^{m\bar{j}} 
= \frac{it}{2}\int_{\Sigma}\{Q_{B},V\} +  \frac{t}{2}\int_{\Sigma}W
\end{equation}
 
with 

\[
V = g_{i\bar{i}}\left(\rho^{i}_{z}\bar{\partial}X^{\bar{i}} + \rho^{i}_{\bar{z}}\partial X^{\bar{i}} \right) 
\]

and

\[
W =  \int_{\Sigma}\left(-\theta_{i}\left(D_{z}\rho^{i}_{\bar{z}} - D_{\bar{z}}\rho^{i}_{z}\right) + R_{i\bar{i}j\bar{j}}\rho^{i}_{z}\eta^{\bar{i}}\rho^{j}_{\bar{z}}\theta_{m}g^{m\bar{j}}\right)
\]

It is a little more difficult than in the A-model, but also here it is possible to show a semi independence by target space moduli. In particular under a change of K\"ahler form the variation of $W$ can be shown to be $Q_{B}$-exact. Obviously there is dependence by the target space complex structure as it is manifest already from (\ref{0d}).

In analogy with what already done for the A-model we can write down the most generic local operator in the cohomology of $Q_{B}$:

\begin{equation}\label{0i}
O_{W} = W_{\bar{i}_{1},\dots,\bar{i}_{n}}^{\;\;\;\;\;\;\;\;\;\;j_{1}\dots j_{m}}\eta^{\bar{i}_{1}}\dots\eta^{\bar{i}_{n}}\theta_{j_{1}}\dots\theta_{j_{m}}
\end{equation}

with $W$ a $(0,n)$ form in the $\bar{\partial}$-cohomology of the target space with values in $\wedge^{m}T^{1,0}X$ where $X$ is our target space and $TX$ its tangent space. Now it is a theorem that, in every Calabi Yau manifold ( we will soon see that for the B-model it is required also the Calabi Yau condition ), it exists an isomorphism between $(0,n)$ forms with values in $\wedge^{m}T^{1,0}X$, $\Omega^{(0,n)}\otimes\wedge^{m}T^{1,0}X$, and $(d-m,n)$ forms, where $d$ is the complex dimension of $X$. This isomorphism is given by the contraction with the, up to a scale factor, unique holomorphic $d$ form existing on the Calabi Yau $X$. So the space of physical operators is in one to one correspondence with $H_{\bar{\partial}}^{d-m,n}(X)$. 

If we consider the twisting procedure (\ref{0b}) and we specialize to the case of the $\frac{1}{z^{2}}$ and $\frac{1}{\bar{z}^{2}}$ coefficients of a Laurent expansion, we find

\[
L_{0} \rightarrow L_{0} + \frac{1}{2}J_{0} \;\;\; \bar{L}_{0} \rightarrow \bar{L}_{0} + \frac{1}{2}\bar{J}_{0}
\]

and this corresponds to the twisting of the generator of Euclidean rotations on the Riemann surface ( the Wick rotated Lorentz group ) by the generator of the axial R-symmetry current. But the $U(1)_{A}$ has an anomaly proportional to the first Chern class and so, while for the $A$-model the Calabi Yau requirement was only for conformal invariance, here it is necessary for the very existence of the $B$-model itself. Finally we have a zero mode condition for correlation functions such that, being $p_{a}$ the number  of $\eta$ zero modes from the operator $a$ and $q_{a}$ the number  of $\theta$ zero modes, then

\[
\sum_{a}p_{a} = \sum_{a}q_{a} = d(1-g) 
\]

where the requirement of equal number of left-right zero modes comes from a discrete left-right ghost number symmetry hidden by the choice of a single $\epsilon$ parameter in (\ref{0d}) and the mixed scalar fermions.
 
\subsection{Fixed point theorem}

We want here to describe a intriguing idea due to Witten \cite{Wittentop2} for showing how the path integral for the $A$ and $B$ model reduces essentially over an integral over, respectively, holomorphic maps and constant maps. This is usually shown starting from the two actions in the shape (\ref{0c}),(\ref{0e}) and going to the classical $t\rightarrow\infty$ limit. But the following derivation is much simpler and more elegant.

Consider some symmetry $Q$ of a theory acting without fixed points. Then being the action and, if any, also the operator insertions $Q$-invariants, you can split the path integral on the functional space $S$ into the one over the fiber of $Q$ times the coset $S/Q$:

\[
\int_{S}O_{1}\dots O_{n} e^{-S} = \int_{Q}\int_{S/Q}O_{1}\dots O_{n} e^{-S}  
\]
  
The volume of the symmetry group $\int_{Q}$ is some general prefactor but, if the symmetry is fermionic, by definition of Grasmannian integral it is zero. So the path integral itself is zero. If now we allow the presence of fixed points the result easily generalizes to the statement that the path integral collapses to the integral over the fixed locus. So the only thing we have to do is to look at the fixed locus of the $Q_{A}$ and $Q_{B}$ symmetries (\ref{0f}),(\ref{0d}). For the $A$ model this is ( other than $\chi^{I} = 0$ ) $\partial X^{\bar{i}} = \bar{\partial}X^{i} = 0$ thus holomorphic maps. While for the $B$ model $\partial X^{i} = \bar{\partial}X^{i} = 0$ ( and $\eta^{\bar{i}} = 0$) so constant maps!

\subsection{Integration on the metric}

Because in the following we will concentrate on the $B$ model let us use that explicit example. In any case, most of the things we are going to say can be translated in $A$ model language. Until now what we have done is a path integral over the embedding maps $X^{I}:\Sigma\rightarrow X$; so just a usual quantum field theory. Now we want to pass to a string theory and for doing so we have to extend the path integral to the worldsheet metric. In bosonic string theory, because of the symmetries of the action, the integral over the metric reduces to an integral, with appropriate measure, over the nonequivalent complex structures of the Riemann surface modulo, on genus zero and one, the conformal Killing vector symmetries. Because of the similarity with the structure of the bosonic string, amplitudes in topological string are defined in the same way, translating the bosonic string objects into the topological ones, following the dictionary given in (\ref{0g}). In particular the Beltrami differentials $\mu$ are now folded with the $b$ and $\bar{b}$ ghosts topological string analogues, that is the spin two supercurrents:

\begin{equation}\label{0h}
\prod_{k=1}^{3g-3}\int_{\Sigma}b\mu_{k} \int_{\Sigma}\bar{b}\bar{\mu}_{k}\leftrightarrow\prod_{k=1}^{3g-3}\int_{\Sigma}G^{-}\mu_{k} \int_{\Sigma}\bar{G}^{-}\bar{\mu}_{k}
\end{equation}
 
We have seen that the $c$ ghost has no analogue. So for fixing the conformal Killing vector symmetries it will be enough to fix some positions for vertex operator insertions without the $c$ ghost contraction. This means that we will have to deal, as usual, with sphere three point amplitudes etc...
In quantum field theory the fact that we have to soak up zero modes can be translated in a $U(1)_{A}$ anomaly statement with charge anomaly of ( on a Calabi Yau ) $d(2-2g)$. For $g>1$ surfaces there is no way to balance it with local operator insertions because they do not contain negative R-charge fermions, the $\rho^{i}$'s. But now, in string theory, our amplitudes are defined with the insertion of (\ref{0h}), and both $G^{-}$ and $\bar{G}^{-}$ carry negative R-charge. In particular, for the case $d=3$ we can have non vanishing amplitudes at whatever genus!

\section{Holomorphic Anomaly Equation}

In this section we are going to describe a series of powerful recursive relations which should be satisfied by topological amplitudes. Both the derivation of these constraints and the physics involved, their actual meaning, the implications and their interpretation are beautiful and rich subjects and describe, as we will see, something novel with respect to the usual string theory.
 
\subsection{Deformations}

We have seen what are the local operators in the cohomology of the topological charge $Q_{B}$, but those bring with them always some positive R-charge so, as long as we want to maintain R-symmetry an actual symmetry, we cannot insert them in higher genus amplitudes. Still there is a way of defining correlations functions but for non-local operators. To this goal consider the following descent equations for $O^{[1]}$ and $O^{[2]}$, where $O^{[0]}$ is a physical operator like (\ref{0i}) with $n=m=1$ and, in the meantime, $d$ is the worldsheet de Rham differential:

\[
0 = [Q_{B},O^{[0]}] 
\]
\[
dO^{[0]} = \{Q_{B},O^{[1]}\} 
\]
\[
dO^{[1]} = [Q_{B},O^{[2]}] 
\]
\[
dO^{[2]} = 0
\]

It should be already clear that  $O^{[1]}$ and $O^{[2]}$ are respectively a one and a two worldsheet forms. The fact that it exist a solution for the equation relies on the topological invariance of the theory with respect to the worldsheet metric which in turns implies independence of the correlation functions by the positions of the local operator insertions. And this is true because the worldsheet energy momentum tensor of the theory is, due to the supersymmetry algebra, $Q_{B}$ ( $Q_{A}$ ) exact; because a derivative with respect to the worldsheet metric of some correlation function means the insertion in the amplitude of an exact operator, that should give zero \footnote{ in fact after the passage to string theory, integrating over metrics, this is no longer true because of the insertions of the Jacobians (\ref{0h}) whose effect is to give contributions from the boundary of the moduli space as we will soon see. Nevertheless the following ideas can be implemented in the quantum field theory framework where the argument is strict }. So

\[
0 = d\langle O^{[0]}(z,\bar{z})\dots \rangle \;\; implies \;\;dO^{[0]} = \{Q_{B},O^{[1]}\} 
\]

The second step of the descent equation can be similarly justified looking at variations of amplitudes after a change in the one chain over which you are integrating $O^{[1]}$.
If now we integrate $O^{[p]}$ for $p=1,2$ over a $p$-cycle $C_{p}$ we obtain

\[
[Q_{B},\int_{C_{p}} O^{[p]}] = \int_{C_{p}} dO^{[p-1]}= \int_{\partial C_{p}} O^{[p-1]} = 0
\]

( we will later see that, even in the case of integration over the whole $\Sigma$, but for an open surface, the result still holds because of the boundary conditions on the $B$-model fields ). So we have some nice nonlocal operators $Q_{B}$ closed. Moreover the difference between two identical operators, only integrated over two different p-cycles whose difference is $C_{p}-C_{p}'=\partial B_{p+1}$, is an exact object in the shape $[Q_{B},\int_{B_{p+1}} O^{[p+1]}]$. It is even nicer the fact that, when $p=2$, the solution to the descent equation is a zero R-charge object: 

\begin{equation}\label{0l}
O^{[2]} = \{Q^{-},[\bar{Q}^{-},O^{[0]}]\} 
\end{equation}

Thus correlation functions of such operators are always possible. 
Not only that but we can also add the same objects directly to the action and obtaining in this way a full spectrum of deformed $B$-model theories. 

\[
S \rightarrow S + x^{i}\int_{\Sigma} \{Q^{-},[\bar{Q}^{-},O^{[0]}_{i}]\} 
\]

Note that a more generic kind of deformations is in principle possible, the ones with $O^{[0]}_{i}$ replaced by generic R-charge physical operators, and not only the marginal $(1,1)$ objects we have used until now. But this class of deformations breaks the R-symmetry of the Lagrangian and destroys conformal invariance, so in the future we will concentrate to the marginal case which will be identified because of the middle roman indexes $i, l,\dots$ parametrizing those operators. The generic case will be labelled by beginning roman indexes $a,b,\dots$.

It is a mathematical statement that the tangent space to the moduli space of complex structures ( or K\"ahler moduli for the $A$ model ) is parametrized by deformations belonging to $H_{\bar{\partial}}^{0,1}\otimes T^{1,0}X$, that is they are in correspondence with the class of operator in the $B$-model we are looking at, the marginal ones ( for the $A$ case the tangent space of deformations is parametrized by $H_{d}^{1,1}$ ). Then it is possible to show \cite{Wittentop2}, but easy to guess, that the deformed action just described corresponds exactly to another $B$ model ( resp. $A$ model ) with the original action but in a target space with deformed complex structure ( resp. deformed K\"ahler class ). 

When we have introduced the twisting of a supersymmetric $\sigma$ model we have considered a standard Lagrangian in the shape (\ref{0a}). However we can think to start from something more complicated with generic F-terms and then twist it. What we end up with is a more generic set of topological theories parametrized by these new chiral pieces; in detail the new Lagrangian we will consider is, before the twist,

\begin{equation}\label{0m}
S + \sum_{i}t^{i}\int_{\Sigma}O^{[2]}_{i} + \sum_{\bar{i}}\bar{t}^{\bar{i}}\int_{\Sigma}\bar{O}^{[2]}_{\bar{i}} 
\end{equation}
 
with $O^{[2]}_{i}$ the solutions (\ref{0l}) to the descent equation already discussed, but before the twist, that is with the two supercurrents still spin $\frac{3}{2}$ objects and with the fermions in (\ref{0l}) still spin $\frac{1}{2}$. The piece $\bar{O}^{[2]}_{\bar{i}}$ is the charge conjugated of $O^{[2]}_{i}$. Clearly, if we are going to twist (\ref{0m}) in the $A$ ( resp. $B$ ) way, we will include in (\ref{0m}) only terms such that $O^{[2]}_{i}$ will be the solution to the descent equation for that twist, that is containing the two novel spin two supercurrents after an $A$ ( resp. $B$ ) twist. The theory is parametrized by $t^{i},\bar{t}^{\bar{i}}$. Now we can apply the procedure described in the previous section and deform this theory with the zero R-charge non local operators. So in general we will have

\begin{equation}\label{0n}
S + \sum_{i}\left(t^{i} + x^{i}\right)\int_{\Sigma}O^{[2]}_{i} + \sum_{\bar{i}}\bar{t}^{\bar{i}}\int_{\Sigma}\bar{O}^{[2]}_{\bar{i}} 
\end{equation}

Note that if we had twisted in the antitopological way the deformations $x^{i}$ would have appeared as $\bar{x}^{\bar{i}}$ beside $\bar{t}^{\bar{i}}$. As $x^{i}$ correspond to an infinitesimal deformation in the point in moduli space around which the theory is, so $t^{i},\bar{t}^{\bar{i}}$ determine that point. That means that we will have a theory whose target space complex structure ( or K\"ahler form ) will end up to be in $t^{i} + x^{i},\bar{t}^{\bar{i}}$. As an aside comment note that while (\ref{0m}) was still an hermitean action, after the twist and the deformation (\ref{0n}) is no longer. In particular if it was true that  $\bar{t}$ was the complex conjugate of $t$, of course $t + x$ is a completely independent variable. 

\subsection{State operator correspondence}

We are going to describe, first in a mathematical abstract way, and then giving the physical interpretation, the correspondence between physical operators and vacua in the theory. Let us start by remembering the correspondence between topological charges and target space differential operators $d$ and $\bar{\partial}$. It exist a powerfull statement call Hodge decomposition, which decomposes differential forms into harmonic, exact and co exact pieces, and that can be applied to our case. In particular consider a fixed vacuum called $|0\rangle$ and a very generic operator, $\phi$, not necessarily in the cohomology of the topological charge. It is true that when we apply $\phi$ to $|0\rangle$ we obtain a state $|\phi\rangle_{0}$ uniquely defined by both the operator and the vacuum. We can apply Hodge decomposition to this state, with respect to the topological charge $Q$ ( either $Q_{A}$ or $Q_{B}$ ) having

\[
|\phi\rangle_{0} = |\phi_{H}\rangle_{0} + Q|\alpha\rangle_{0} + \bar{Q}|\beta\rangle_{0}
\]

where $\bar{Q}$ is the hermitean conjugate of $Q$, $|\phi_{H}\rangle_{0}$ is some harmonic state, that is annihilated by both $Q$ and $\bar{Q}$, and $|\alpha\rangle_{0}$ and $|\beta\rangle_{0}$ are generic states. If we now select $|\phi\rangle_{0}$ to be $Q$-closed, that is $[Q,\phi] = Q|0\rangle = 0$, it is true that 

\begin{equation}\label{0o}
0 = \left(Q|\phi\rangle_{0},|\beta\rangle_{0}\right) = \left(Q\bar{Q}|\beta\rangle_{0},|\beta\rangle_{0}\right) = \left(\bar{Q}|\beta\rangle_{0},\bar{Q}|\beta\rangle_{0}\right) 
\end{equation}

where $\left(\dots,\dots\right)$ is some norm with respect to which $\bar{Q}$ is the hermitean of $Q$. So, because of the properties of the norm, $\bar{Q}|\beta\rangle_{0} = 0$ and the decomposition of the state $|\phi\rangle_{0}$ is made out of an harmonic object plus a $Q$ exact one, that is its cohomology class is represented by a unique harmonic representative. The point is that an harmonic state is, by definition, annihilated by both $Q$ and $\bar{Q}$ and so, because of the supersymmetry algebra, it has zero energy. It is a vacuum. So the conclusion is that physical operators in the theory are in one to one correspondence with vacua. There is an explicit corresponding physical construction which we will now describe.

\begin{figure}[h!]
  \centering
\psfrag{SS}[][][0.8]{$\phi_{a}$} \psfrag{SS1}[][][0.8]{} \psfrag{SS2}[][][0.8]{$|a\rangle_{0}$}
\psfrag{T}[][][0.8]{$T\rightarrow\infty$}
  \includegraphics[width=0.5\textwidth]{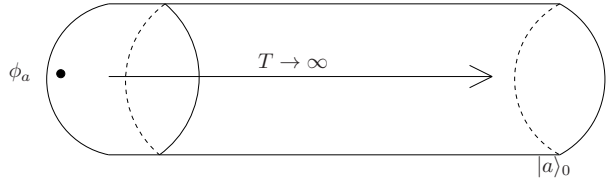}
\caption{State-operator correspondence via a geometrical construction}\label{0fig1}
\end{figure}

Consider a hemisphere as a local patch of some Riemann surface. On this hemisphere you can insert some vertex operator which will propagate along the worldsheet time and will describe a string state. In order to produce a vacuum we stretch the ending piece of the hemisphere as a long tube of length $T$; associated to it there will be a long time evolution of the state which can be described by the usual exponential of the worldsheet Hamiltonian $e^{-HT}$. In the limit $T\rightarrow \infty$ everything with non zero energy is suppressed and the empty hemisphere can be interpreted as a vacuum state $|0\rangle$. Instead applying the $Q$-closed operator $\phi_{a}$ on the tip of the hemisphere acts producing a state $|a\rangle_{0}$ $Q$-closed as well. This is shown in picture (\ref{0fig1})  

It seems at first that it exists a problem with this definition, that is if you consider for example marginal operators, these have one right moving and one left moving fermion. So natural boundary conditions require the state to be in the NS-NS sector while we would like a vacuum to be R-R. But here is the place where the topological twist enters to save the day and, magically, the two fermions, while conserving the statistic, become worldsheet bosons. 

Going back to our previous notation we can say that every $O_{a}$ is associated with a vacuum $|a\rangle$. As special cases we have marginal operators $O_{i}\rightarrow|i\rangle$ and the identity operator,$Id$, whose insertion on the Riemann surface is obviously irrelevant, and which is mapped onto the originally chosen vacuum $|0\rangle$. From now on we will drop the ``$0$`` from the notation leaving implicit a choice of some original vacuum corresponding to $Id$. Out of the states so created we can construct two metrics. The first one is defined as

\begin{equation}
\eta_{ab} = \langle a|b \rangle 
\end{equation}

and can be seen as a sphere ( two sewn hemispheres ) with two operator insertions. The norm $\langle\dots|\dots\rangle$ is different from the one in (\ref{0o}) as moving around a scalar charge in this case does nothing, so that $Q$ can pass untouched from one side to the other, while in the previous case it would have changed, by definition, to $\bar{Q}$.  Because of this property $\eta_{ab}$ is well defined as a metric between the two vacua also without the necessity of an infinitely long tube in the middle; this comes from the fact that every $Q$ exact piece eventually added on one side is killed by the closed state on the other. Being a sphere with two insertions the R-symmetry selection rules forces the left $+$ right charge to be equal to $3$, so two marginal insertions are not possible. We can construct also another metric \cite{Cecotti:1991me} made out sewing an hemisphere topologically twisted, that is over which the scalar charge is $Q$, and another antitopologically twisted, with scalar charge $\bar{Q}$. The equator will correspond to the ''border'' separating the topological from the antitopological part. If one is asking how practically does it work the twisting procedure on a patch of Riemann surface, the answer can be visualized in term of (\ref{0t}), that is we can imagine the presence of a background gauge field appropriately coupled with charged objects. When the twist changes this coupling changes. Obviously the scalar topological charge on one side becomes a spin one object on the other so it can move freely only on its side, and vice versa. Thus for defining a metric between two vacua here we need to have a long neck in the middle. Further these strange objects have no reason to satisfy the usual R-symmetry rules for the sphere, and in fact they do not. Only it is required the same opposite charge on both sides. The definition is as follow

\begin{equation}
g_{a\bar{b}} = \langle a|\bar{b} \rangle 
\end{equation}  

In fact there is no reason to prefer the topological theory to the antitopological one so there should be a linear transformation between the topological base of all vacua $|a\rangle$ and the antitopological one $|\bar{b}\rangle$:

\begin{equation}
|a\rangle = M^{\bar{b}}_{a}|\bar{b}\rangle
\end{equation}

CPT transformations in 2 dimensional theories exchanges $|a\rangle$ with $|\bar{b}\rangle$ and expresses $|\bar{b}\rangle = \bar{M}^{c}_{\bar{b}}|c\rangle$ so we have $M\bar{M} = 1$. Playing with $\eta$,$g$ and $M$ it is easy to find the expression

\begin{equation}
g^{\bar{a}b}\eta_{bc} = M^{\bar{a}}_{c} 
\end{equation}

The last element to introduce is the matrix of the chiral ring $C$. From the requirement that the OPE of two operators in the cohomology of $Q$ still belongs to it, we have

\begin{equation}\label{0cc}
O_{a}O_{b} = C^{c}_{ab}O_{c} + Q-exact \;\;pieces
\end{equation}

or, on the states,

\begin{equation}
O_{a}|b\rangle = C^{c}_{ab}|c\rangle
\end{equation}

So clearly it is correct the interpretation of $C_{abc}$ as a three point function on the sphere ( once the R-charge anomaly is satisfied ), because

\[
C_{abc} \equiv \eta_{ad}C^{d}_{bc} =  \langle a|d\rangle C^{d}_{bc} = \langle a|O_{b}|c\rangle = \langle 0|O_{a}O_{b}O_{c}|0\rangle
\]

We can now try to define a connection describing how the vacuum states $|a\rangle$ and $|\bar{a}\rangle$ behave when we move around the space of moduli, that is we are fibering the vacuum bundle over a base space which is the moduli space of the theory, this last one parametrized by coordinates $t,\bar{t}$. The standard definition is

\begin{equation}\label{0conntt}
A_{\alpha\beta\gamma} = \langle \beta |\partial_{\alpha}|\gamma\rangle 
\end{equation}

where Greek indexes can run on both barred, $a$, and unbarred, $\bar{a}$, ones. If we want to raise indexes in the connection we should use the appropriate metric to the twist we have locally decided to do. So for example

\[
A^a_{bc} = g^{a\bar{d}}A_{b\bar{d}c} = \eta^{ad}A_{bdc} 
\]

This connection has the property of giving a covariant derivative whose action on some vacua is orthogonal to every other vacua ( of true zero energy ), because

\[
\langle a |D_{b}|c\rangle = \langle a |\partial_{b} - A_{b}|c\rangle = 0
\]
  
This is also equivalent to the property of being a metric connection, that is such that the metric is covariantly constant. This can be explicitly checked, for example for the metric $g_{a\bar{b}}$:

\[
D_{a}g_{b\bar{c}} = \partial_{a}g_{b\bar{c}} - A^{d}_{ab}g_{d\bar{c}} - A^{\bar{d}}_{a\bar{c}}g_{\bar{d}b}  = \partial_{a}g_{b\bar{c}} - (\partial_{a}\langle b|)|\bar{c}\rangle - \langle b|\partial_{a}|\bar{c}\rangle = 0
\]

where the last equality follows from the very definition of $g_{b\bar{c}} = \langle b|\bar{c} \rangle $ and remembering that the derivative effectively follows the usual Leibniz rule also for $\langle b|\bar{c} \rangle $, because deriving means the insertion of a state on the whole worldsheet and so, in particular, on the sum of the left and right hemispheres of $g$. Another property we will use is that mixed indexes connections are zero. For example

\[
A^{\bar{c}}_{\bar{a}b} = g^{\bar{c}d}\langle d |\partial_{\bar{a}}|b\rangle  
\]
 
but 

\[
\langle d |\partial_{\bar{a}}|b\rangle = \langle d |[Q,O_{a}]|b\rangle = \langle d |QO_{a}|b\rangle = 0
\]

because $Q$, being scalar on both the hemispheres, can be moved on the other side and killed by $|d\rangle$. If we were going to analyse $A^{c}_{\bar{a}b}$ it would have sufficed to lower the index $c$ with $\eta$ and apply the same argument. Clearly this does not work for $\partial_{a}|b\rangle$ being $\bar{Q}$ a charge one object we cannot move around.
Now, as usual for metric connections, we have formulas for the non vanishing components:

\[
A^{c}_{ab} = g^{c\bar{d}}\partial_{a}g_{b\bar{d}}\;\;\;A^{\bar{c}}_{\bar{a}\bar{b}} = g^{\bar{c}d}\partial_{\bar{a}}g_{\bar{b}d}
\]

The key property about the geometry of the moduli space relies on a set of equations called the tt* equations, \cite{Cecotti:1991me}, and relating commutators of the covariant derivatives analysed with the ones for the matrices $C$ of (\ref{0cc}). These equations are

\begin{equation}\label{0tt}
\left[D_{a},D_{\bar{b}}\right] = -\left[C_{a},\bar{C}_{\bar{b}}\right]
\end{equation}
\[
\left[D_{a},D_{b}\right] = \left[D_{\bar{a}},D_{\bar{b}}\right] = 0
\]

with $C^{a}_{\bar{b}c}=g^{c\bar{c}}\bar{C}^{\bar{a}}_{\bar{b}\bar{c}}g_{\bar{a}a}$. We will not derive them, computations are clear enough in the original paper, but only explain one key point needed in the proof. In fact the basic idea is simply to interpret derivatives as insertion of operators. Then you have to play with the various supercharges involved and the supersymmetry algebra in order to obtain the result. The key point that differentiate this from analogous computations we will do later for the holomorphic anomaly equation is that here we will discard eventual contact terms. Contact terms are singular coefficients that can appear in the OPE of two operators and that should be appropriately regulated. In fact, when you insert an operator integrated over the entire Riemann surface, you should regularize the procedure avoiding it approaches too much other operator insertions. And this is part of the definition for an operator insertion. Here we will not care about this point because of the long tube limit we are implicitly taken for $g$ that can spread apart whatever close contact term present before the limit.

The very last point for this section is the relation between the metric $g_{i\bar{j}}$ and the K\"ahler metric on the moduli space we will soon define. We will restrict ourself on marginal directions because we want to look at the tangent space of the moduli space preserving conformal invariance. Let us start from the definition

\begin{equation}\label{0gg}
G_{i\bar{j}}=\frac{g_{i\bar{j}}}{g_{0\bar{0}}} 
\end{equation}

The rational behind this definition is to have a metric independent of whatever redefinition of the vacuum $|0\rangle$ through an holomorphic function of the moduli space parameters, $|0\rangle\rightarrow f(t^{i})|0\rangle$, ( we can think to have fixed the antiholomorphic ambiguity after having chosen a topological twist because the vacua corresponding to $\bar{t}^{\bar{i}}$ coordinates are the antitopological ones and they do not mix with the others ). In fact the vacuum $|0\rangle$ can be seen as a section of a line  bundle $\cal{L}$ over the moduli space, the sphere amplitude given by $\langle 0| 0\rangle$ is a section of ${\cal{L}}^{2}$ and generic genus g amplitudes, because of their reduction to lower genus ones through sewing principles, as sections of ${\cal{L}}^{2 - 2g}$. In the same way the antitopological vacuum $|\bar{0}\rangle$ is considered a section of $\bar{\cal L}$ and similarly the amplitudes. In addition we define the function $K$ such that

\begin{equation}
e^{-K} = g_{0\bar{0}} 
\end{equation}

A chain of equalities says

\[
G_{i\bar{j}} =  \frac{g_{i\bar{j}}}{g_{0\bar{0}}} = g^{0\bar{0}}g_{i\bar{j}} = C^{l}_{i0}g^{0\bar{0}}C^{\bar{k}}_{\bar{j}\bar{0}}g_{l\bar{k}}
\]

where we have used $C^{i}_{j0} = \delta^{i}_{j}$ which is clear from the very definition (\ref{0cc}) remembering that the operator associated with $|0\rangle$ is the identity. Now we use the tt* equations and in particular the zero-zero component of $\left[D_{i},D_{\bar{j}}\right] = -\left[C_{i},\bar{C}_{\bar{j}}\right]$ which reads, after deleting all zero pieces and remembering that for charge conservation $g_{0\bar{k}} = g_{0\bar{0}}$

\[
C^{l}_{i0}g^{0\bar{0}}C^{\bar{k}}_{\bar{j}\bar{0}}g_{l\bar{k}} = \left[C_{i} ,g^{-1}\bar{C}_{\bar{j}}g\right]^{0}_{0} = -\partial_{\bar{j}}A^{0}_{i0} = -\partial_{\bar{j}}\left(g^{0\bar{0}}\partial_{i}g_{0\bar{0}}\right) = -\partial_{\bar{j}}\partial_{i}log(g_{0\bar{0}}) = \partial_{\bar{j}}\partial_{i}K
\]
 
So we have the result

\begin{equation}
\bar{\partial}_{\bar{j}}\partial_{i}K = G_{i\bar{j}}=g_{i\bar{j}}e^{K} 
\end{equation}

where of course $K$ is interpreted as the K\"ahler potential on the moduli space.

\subsection{Closed case}

\begin{figure}[h!]
  \centering
\psfrag{F1}[][][0.8]{$\bar{O}_{\bar{i}}$} \psfrag{G1}[][][0.8]{$G^-$} \psfrag{G2}[][][0.8]{$\bar{G}^+$}
\psfrag{G3}[][][0.8]{$G^+$}
  \includegraphics[width=0.8\textwidth]{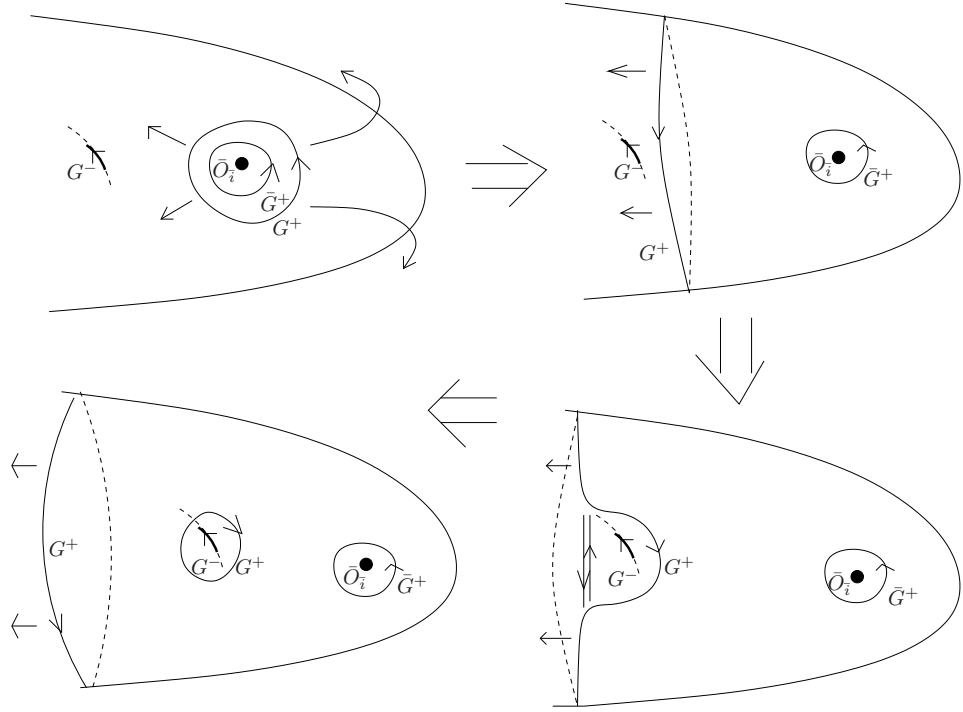}
\caption{How to implement a commutator on amplitudes using deformations of the corresponding currents on the Riemann surface}\label{00fig}
\end{figure}

We want now to discuss what is the dependence of the theory by the antiholomorphic parameters $\bar{t}^{\bar{i}}$ corresponding to the marginal directions. Naively deriving a path integral containing the action (\ref{0n}) with respect to $\bar{t}^{\bar{i}}$, gives as a result the insertion of a $Q$-exact object in the correlation function, and we know it is zero. However here we are really talking about topological string theory and the path integral contains, by definition, the insertions (\ref{0h}) as well. And these insertions will make the story more interesting. So let us derive a genus $g$ amplitude, for now without vertex operator insertions, with respect to $\bar{t}^{\bar{i}}$. As already said it corresponds to the insertion of the operator 

\[
\int_{\Sigma}\bar{O}^{[2]}_{\bar{i}} = \int_{\Sigma}[Q^+,\{\bar{Q}^+,\bar{O}_{\bar{i}}\}]
\]

and we can substitute the charge commutators with circle integrals of the corresponding currents around the position of the operator $\bar{O}_{\bar{i}}$. So we have

\[
\bar{\partial}_{\bar{i}}{\cal F}^{g}= \int_{{\cal M}_{g}}[dm]\langle \int d^{2}z\oint_{C_{z}} G^+\oint_{\tilde{C}_{z}}\bar{G}^+\bar{O}_{\bar{i}}(z)\prod_{k=1}^{3g-3}\int_{\Sigma_{g}}G^{-}\mu_{k} \int_{\Sigma_{g}}\bar{G}^{-}\bar{\mu}_{k}\rangle_{\Sigma_{g}}
\]

where $\langle\dots\rangle_{\Sigma_{g}}$ is the usual quantum field theory amplitude computed on $\Sigma_{g}$, and ${\cal M}_{g}$ is the moduli space of its nonequivalent complex structures. If we want to kill $Q^+$ and $\bar{Q}^+$ on the vacuum, that is geometrically deform $G^+$ and $\bar{G}^+$ till they reduce to circle integrals without anything with singular OPE inside, we should make them commuting with $G^-$ and $\bar{G}^-$ contained in the Jacobian measure. Again the geometrical interpretation is that $G^-$ and $\bar{G}^-$ are inserted all over $\Sigma_{g}$ and making $G^+$ and $\bar{G}^+$ pass across them gives as a result ( minus ) the commutator of the corresponding charges. This has been represented in (\ref{00fig}) 

and holds, after appropriate cutting and sewing of the currents, also for surfaces with handles. The result of this double commutator gives, because of the supersymmetry algebra, the holomorphic and antiholomorphic part of the energy momentum tensor $T = \frac{\partial S}{\partial g}$, contracted with the Beltrami differentials around which $G^-$ and $\bar{G}^-$ were integrated. And being $\mu_{\tilde{k}} = \frac{\partial g}{\partial m^{\tilde{k}}}$ then $T\mu_{\tilde{k}} = \frac{\partial S}{\partial m^{\tilde{k}}}$, that is a moduli derivative of the whole amplitude.

\[
\bar{\partial}_{\bar{i}}{\cal F}^{g}= \int_{{\cal M}_{g}}[dm] \sum_{\tilde{k},\bar{\tilde{k}}=1}^{3g-3} \frac{\partial}{\partial m^{\tilde{k}}}\frac{\partial}{\partial \bar{m}^{\bar{\tilde{k}}}}\langle \int_{\Sigma_{g}} \bar{O}_{\bar{i}}\prod_{k\neq \tilde{k}}\int_{\Sigma_{g}}G^{-}\mu_{k}\prod_{\bar{k}\neq \bar{\tilde{k}}} \int_{\Sigma_{g}}\bar{G}^{-}\bar{\mu}_{\bar{k}}\rangle_{\Sigma_{g}}
\]

So we see how the antiholomorphic derivative of ${\cal F}^{g}$ reduces to a contribution coming from the boundary of the moduli space. This boundary is nothing but the degeneration of $\Sigma_{g}$ when one non trivial cycle shrinks to zero size or, a conformally equivalent picture, when an handle ( with complex modulus $\tau$ ) goes in the limit of a long tube ( $Im\tau \rightarrow \infty$ ) \footnote{This is clear from the fundamental region of integration of $\tau$ on a torus, $-\frac{1}{2}\leq Re\tau \geq \frac{1}{2}$, $|\tau|\geq 1$ but with the lines at $\pm\frac{1}{2}$ and the two lines of the low arch identified, so that the only boundary is really at $Im\tau \rightarrow \infty$}. There are two cases: the cycle can be dividing or not, that is cutting along it you end up either with two surfaces with genus $g_{i} + g_{2} = g$ or with one surface with genus $g-1$. 

\begin{figure}[h!]
  \centering
\includegraphics[width=0.5\textwidth]{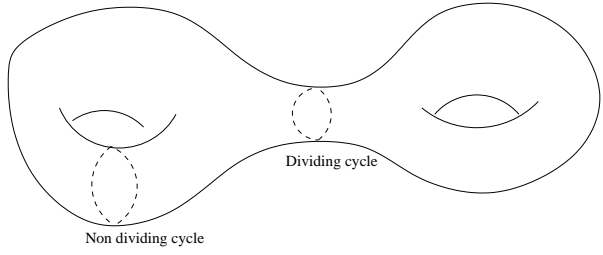}
\caption{Dividing and non dividing cycles.}\label{00figc}
\end{figure}

Let us analyse the second case first. When the limit is taken the degenerated Riemann surface has $3g-6$ complex moduli on the non degenerated cycles and two for the degenerated one, placed on the two insertion points $z^{1}$,$z^{2}$ of the long tube on the surface \footnote{see the discussion around (\ref{3f1})}; these two carry a Beltrami differential written as \cite{Polchinski_I} 

\begin{equation}\label{0bel}
 \oint_{C_{z^{i}}}G^{-}\oint_{\tilde{C}_{z^{i}}}\bar{G}^{-} 
\end{equation}

Having a double derivative in our computation, one of them will force us on the previously described boundary on the moduli space and we still remain with the other. This will be a derivative normal to the boundary and so, for the particular piece of moduli space we are considering, written as $\frac{\partial}{\partial Im\tau}$. As will be better explained in (\ref{sec:cap3}) there is a general way of sewing and cutting Riemann surfaces where you mimic the cut and paste along some non trivial cycle as a sum over a complete set of states times some metric. In the case of topological string it is enough to use the states $O_{j}$ ( for an explanation see (\ref{sec:cap3}) ) so that we will replace the long tube with an elongated sphere with two operator insertions, two additional insertions on $z^{1}$ and $z^{2}$, and two metrics. The (\ref{0bel}) Jacobian will be integrated around the operators $O_{j}$ while the two insertions on the sphere will be fixed. It remains to analyse the various positions of the operator $\bar{O}_{\bar{i}}$ originally integrated on the full Riemann surface. It can be shown that the contribution, when the domain of integration is restricted to the degenerate surface but not on the long handle, is vanishing; this both from a zero mode counting and because of the action of $\frac{\partial}{\partial Im\tau}$ on an amplitude independent by $Im\tau$. So we remain with ( $[dm']$ is the measure for the moduli space of the $g-1$ Riemann surface without the long handle)

\[
\bar{\partial}_{\bar{i}}{\cal F}^{g}= \frac{1}{2}\sum_{all\;\;handles}\int_{{\cal M}_{g-1}}[dm']\int d^{2}z^{1}d^{2}z^{2} \frac{\partial}{\partial Im\tau}  
\]
\[
\langle \int_{\Sigma_{g}^{Im\tau\rightarrow\infty}} \bar{O}_{\bar{i}}\oint_{C_{z^{1}}}G^{-}\oint_{\tilde{C}_{z^{1}}}\bar{G}^{-}\oint_{C_{z^{2}}}G^{-}\oint_{\tilde{C}_{z^{2}}}\bar{G}^{-}\prod_{k=1}^{3g-6}\int_{\Sigma_{g-1}}G^{-}\mu_{k} \int_{\Sigma_{g-1}}\bar{G}^{-}\bar{\mu}_{\bar{k}}\rangle_{\Sigma_{g}^{Im\tau\rightarrow\infty}} 
\]

with the $ \frac{1}{2}$ coming from the invariance of the moduli under exchange of $z^{1}$ with $z^{2}$. The second line becomes

\[
\langle\oint G^{-}\bar{G}^{-}O_{j}(z^{1})\oint G^{-}\bar{G}^{-}O_{k}(z^{2})\prod_{k=1}^{3g-6}\int_{\Sigma_{g-1}}G^{-}\mu_{k} \int_{\Sigma_{g-1}}\bar{G}^{-}\bar{\mu}_{\bar{k}}\rangle_{\Sigma_{g-1}}\cdot
\]
\[
\cdot g^{j\bar{j}}g^{k\bar{k}}\langle \bar{O}_{\bar{j}}\bar{O}_{\bar{i}} \bar{O}_{\bar{k}}\rangle_{sphere} 
\]

Because of the presence of $\bar{O}_{\bar{i}}$ on the sphere, the only non vanishing case is with the remaining two operators to be marginal and antitopological. So that the metric contracting them with $O_{j}$ and $O_{k}$ on the $g-1$ surface is the topological-antitopological one. Finally by definition the sphere three point insertion is our $\bar{C}_{\bar{j}\bar{i}\bar{k}}$. It remains to be analysed the situation when you have a dividing cycle. Working in the same way you end up with two Riemann surfaces of genus $g_{1}$ and $g_{2}$ such that $g_{1}+g_{2}=g$, each with one operator insertion, still the two metrics and the antitopological sphere with three insertions.

Having expressed $\partial_{\bar{i}}{\cal F}^{g}$ as a sum other lower genus amplitudes with insertions of marginal operators, we can think to rewrite these in terms of holomorphic derivatives of the empty amplitudes. We know that a derivative with respect to $t^{i}$ brings down the operator in the corresponding deformation, exactly in the shape it appears here: $\int d^{2}z\oint G^{-}\bar{G}^{-}O_{j}(z) = \int d^{2}z O^{[2]}(z) $, so it seems reasonable to write $\langle \int d^{2}z\oint G^{-}\bar{G}^{-}O_{j}(z) \dots \rangle_{\Sigma_{g}} = \partial_{j}\langle \dots \rangle_{\Sigma_{g}}$. However we have to take care of contact terms. Contact terms have a double explanation. Physically they are divergent terms in the OPE of two operators approaching each other that should be subtracted for a regularized amplitude. Mathematically they arise as connections for derivatives on fiber bundles. The computation is local and can be performed on a small patch of a Riemann surface that can be seen, because of the state-operator correspondence, as a state. The contact term will be given by the difference between the situation with one operator fixed on the local piece of the Riemann surface  and the other integrated all around, also close to the fixed one, and the case when the fixed operator is excluded by the close region of integration domain as should be in an already regularized amplitude. 

\[
\partial_{i}\left(O_{j}|0\rangle \right)- O_{j}\partial_{i}|0\rangle
\]

the first piece is

\[
\partial_{i}\left(O_{j}|0\rangle \right) = \partial_{i}|j\rangle = |\bar{k}\rangle g_{\bar{k}k}g^{k\bar{l}}\langle\bar{l}|\partial_{i}|j\rangle = A^{k}_{ij}g_{\bar{k}k}|\bar{k}\rangle = A^{k}_{ij}|\bar{k}\rangle \langle \bar{k}|k \rangle = A^{k}_{ij}O_{k}|0 \rangle
\]
 
and the second

\[
- O_{j}\partial_{i}|0\rangle = - O_{j}g^{l\bar{k}}g_{l\bar{n}}|\bar{n} \rangle \langle\bar{k} |\partial_{i}|0\rangle =- O_{j}g_{l\bar{n}}A^{l}_{i0}|\bar{n} \rangle
\]

but $A^{l}_{i0} = g^{l\bar{l}}\partial_{i}g_{0\bar{l}}= g^{0\bar{0}}\partial_{i}g_{0\bar{0}} = A^{0}_{i0}$ so

\[
- O_{j}g_{l\bar{n}}A^{l}_{i0}|\bar{n} \rangle = -A^{0}_{i0}O_{j}|0 \rangle
\]

and the difference is

\begin{equation}\label{0conn1}
\left(A^{k}_{ij}O_{k}-A^{0}_{i0}O_{j}\right)|0 \rangle = \left(g^{k\bar{j}}\partial_{i}g_{\bar{j}j} + \partial_{i}K\delta^{k}_{j} \right)O_{k}|0 \rangle = \Gamma^{k}_{ij}O_{k}|0 \rangle
\end{equation}

where $\Gamma^{k}_{ij}$ is the metric connection for $G_{i\bar{j}} = g_{i\bar{j}}e^{K}$ defined as $\Gamma^{k}_{ij} = G^{k\bar{l}}\partial_{i}G_{\bar{l}j}$.
So we conclude that the contact term of two operators approaching each other is given by the connection $\Gamma$. What does it happen if we try to apply a similar argument to the operator identity in itself? Then the contact term is given by 

\[
\partial_{i}|0 \rangle = A^{0}_{i0} = -\partial_{i}K 
\]

and the interpretation is in terms of contact terms with the background gauge field $A$ coupled to the R-symmetry currents in order to locally twist the theory. This background field is seen as part of the definition of the topological vacuum $|0\rangle$ dual to the identity operator. In this last case, when you subtract the contact term on a genus $g$ surface, there is an additional relative normalization of $2-2g$ to apply \cite{BCOV}. So all in all the regularized insertion of an operator $O_{i}$ on a genus $g$ amplitude with other operator insertions, is given by the covariant derivative

\begin{equation}\label{0cov}
D_{i} = \partial_{i} - \Gamma_{i} -(2 - 2g)\partial_{i}K 
\end{equation}

with as many contraction of $\Gamma_{i}$ as operators that are already present.\footnote{In the case of a Riemann surface with $h$ boundaries ( and even $c$ crosscaps ) the factor $2 - 2g$ will generalize to $2 - 2g -h -c$ due to contact terms with the background gauge field $A$ on the boundary of the Riemann surface ( or on the crosscaps ).} The mathematical interpretation of this is that the amplitude given by $n$ operator insertions on a genus $g$ surface is really a section of a bundle fibered over the moduli space of the theory. Fibers are given by ${\cal L}^{2g-2}\otimes Sym T^{n}$ where ${\cal L}^{2g-2}$ was already explained in terms of the ambiguity in the definition of $|0\rangle$ while $Sym T^{n}$ is the symmetrized product of the tangent to the moduli space, as it should be clear because of the idea that operators really parametrize deformations of the moduli space ( read tangent vectors ) and the amplitude is symmetric under exchange of the position.

Let us here clarify a point; the connection appearing in (\ref{0cov}) is obviously very similar to the one introduced in (\ref{0conntt}), and in fact the difference arises basically in the contact term with the ``vacuum``. Still we can borrow the commutator structure from the tt* equations (\ref{0tt}), as the part in $0$ index basically decoulpes, and we can ask ourselves: is it possible, because of the vanishing of the $(2,0)$ and $(0,2)$ part of the curvature, to chose a gauge where, when restricting ourselves to only holomorphic ( or antiholomorphic ) derivatives, the corresponding covariant derivative is replaced by an ordinary one? Indeed it is and the choice of coordinates where it holds are called Canonical. In term of the action deformations (\ref{0n}) this can be naturally achieved when $\bar{t}$ stays fixed. In other words the deformations parametrized by $x^{i}$ are only by topological operators and shift uniquely the value of $t^{i}$ leaving fixed $\bar{t}^{\bar{i}}$. In this case it always exist an appropriate choice of coordinates for the point $t,\bar{t}$ such that the insertion on the worldsheet of the operator $\int_{\Sigma}O^{[2]}_{i}$ can be obtained exactly deriving the amplitude with action (\ref{0n}) with respect to $x^{i}$ \cite{BCOV}. And in general we will assume that that choice has been done. Obviously, when we are considering the holomorphic anomaly equation where both $t$ and $\bar{t}$ are shifted, we are obliged to write covariant derivatives. Still the antiholomorphic one $D_{\bar{i}}$ is written as $\bar{\partial}_{\bar{i}}$ because there are no other antitopological operators on the worldsheet ( and this kills $\Gamma_{\bar{i}}$ ) and the Riemann surface is topologically twisted, so a section of ${\cal L}$ and not $\bar{\cal L}$ ( and this kills $\partial_{i}K$ being $A_{\bar{i}0}^{0} = 0$ ). 

We can then conclude with the Holomorphic anomaly equation for closed amplitudes

\begin{equation}\label{0hae}
\bar{\partial}_{\bar{i}}{\cal F}^{g} = \frac{1}{2}\bar{C}_{\bar{i}\bar{j}\bar{k}}g^{\bar{j}j}g^{\bar{k}k}\left(D_{j}D_{k}{\cal F}^{g-1} + \sum_{s=1}^{g-1}D_{j}{\cal F}^{s}D_{k}{\cal F}^{g-s} \right)
\end{equation}

This set of equations can be summarized by a single master equation. In order to do this let us introduce the generating functional

\begin{equation}
W(\lambda;t,\bar{t}) = \sum_{g=0}^{\infty}\lambda^{2g-2}{\cal F}^{g}
\end{equation}

Expanding in powers of $\lambda$ it is easy to check that 

\begin{equation}\label{0mast}
\bar{\partial}_{\bar{i}}e^{W} = \left( \frac{\lambda^{2}}{2}\bar{C}_{\bar{i}\bar{j}\bar{k}}g^{\bar{j}j}g^{\bar{k}k}D_{j}D_{k}\right)e^{W} 
\end{equation}

is equivalent to (\ref{0hae}).

One comment is in order: in (\ref{01}) we have seen a parallel between the structure of the bosonic string and the one of topological string. The natural question to ask now is, what is the corresponding equation in bosonic string theory? The answer is, nothing; or better, you do not have any antiholomorphic contribution. To understand why it is enough to analyse one of the differences between bosonic string and topological string, that is the non trivial cohomology of what correspond to the $b$ ghost, the spin two supercurrent $G^{-}$. If we had tried to reproduce the holomorphic anomaly equation for bosonic string at first it seems we would have succeeded. We can insert in the amplitudes the objects corresponding to $\int d^{2}z\oint_{C_{z}} G^+\oint_{\tilde{C}_{z}}\bar{G}^+\bar{O}_{\bar{i}}(z)$, given by some BRST exact operator and work in the same way. The commutator of the charge from the $b$ ghost current with the BRST charge gives again the energy momentum tensor, the amplitudes are still defined with the right Beltrami differentials insertions, and you can easily deduce that the contribution has again to come from the boundary of the moduli space. But remember that in topological string the requirement for $\bar{O}_{\bar{i}}$ is to be $\bar{Q}$-closed ( because it is the antitopological physical operator ), and it translates in bosonic string in being $\oint b$-closed. But being the $b$-cohmology vanishing this means that it should be $b$-exact. And, because the only nonvanishing contribution come from the case when the $\bar{O}_{\bar{i}}$ insertion is on the degenerated handle, it is projected to the vacuum state, which is vanishing being every state with zero energy momentum tensor also vanishing against the $b$ ghost. So the contribution is zero. 

It is possible to repeat all the procedure for the most generic case of the holomorphic anomaly from closed amplitudes with $n$ previous marginal operator insertions of the usual conformal symmetry preserving type: 

\[
\bar{\partial}_{\bar{i}}D_{i_{1}}\dots D_{i_{n}}{\cal F}^{g} = \partial_{\bar{i}}{\cal F}^{g}_{i_{1}\dots i_{n}}
\]

The novel point is a second type of moduli boundary which arises when the operator $\bar{O}_{\bar{i}}$ approaches one of the already existing operators $O_{j}$. There is then a contact term of the type

\[
\bar{O}_{\bar{i}}(z)O_{j}(w) \sim \frac{G_{\bar{i}j}}{|z-w|^{2}}
\]

different from the usual kind of contact terms between two topological operators. That one is given by the connection $\Gamma$ which here vanishes because $\Gamma_{j\bar{i}}^{k} = 0$, while this one would vanish in that case because $G_{ij}=0$. In any case generalizing the same kind of arguments already analysed you can conclude with the equation ( $\sigma$ belongs to the permutation group of $n$ elements )

\[
\bar{\partial}_{\bar{i}}{\cal F}^{g}_{i_{1}\dots i_{n}} = \frac{1}{2}\bar{C}_{\bar{i}\bar{j}\bar{k}}g^{\bar{j}j}g^{\bar{k}k}\left({\cal F}^{g-1}_{jki_{1}\dots i_{n}} + \sum_{s=0}^{g}\sum_{r=0}^{n}\frac{1}{r!(n-r)!}\sum_{\sigma}{\cal F}^{s}_{ji_{\sigma(1)}\dots i_{\sigma(r)}}{\cal F}^{g-s}_{ki_{\sigma(r+1)}\dots i_{\sigma(n)}}\right) -
\]
\begin{equation}\label{0haei}
- \left(2g-2+n-1 \right)\sum_{r=0}^{n}G_{\bar{i}i_{r}}{\cal F}^{g}_{i_{1}\dots i_{r-1}i_{r+1}\dots i_{n}}
\end{equation}

where you can immediately distinguish the novel piece with the right normalization ( which can be worked out in the same way it was done for the vacuum contact term ) and the fact that now the index $s$ in the sum runs also on the value zero, because with operator insertions we can have non vanishing genus zero amplitudes, and in detail

\[
{\cal F}^{0}={\cal F}^{}_{i_{1}}={\cal F}^{0}_{i_{1}i_{2}}=0
\]
\[
{\cal F}^{0}_{i_{1}\dots i_{n}}=D_{i_{1}}\dots D_{i_{n-3}}{\cal F}^{0}_{i_{n-2}i_{n-1}i_{n}}=D_{i_{1}}\dots D_{i_{n-3}}C_{i_{n-2}i_{n-1}i_{n}}
\]
 
Trying to apply equation (\ref{0hae}) to the case at genus one doesn't work correctly. First because one loop amplitudes are well defined with one vertex operator already fixed, second because of the contact terms between this vertex operator and the one brought down from derivation, which are similar to the ones encountered in (\ref{0haei}) but with different coefficient due to the particular structure of one loop amplitudes. The result is \cite{Bershadsky:1993ta}

\begin{equation}\label{0haeo}
\bar{\partial}_{\bar{i}}\partial{i}{\cal F}^{1} = \frac{1}{2}\bar{C}_{\bar{i}\bar{j}\bar{k}}g^{\bar{j}j}g^{\bar{k}k}C_{ijk} - \frac{\chi}{24}G_{i\bar{j}} 
\end{equation}

where $\chi$ is the Euler characteristic of the Calabi Yau\footnote{and we have used $\partial_{i}$ instead of the covariant derivative because without operator insertions and at genus one the connection in (\ref{0cov}) vanishes.}.

The very last point to see is a way of rewriting the full set of equations (\ref{0haei}) and (\ref{0haeo}) in one single expression generalizing (\ref{0mast}). Later we will slightly change the way it looks but for now let us define, in its original form, the following object:

\begin{equation}
W(\lambda,x;t,\bar{t}) = \sum_{g,n =0}^{\infty}\frac{\lambda^{2g-2}}{n!}{\cal F}^{g}_{i_{1}\dots i_{n}}x^{i_{1}}\dots x^{i_{n}} + \left(\frac{\chi}{24}-1 \right)log\lambda 
\end{equation}
  
where $x^{i}$ is an expansion parameter and $t,\bar{t}$ are the same as in (\ref{0n}); $\lambda$, weighting the sum of the various amplitudes, is the string coupling constant. The interpretation of $W(\lambda,x;t,\bar{t})$ as the partition function computed in the moduli space point $t + x,\bar{t}$ is natural in canonical coordinates. We can now reproduce both (\ref{0haei}) and (\ref{0haeo}) in a single master equation just expanding in powers of $x^{i}$ and $\lambda$ the following

\begin{equation}\label{0haem}
\bar{\partial}_{\bar{i}}e^{W} = \left( \frac{\lambda^{2}}{2}\bar{C}_{\bar{i}\bar{j}\bar{k}}g^{\bar{j}j}g^{\bar{k}k}\frac{\partial^{2}}{\partial x^{i}\partial x^{j}} - G_{\bar{i}j}x^{j}\left[\lambda\frac{\partial}{\partial\lambda} + x^{k}\frac{\partial}{\partial x^{k}}\right]\right)e^{W}
\end{equation}

\subsection{An interpretation}
 
In this section we want to briefly review an utmost interesting interpretation given by Witten in \cite{witten-ind} to the holomorphic anomaly equation. We will be very brief as the original paper is already clear and well written, and we just need to roughly describe the idea in order to look for the possible analogue in the open case. The discussion is limited to the B-model and cannot directly be applied to the A-model: it will be soon clear why.

Let us start saying that, if we want to quantize a theory containing a symplectic linear space of variables ${\cal W}$, of dimension $2n$, in general quantum mechanics enters in the shape of the uncertainty principle and tells us that the Hilbert space will depend only by $n$ of these. Which ones is our choice of the initial conditions and mathematically we can translate this in a choice of complex structure over ${\cal W}$. This should be clear as a choice of complex structure reflects in a splitting in an equal number of holomorphic and antiholomorphic coordinates. However, if the theory itself depends on the complex structure, moving in the moduli space reflects in changing the quantization, and we should find a way for relating the different possibilities. In detail we need a flat connection on the quantum Hilbert space fibered over the moduli space of complex structures, such that we can parallel transport the fibers from one point to the other. The requirement for the physical states to be killed by this connection ( parallel transport ) will turn out to be, for our specific case, exactly the holomorphic anomaly equation! 

A little more in detail let us introduce the prequantum line bundle ${\cal H}_{0}$ consisting of squared integrable functions of the $2n$ variables in ${\cal W}$, and the quantum Hilbert space ${\cal H}_{J}$ whose elements are sections of only $n$ variables, identified after a choice of complex structure $J$ in the moduli space the theory depends on: ${\cal M}$. We also define ${\cal H}_{Q}$ as a fiber bundle over the base space ${\cal M}$, whose fiber is really ${\cal H}_{J}$ itself. Obviously ${\cal H}_{Q}$ will be a subbundle of ${\cal M}\times{\cal H}_{0}$ and we want to find a flat connection $\tilde{\cal D}$ on it, as a projection from a connection on ${\cal M}\times{\cal H}_{0}$, such that it will annihilate the physical states allowing us to identify different ${\cal H}_{J}$ by parallel transport. We can be more specific and write down in a general case what this connection is supposed to be. A flat connection on ${\cal M}\times{\cal H}_{0}$ is clearly

\[
{\cal D} = \sum_{I,J} dJ^{I}_{J}\frac{\partial}{\partial J^{I}_{J}} 
\]

to ask that this object project to a connection on ${\cal H}_{Q}$ we should require that its action on a section gives another section still belonging to ${\cal H}_{Q}$. Let us chose the fiber ${\cal H}_{J}$ to be the one of holomorphic functions $\psi$,

\[
D_{\bar{i}}\psi|_{J} = \left(\frac{\partial}{\partial \bar{x}^{\bar{i}}} + \frac{i}{2}w_{\bar{i}j}x^{j}\right)\psi|_{J} = 0 
\]
 
with $D_{\bar{i}}$ the covariant derivative from the symplectic two form $w$. We require its commutator with $\tilde{{\cal D}}$ to be at most a linear combination of $D_{\bar{i}}$ itself so that 

\[
0 =  \tilde{\cal D}(D_{\bar{i}}\psi) = D_{\bar{i}}(\tilde{\cal D}\psi) + c^{\bar{j}}D_{\bar{j}}\psi = D_{\bar{i}}(\tilde{\cal D}\psi)
\]
 
and $\tilde{\cal D}\psi$ still belongs to ${\cal H}_{Q}$. Splitting ${\cal D}$ in two pieces

\[
{\cal D}^{(1,0)} = \sum_{i,\bar{j}} dJ^{i}_{\bar{j}}\frac{\partial}{\partial J^{i}_{\bar{j}}}
\]
\[
{\cal D}^{(0,1)} = \sum_{\bar{i},j} dJ^{\bar{i}}_{j}\frac{\partial}{\partial J^{\bar{i}}_{j}}
\]
 
and checking the commutation relations with $D_{\bar{i}}$ it turns out that only the $(0,1)$ part commutes. The solution is to modify the $(1,0)$ part leaving untouched the $(0,1)$ piece, and thus defining

\begin{eqnarray}\label{0con}
\tilde{{\cal D}}^{(1,0)} &=& {\cal D}^{(1,0)} - \frac{1}{4}\left(dJw^{-1} \right)^{ij}D_{i}D_{j} \nonumber \\
\tilde{{\cal D}}^{(0,1)} &=& {\cal D}^{(0,1)}
\end{eqnarray}

It can be checked that, even after the changing, the connection remains flat ( or better projectively flat ). 
Using as a linear space to be quantized the ensamble of all the physical operators of our theory, and so allowing among the deformations (\ref{0n}) every R-charge object and not only the marginal ones, we can see how the first of the equations (\ref{0con}) gets translated exactly in the master equation (\ref{0haem}) simplified to the case with no operator insertions ( while the second is the requirement of holomorphicity in $\bar{t}$, that is independence by $c.c.(\bar{t}) = t$ which is different by the $t + x$ the theory depends on; see \cite{Verlinde:2004ck} and \cite{Gunaydin:2006bz} for a discussion on a different formalism for the HAE in which holomorphicity is manifest ). 

This last statement can be verified after considering on ${\cal W} =$ ``space of physical operators`` $= H_{\bar{\partial}}^{3,0}(X) \oplus H_{\bar{\partial}}^{2,1}(X) \oplus H_{\bar{\partial}}^{1,2}(X) \oplus H_{\bar{\partial}}^{0,3}(X) = H_{\bar{\partial}}^{3}(X,{\mathbb R}) $ the symplectic form defined as

\[
w(O_{a},O_{b}) = \int_{X}W_{a}\wedge W_{b} 
\]
  
with $W_{a/b}$ the differential forms associated to the physical operators $O_{a/b}$. Defining a complex structure over ${\cal W}$ induced by the complex structure on $X$, the only issue remaining is to compute the expression $\frac{1}{4}\left(dJw^{-1} \right)^{ij}$ in (\ref{0con}) and to rearrange a bit the notation. The result turns out to be our holomorphic anomaly equation. Thus the moral of the story is that we can reinterpret those equations as a request for the theory in order to be in some sense independent by the moduli space, or better to develop a flat connection such that different quantizations in different points can be identified through parallel transport.

\subsection{Open case}\label{0trallalla}

All the discussion about closed strings is fine, but what does it happen when you try to generalize to open strings? What are the differences? When you have open strings different new things happen; because your Riemann surfaces develop boundaries all the formulae involving the Euler number $2g-2$ are generalized to $2g-2+h$ where $h$ from now on will be the number of boundaries ( and for unoriented surfaces you have to add also $c$, the number of crosscaps ). This means different rules for counting zero modes, moduli and so on. Boundaries also mean boundary conditions and D-branes; depending on the model you chose D-branes should, for consistency, stay on different geometric loci in the target space. Specifically in the A model they should be wrapped on Lagrangian submanifolds, and so they are required to be three dimensional, while in the B-model they stay on holomorphic cycles, that is $D_{0},D_{2},D_{4}$ and $D_{6}$ branes are permitted. This means that we are bringing in the game a set of additional geometric information. More than this, open strings mean Chan Paton factors which in turn mean a possible background gauge field, and so, on nontrivial cycles, the appearance of Wilson lines. So in general the space of operators as well as the space of moduli in the theory is enlarged; the question to ask now is, can I apply the same way of reasoning used to develop the holomorphic anomaly equation also in the case of open strings? The answer is of course affirmative and was first given in \cite{Wal1} and later generalized allowing purely open moduli in \cite{BT}. This will be shortly reviewed in this section. But a few additional questions can be asked and these will be the subjects of later chapters.

Again we will work in the ''easy`` case of the B-model but most of what we are going to say can be extended. Let us consider the boundary part of the open topological string action. As derived in \cite{W1}, we can add a Wilson line under the condition that should be supersymmetrized. In addition the field strength of background gauge field has to satisfy $F_{A}^{(0,2)} = 0$. Under these assumptions the boundary action

\begin{equation}\label{0bou}
S_{bound.} = i\int_{\partial\Sigma_{g,h}}\left(X^{*}(A) + F_{A\;i\bar{j}}\rho^{i}\eta^{\bar{j}} \right)
\end{equation}

is $Q_{B}$-invariant ( from now on simply $Q$ ) and it can be rewritten as

\[
S_{bound.} = \int_{\partial\Sigma_{g,h}}QA_{i}\rho^{i} + \int_{\partial\Sigma_{g,h}}\bar{Q}A_{\bar{i}}\eta^{\bar{i}}
\]

The principal point here is the fact that, in addition to the deformations (\ref{0n}), we can add two new classes of terms: the first one still given by complex structure deformation, and so expanded in $t$-parameters, but inducing changes in (\ref{0bou}) and not in the bulk action, and the second class corresponding to deformation for the value of the background gauge field itself. They are written, in analogy with (\ref{0n}):

\begin{equation}\label{0defb}
\delta S_{bound.}=\int_{\partial\Sigma_{g,h}}Q\left(\bar{u}^{\bar{\alpha}}\bar{\Theta}_{\bar{\alpha}} + \bar{t}^{\bar{i}}\bar{\Psi}_{\bar{i}}\right) + \int_{\partial\Sigma_{g,h}}\bar{Q}\left((u^{\alpha}+\delta u^{\alpha})\Theta_{\alpha} + (t^{i}+x^{i})\Psi_{i}\right)
\end{equation}

with 

\begin{eqnarray}\label{0op}
\Theta_{\alpha} &=& \delta_{\alpha}A^{(0,1)}_{\bar{i}}\eta^{\bar{i}}\;\;\;\;\;\;\;\;\;\; \bar{\Theta}_{\bar{\alpha}} = \delta_{\bar{\alpha}}A^{(1,0)}_{i}\left(\rho^{i}dz+\rho^{i}d\bar{z}\right) \nonumber \\
\Psi_{i} &=& \left[W_{i}\right]_{\bar{k}}^{j}A^{(1,0)}_{j}\eta^{\bar{k}}\;\;\;\;\bar{\Psi}_{\bar{i}} = \left[W_{\bar{i}}\right]_{k}^{\bar{j}}A^{(0,1)}_{\bar{j}}\left(\rho^{k}dz+\rho^{k}d\bar{z}\right)
\end{eqnarray}

The notation can be misleading as target space and moduli space indexes are indicated with the same letter. In the following we will avoid, where possible, to use target space indexes. The letter $\alpha$ refers to the tangent to the open moduli space. Open marginal operators are $(0,1)$ forms in the $\bar{\partial}$ cohomology and have values in the Lie algebra matrices, $\Lambda$, from the Chan Paton factors: $\Theta_{\alpha} \in H^{(0,1)}_{\bar{\partial}}\otimes\Lambda$; the index $\alpha$ parametrizes a base for the full spectrum of these boundary marginal operators. Also $\Psi$ has value in the lie algebra matrices, but in this case the index $i$ is blind to them. 
It is evident that, deriving with respect to $\bar{t}^{\bar{i}}$, we will bring on the Riemann surface a new full set of operators, $\bar{\Psi}_{\bar{i}}$, while we are also allowed to derive with respect to new parameters, $\bar{u}^{\bar{\alpha}}$, with corresponding insertions by $\bar{\Theta}_{\bar{\alpha}}$. Before going on we state the boundary conditions \cite{new,Wal1} for the supercurrents and R-symmetry current ( $\chi^{z}$ and $\bar{\chi}^{\bar{z}}$ are vectors parallel to the boundary directions ):

\begin{eqnarray}\label{0bcond}
&&\left(G^+dz + \bar{G}^{+}d\bar{z}\right)|_{\partial \Sigma} = 0 \nonumber \\
&&\left(G^-\chi dz + \bar{G}^{-}\bar{\chi}d\bar{z}\right)|_{\partial \Sigma} = 0 \\
&&\left(Jdz - \bar{J}d\bar{z}\right)|_{\partial \Sigma} = 0 \nonumber 
\end{eqnarray}

and we can see how the topological charge, as well as the antitopological one ( which will appear folded with the Beltrami differentials ) and the axial current are all annihilated at the boundary. For spacefilling branes, instead, the boundary conditions for the fermions are \cite{W1}

\begin{eqnarray}\label{0bcond2}
&&\theta_{i}|_{\partial \Sigma} = 0\nonumber \\
&&\left(\rho^{i}dz-\rho^{i}d\bar{z}\right)|_{\partial \Sigma} = 0
\end{eqnarray}
 
The only missing conceptual ingredient for the derivation of the holomorphic anomaly equation are now the following: 
\begin{itemize}
\item{ A definition of higher genus and boundary amplitudes generalizing (\ref{0h})}
\item{ An analysis on the possible additional moduli space boundary contribution }
\item{ A generalization of the cut and sew procedure for degenerating handles, now to be applied to degenerating strips }
\item{ An analysis on the new building block amplitudes entering the holomorphic anomaly equation, other then $\bar{C}_{\bar{i}\bar{j}\bar{k}}$}
\end{itemize}

The first point is easily clarified; we define the genus $g$ and boundary $h$ amplitude as

\begin{equation}\label{0ciccio}
{\cal F}^{g,h} = \int_{{\cal M}_{g,h}}\prod_{k=1}^{3g-3+h}\int_{\Sigma_{g,h}}G^{-}\mu_{k} \int_{\Sigma_{g,h}}\bar{G}^{-}\bar{\mu}_{k}\prod_{l=1}^{h}\int_{\Sigma_{g,h}}\left(\lambda_{l} G^{-} + \bar{\lambda}_{l}\bar{G}^{-}\right)
\end{equation}

having defined $\lambda$ and $\bar{\lambda}$ as the Beltrami differentials corresponding to variations of the length of the boundary component with respect to ( a part ) of the open moduli; the additional $h$ appearing in the number of closed differentials takes care of the open moduli associated to the position of the holes.

Also a ( maybe rough ) description of point two can be given understanding that, on top of the co dimension two boundaries obtained in the purely closed case by the commutator of the supercharges in $\bar{O}_{\bar{i}}^{[2]}$ with the "closed" measure in (\ref{0ciccio}) we have an additional contribution ( still of co dimension two ) represented by a boundary moving far apart from the rest of the surface or, conformally equivalent, shrinking to zero size, from the commutation relations with the ''open" measure in (\ref{0ciccio}). Finally we can have also co dimension one pieces from the commutator of a single combination of supercharges as in (\ref{0defb}). We will add the following cases: two boundaries colliding, one boundary closing on itself along a nontrivial dividing path and one boundary closing on itself along a nontrivial nondividing path. These cases are shown pictorially.

\begin{figure}[h!]
  \centering
 \includegraphics[width=0.8\textwidth]{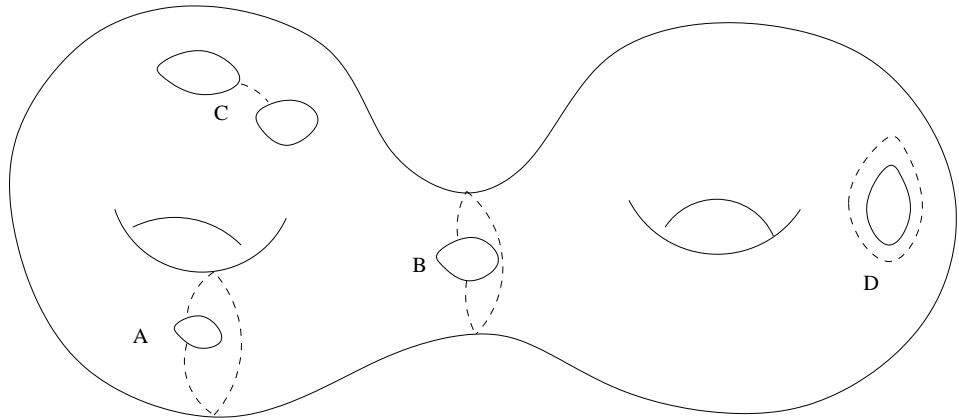}
\caption{Here are shown some open and closed paths which describe a co dimension one and two boundary in the moduli space through their collapsing to zero size. A non dividing open path corresponding to a boundary closing on itself (A); similar situation but along a dividing open path (B); two boundary colliding (C); one boundary shrinking to zero size (D)}\label{00figc2}
\end{figure}

Point three is simply the observation that, in cases A,B,C of picture (\ref{00figc2}), we remain with a narrow strip instead of a long tube. This strip can be replaced by a complete set of boundary states, much in the same way we had done for the handle, contracted with an appropriate open metric, $g_{\bar{\alpha}\beta}$, given by a topological-antitopological disk amplitude.
The fourth point will be fully developed only in (\ref{sec:cap3}) that is we will give explicit formulae in term of target space quantities only there. For now we limit ourselves to a list of the new basic amplitudes entering the full open extended holomorphic anomaly equation ( the notation is $\langle \dots \rangle_{genus,holes}$):

\[
\Delta_{ij} = \langle O_{i}O_{j}^{[1]}\rangle_{0,1} \;\;\;\;\Delta'_{ij} = \langle O_{i}\Psi_{j}^{[1]}\rangle_{0,1} 
\]
\[
\Delta_{\alpha\beta\gamma} = \langle \Theta_{\alpha}\Theta_{\beta}\Theta_{\gamma}\rangle_{0,1}\;\;\;\;\Pi_{\alpha i} = \langle \Theta_{\alpha}O_{i}\rangle_{0,1}\;\;\;\;\Delta'_{\beta i \alpha } = \langle \Theta_{\beta}\Psi_{i}^{[1]}\Theta_{\alpha}\rangle_{0,1} 
\]

Inserting the operator corresponding to $\bar{t}^{\bar{i}}$ and $\bar{u}^{\bar{\alpha}}$ derivatives, working out the commutators, taking care of the boundary conditions (\ref{0bcond}) and remembering the three points discussed, we can give the result:

\[
\bar{\partial}_{\bar{\alpha}}{\cal F}^{g,h} = \frac{1}{2}g^{\beta\bar{\beta}}g^{\alpha\bar{\alpha}}\bar{\Delta}_{\bar{\beta}\bar{\alpha}\bar{\gamma}}\left(D_{\beta}D_{\gamma}{\cal F}^{g-1,h+1} +D_{\beta}D_{\gamma}{\cal F}^{g,h-1}+ \sum_{\substack{g_{1}+g_{2}=g\\h_{1}+h_{2}=h+1}}D_{\beta}{\cal F}^{g_{1}h_{1}}D_{\gamma}{\cal F}^{g_{2}h_{2}}   \right) +
\]
\begin{equation}\label{0haeo1}
 + g^{i\bar{i}}\bar{\Pi}_{\bar{\alpha}\bar{i}}D_{i}{\cal F}^{g,h-1}
\end{equation}
 
\[
\bar{\partial}_{\bar{i}}{\cal F}^{g,h} = \frac{1}{2}g^{j\bar{j}}g^{k\bar{k}}\bar{C}_{\bar{i}\bar{j}\bar{k}}\left(\sum_{\substack{g_{1}+g_{2}=g\\h_{1}+h_{2}=h}}D_{j}{\cal F}^{g_{1}h_{1}}D_{k}{\cal F}^{g_{2}h_{2}}  +  D_{j}D_{k}{\cal F}^{g-1,h}\right) - \left(\bar{\Delta}_{\bar{i}\bar{j}} + \bar{\Delta'}_{\bar{i}\bar{j}}\right)g^{\bar{j}j}D_{j}{\cal F}^{g,h-1} +
\]
\begin{equation}\label{0haeo2}
\frac{1}{2}\left(\bar{\Delta'}_{\bar{i}\bar{\beta}\bar{\gamma}} + \bar{B}_{\bar{i}\bar{\beta}\bar{\gamma}}\right)g^{\bar{\beta}\beta}g^{\bar{\gamma}\gamma}\left(D_{\beta}D_{\gamma}{\cal F}^{g-1,h+1} +D_{\beta}D_{\gamma}{\cal F}^{g,h-1}+ \sum_{\substack{g_{1}+g_{2}=g\\h_{1}+h_{2}=h+1}}D_{\beta}{\cal F}^{g_{1}h_{1}}D_{\gamma}{\cal F}^{g_{2}h_{2}}\right)
\end{equation}

where the only amplitude not yet given is $\bar{B}_{\bar{i}\bar{\beta}\bar{\gamma}}$ 

\[
\bar{B}_{\bar{i}\bar{\beta}\bar{\gamma}} = \int_{0}^{1} dr \langle \bar{\Theta}_{\bar{\beta}}\int_{\partial\Sigma_{g,h}}\left[G^{+}-\bar{G}^{+}\right]\bar{\Theta}_{\bar{\gamma}}\bar{O}_{\bar{i}}(r)\rangle 
\]

which arises because the boundary conditions do not kill the combination $\left[G^{+}-\bar{G}^{+}\right]$.
It is possible now a comparison with the results of \cite{Wal1}. There one strong statement is assumed that is the non contribution by the continuous value of open string moduli. Translated in our case we don't have (\ref{0bou}) because we cannot create a gauge field background. This means we have to suppress all the contribution from "Greek letter" operators and $\Psi_{i}$ as well. So equation (\ref{0haeo1}) no longer exists while (\ref{0haeo2}) is reduced to contain only the extra contribution by $\bar{\Delta}_{\bar{i}\bar{j}}$. In this case \cite{Wal1} gives also the one loop HAE ( $N$ is the number of branes ):

\begin{equation}\label{0oneloop}
\bar{\partial}_{\bar{i}}\partial_{j}{\cal F}^{0,2}=-\Delta_{jk}\Delta^{k}_{\bar{i}} + \frac{N}{2}G_{j\bar{i}}  
\end{equation}

In analogy with the closed case, and using canonical coordinates, we can also rewrite (\ref{0haeo1}) and (\ref{0haeo2}) in term of a master equation. This will be the analogous of (\ref{0mast}) with open parameters in the game ( but restricted to higher genus-boundary amplitudes ). To this goal let us introduce a change of variables ( for a precise definition see (\ref{3z}) ), that is 

\[
\tilde{t}^{i} = t^{i} + \lambda\mu\bar{\Delta}^{i} 
\]

with $\bar{\Delta}^{i}$ the primitive of $\bar{\Delta}^{i}_{\bar{i}} ( = g^{i\bar{j}}\bar{\Delta}_{\bar{j}\bar{i}} )$, $\bar{\Delta'}^{i}_{\bar{i}}$ and $\bar{\Pi}^{i}_{\bar{\alpha}}$ ( note that deriving with respect to $\bar{t}^{i}$ gives rise to both $\bar{\Delta}$ and $\bar{\Delta'}$ ). In the new variables the single derivative terms in (\ref{0haeo1}) and (\ref{0haeo2}) are generated by the chain rule, and the master equation is simply:

\[
\left(\frac{D}{D \bar{t}^{\bar{i}}} + \frac{D}{D \bar{u}^{\bar{\alpha}}}\right)e^{W(\tilde{t})} = \left(\bar{\partial}_{\bar{i}} + \lambda\mu\frac{\bar{\partial} \bar{\Delta}^{j}}{\bar{\partial} \bar{t}^{\bar{i}}}D_{j} + \bar{\partial}_{\bar{\alpha}} + \lambda\mu\frac{\bar{\partial} \bar{\Delta}^{j}}{\bar{\partial} \bar{u}^{\bar{\alpha}}}D_{j}\right)e^{W(\tilde{t})} = 
\]
\begin{equation}\label{0mast2}
\left( \frac{\lambda^{2}}{2}\bar{C}_{\bar{i}\bar{j}\bar{k}}g^{\bar{j}j}g^{\bar{k}k}D_{j}D_{k} + \frac{1}{2}\left[ \bar{\Delta'}_{\bar{i}\bar{\beta}\bar{\gamma}} + \bar{B}_{\bar{i}\bar{\beta}\bar{\gamma}} + \bar{\Delta}_{\bar{i}\bar{\beta}\bar{\gamma}}\right]g^{\bar{\beta}\beta}g^{\bar{\gamma}\gamma}\left[\frac{\lambda}{\mu}\sum_{a\neq b = 1}^{h}D^{a}_{\beta}D^{b}_{\gamma} + \lambda\mu\sum_{a = 1}^{h}D^{a}_{\beta}D^{a}_{\gamma} \right]\right)e^{W(\tilde{t})}
\end{equation}

with

\[
W = \sum_{g,h}\lambda^{2g-2+h}\mu^{h}{\cal F}^{g,h} 
\]

The chain rule for the covariant derivative $D_{\bar{i}}$ is explained as follows ( for indexes $\bar{\alpha}$ works similarly ): 

\[
\frac{D}{D \bar{t}^{\bar{i}}} =  \frac{\bar{d}}{\bar{d}\bar{t}^{\bar{i}}} + A_{\bar{i}\bar{0}}^{\bar{0}} = \bar{\partial}_{\bar{i}} + \frac{\bar{\partial}t^{j}}{\bar{\partial}\bar{t}^{\bar{i}}}\partial_{j} + \frac{\bar{\partial}t^{j}}{\bar{\partial}\bar{t}^{\bar{i}}}A_{j0}^{0} = \bar{\partial}_{\bar{i}} + \frac{\bar{\partial}t^{j}}{\bar{\partial}\bar{t}^{\bar{i}}}D_{j}
\]

where the connection contains only the piece $A_{\bar{i}\bar{0}}^{\bar{0}} \sim \bar{\partial}_{\bar{i}}K$ as there is any antiholomorphic index on $W(\tilde{t})$. Moreover this connection exist only because of the antitopological vacuum dependence of the shifted variables $\tilde{t}$, so it acts only on them, as shown in the second equality above.
The notation $D^{a}$ means a boundary derivative acting only on the $a$ connected component of the total boundary, that is inserting an operator only on the hole number $a$. From the point of view of (\ref{0defb}) we should say we had hidden the proper notation for a Wilson line that should have been a product over all the connected boundary components, each term with operators labelled by an index $a$. It is important to take care of this in the master equation because it suppresses the case that would arise with $D^{a}_{\beta}D^{b}_{\gamma}$ acting twice on $W$ ( instead of two single derivatives ) and that would generate a term in  (\ref{0haeo1}) and (\ref{0haeo2}) which is topologically impossible. The last important point is the additional expansion parameter $\mu$ used for an independent counting of the boundaries.

\subsection{What interpretation?}

We have seen how the closed holomorphic anomaly equation (\ref{0mast}) can be interpreted in terms of a ''quantum background independence`` requirement for the B-model; is it possible to extend this idea to (\ref{0mast2})? In \cite{Wal2} they give a partial answer for the special case when open moduli are frozen. In that situation the open holomorphic anomaly equation reduces to almost the closed one, with the addition of a single term, as discussed after (\ref{0haeo2}). In that situation the master equation (\ref{0mast2}) looks exactly as the closed one, with the $\Delta$ term absorbed in the shift of variables for $t$. However Neitzke and Walcher noticed that the shift used is not fully appropriate ( this point will be much better discussed in later chapters ) for properly tranlating the constrain over the open amplitudes to the closed ones and vice versa, that is the requirement that closed covariant derivatives are equivalent to operator insertions, applied for every genus and number of holes. Instead we have essentially to divide $\Delta^{i}$ in two parts, one part still used to rewrite $t^{i}$ and the second one involved in a global phase redefinition of $W$. So the moral is that the open master equation indeed reduces to the closed one, but the open geometrical data are not buried completely in the shift of $t$, while instead they partially survive labelling the $W$'s entering in the open equation. So Witten's argument can still be used with the caveat that our Hilbert space will be much richer than the closed one. But what about the generic case (\ref{0mast2})? the idea in \cite{Wal2} can be in principle extended to take care of the additional one derivative terms that would have appeared in the open master equation without the shift of $t$, but the hard part is to justify the derivatives with respect to the purely open moduli. So the question is, is there an analogue of the complex structure moduli space, with respect to which we can introduce flat covariant derivatives that identify, trough parallel transport, different quantizations? Of course a choice of complex structure is a nice tool for choosing a quantization, as it naturally splits the number of variables in two, but can something similar be done using a choice of open moduli? Quantization ( of Chern-Simons theory ) was already done using flat holomorphic connections, see \cite{Axelrod:1989xt}, and we can think to apply similar ideas to the present case. However that quantization is really induced by a choice of complex structure for defining the holomorphicity, while here we are looking for some ''pure open moduli`` quantization, that is we are asking: is it there a natural splitting of variables from a choice of open moduli base space not induced, that is independent, by a choice of complex structure? In case of affirmative answer then the procedure would be simply to derive a flat connection such that $\frac{1}{2}\left[ \bar{\Delta'}_{\bar{i}\bar{\beta}\bar{\gamma}} + \bar{B}_{\bar{i}\bar{\beta}\bar{\gamma}} + \bar{\Delta}_{\bar{i}\bar{\beta}\bar{\gamma}}\right]g^{\bar{\beta}\beta}g^{\bar{\gamma}\gamma}$ times the double open derivatives, is nothing but the modification you have to do to the natural flat connection on the ''prequantum`` Hilbert space fibered bundle, much in the same way it was obtained the term $\frac{\lambda^{2}}{2}\bar{C}_{\bar{i}\bar{j}\bar{k}}g^{\bar{j}j}g^{\bar{k}k}D_{j}D_{k}$. A final answer is still waiting to be given.

    
\chapter{Tadpole cancellation}
\label{sec:cap1}
\noindent

\section{Tadpole cancellation in topological strings}

It is a classical result in open superstring theories that the condition of tadpole cancellation ensures their consistency by implementing the cancellation of gravitational and mixed anomalies \cite{Polchinski_II}. It is also well known that topological string amplitudes calculate BPS protected sectors of superstring theory \cite{BCOV,AGNT,ant}. It is therefore natural to look for a corresponding consistency statement in the open topological string. At the same time we have seen as topological strings are very similar in shape to bosonic strings where tadpole cancellation is not a real physical requirement, as the infrared divergence involved automatically cancels when you expand the theory around the correct equation of motion \cite{Polchinski_I}. So it is interesting to ask if is it there some real physical reason for asking tadpole cancellation, as it should be inherited by superstrings, or not, as would suggest the similitude with bosonic string theory. Already in the literature were present some hints suggesting to prefer the first answer. 

It is true that closed topological strings on Calabi-Yau threefolds provide a beautiful description 
of the K\"ahler and complex moduli space geometry via the A- and B-model respectively \cite{Witten}. However, D-branes naturally couple in these models to the wrong moduli \cite{OOY}, namely A-branes to complex and B-branes to K\"ahler moduli. So it is natural to ask if tadpole cancellation can provide a way of reinforcing the decoupling by the wrong moduli. The first observation in this direction came from a different perspective in \cite{Wal3}, where it was observed that the inclusion of unorientable worldsheet contributions is crucial to obtain a consistent BPS
states counting for some specific geometries in the open A model \cite{Wal3,kw1}. From this it was inferred that tadpole cancellation would ensure the decoupling of A and B model in loop amplitudes. Also an analysis of open oriented amplitudes leads to new anomalies in the topological string due to boundary terms, as observed in \cite{new}, and again suggests a kind of non trivial dependence by the ``wrong moduli''. Moreover it constitutes an obstruction to mirror symmetry and to the realization of open/closed string duality in generic Calabi-Yau targets.

Some analysis on the unoriented sector of the topological string have been performed in
\cite{SV,bobby,Diaco,Bou} for local Calabi-Yau geometries. In these cases the issue of tadpole cancellation gets easily solved by 
adding anti-branes at infinity, as already noticed also in \cite{new}.
However, a more systematic study of this problem is relevant in order to analyze mirror symmetry with D-branes \cite{vafa} 
and open/closed string dualities in full generality.

More in general, it is expected that the topological string captures
D-brane instanton non perturbative terms upon Calabi-Yau compactifications to four dimensions \cite{Ma,u}.
Therefore, the study of the geometrical constraints following from a consistent wrong moduli decoupling could shed light 
on the properties of BPS amplitudes upon wall crossing \cite{KS,JM,OY,NN,GM}.

The following chapter is based on \cite{nostro}.

\section{Wrong moduli dependence in oriented open amplitudes}\label{1borg}

If in the closed topological theory you look for a dependence by the wrong moduli, in a way similar to how it is derived the dependence by the antiholomorphic moduli, you find out that indeed there is none \cite{BCOV} ( for definiteness let us work with the B-model but, redefining $\bar{J}\rightarrow-\bar{J}$, everything can be translated to the A-case ). This can be explicitly seen as follows: let us suppose to add to the action deformations of the kind

\begin{equation}\label{1defy}
y^{p}\int_{\Sigma_{g}}\left\{ \bar{Q}^{+},\left[Q^{-},{\cal O}_{p}\right]\right\} 
\end{equation}

with $\left[Q^{+},{\cal O}_{p}\right]=0$ and $\left[\bar{Q}^{-},{\cal O}_{p}\right]=0$ the physical operators from the ``other'' model ( in our notation the A-model ). These deformations are still $Q = (Q^{+} + \bar{Q}^{+})$-closed because

\[
\int_{\Sigma_{g}} \left[ Q^{+} + \bar{Q}^{+},\left\{ \bar{Q}^{+},\left[Q^{-},{\cal O}_{p}\right]\right\}\right] = \int_{\Sigma_{g}} \left[ Q^{+},\left\{ \bar{Q}^{+},\left[Q^{-},{\cal O}_{p}\right]\right\}\right] = \int_{\Sigma_{g}} \left[ \left\{ Q^{+},\left[ Q^{-},{\cal O}_{p}\right]\right\},\bar{Q}^{+}\right]
\]

and

\[
\left\{ Q^{+},\left[ Q^{-},{\cal O}_{p}\right]\right\} = \left[\left\{Q^{+} ,Q^{-} \right\},{\cal O}_{p}\right] = \left[T,{\cal O}_{p}\right] = \partial {\cal O}_{p}
\]

so inside an integral on a closed Riemann surface it vanishes \footnote{Notice that our action deformation is a two form, but the two worldsheet indexes do not come from the ${\cal O}_{p}$, as it was for the antitopological case. Instead only one is from the operator while the other is from $Q^{-}$. In the topological case both indexes instead where from the charges and the operator was a scalar.}. 
If you now derive an amplitude with respect to $y^{p}$ you bring down, as usual, the corresponding deformation. Then the current $\bar{G}^{+}$ is pushed, as usual, against one of the antiholomorphic Beltrami differentials, giving a derivative in moduli space. But $G^{-}$ is a two form and should remain where it is, so that all the holomorphic differentials are untouched. In particular one of them has to be positioned on the long handle it forms when on the boundary in the moduli space, and such an object is clearly ( because of the supersymmetry algebra ) killed when you project to the ground states ( because of the infinite length of the tube ). So the closed theory is independent by $y^{p}$ and, working in the same way, by $\bar{y}^{\bar{p}}$ as well. What instead goes wrong when you allow boundaries in the worldsheet ( for now we limit ourselves to the case with frozen open moduli )? First of all (\ref{1defy}) is no longer $Q$-closed as the action of the topological charge gives an integral on the boundary $\partial \Sigma_{g,h}$ which now exists. So, in order to deform the action with wrong moduli, we should add to (\ref{1defy}) a piece that makes it $Q$-closed. And the deformation associated to $y^{p}$ becomes

\begin{equation}
y^{p}\left( \int_{\Sigma_{g,h}}\left\{ \bar{Q}^{+},\left[Q^{-},{\cal O}_{p}\right]\right\} + \int_{\partial\Sigma_{g,h}}{\cal O}_{p}\right)
\end{equation}

The additional term is called the Warner term and is not necessary when we have $t^{a}$ deformations because, in that case, are really the boundary conditions (\ref{0bcond}) and (\ref{0bcond2}) to take care of the arising boundary contribution. The second difference is that there is an additional contribution to the boundary of the moduli space coming from the degeneration where a boundary is moving far from the rest of the surface. In that case the long tube connecting the boundary to the rest of the surface has no twisting ( because rotating the disk does not have influence ), so we don't have Beltrami differentials on it. The result \cite{new} is that the derivative with respect to $y^{p}$ of a topological open amplitude is nonvanishing and, moreover, R-charge counting says that the long tube should be replaced by a complete set of states corresponding to marginal deformations for the ``other'' model. The claim we are going to develop in the rest of the chapter is that tadpole cancellation mantains a complete decoupling from the wrong moduli in case of open amplitudes, or, vice versa, that the independence by the wrong moduli is the physical reason we were looking for to ask for tadpole cancellation.   

\section[One loop amplitudes on the torus, tadpole cancellation and analytic torsion ]{One loop amplitudes on the torus, tadpole cancellation and Ray-Singer analytic torsion}\label{1tad1}

In this section we investigate tadpole cancellation
for open unoriented topological string amplitudes at zero Euler characteristic
considering, as a warm up example, the B-model case when the target space is a $T^2$. We conclude by rewriting the amplitudes as Ray-Singer analytic torsions. We will work with open moduli so, before entering in the computation, let us briefly recall how Chan Paton factors work in the case of unoriented amplitudes.

\subsection{Chan Paton factors}\label{1chan}

The notation follows from \cite{Polchinski_I}. Chan Paton factors are introduced assuming an open string state to carry two indices at the two ends, 
each one running on the integers from $1$ to $N$. This additional state is indicated as $\left|i,j\right\rangle$ $ (i,j = 1 ... N)$.

The worldsheet parity $\Omega$ is defined to act exchanging $i$ with $j$ and rotating them with a $U(N)$ transformation $\gamma$. 
This rotation is added simply because it is still a symmetry for the amplitudes. Thus we have
\begin{equation} \label{1aa}
\Omega\left|i,j\right\rangle \equiv \gamma_{j\:l}\left|l,k\right\rangle\gamma^{-1}_{k\:i}
\end{equation}
Asking $\Omega^{2} = 1$ \cite{Polchinski_I} means requiring
\begin{equation} \label{1ab}
\gamma^{T} = \pm \gamma
\end{equation}
Now if we do a base change of the kind
$\left|i,j\right\rangle \rightarrow \left|i',j'\right\rangle = U^{-1}_{i'\:k}\left|k,l\right\rangle U_{l\:j'}$
it transforms $\gamma$ in the new primed base so that
$\gamma \rightarrow U^{T}\gamma U$.
In particular choosing an appropriate $\left|i',j'\right\rangle $ base one can always transform $\gamma$ so that
\begin{equation} \label{1ae}
\gamma = 1 \:\:\:\:\: or \:\:\:\:\: \gamma = \left( \begin{array}{cc} 0 & i \\ -i & 0 \end{array}\right)
\end{equation}
respectively in the $+$ or $-$ case of (\ref{1ab}).
We start from the first case. There we can create the new base
$\left|a\right\rangle = \Lambda^{a}_{i\:j}\left|i,j\right\rangle$
using $N^{2}$ independent matrices, in our case the $N \times N$ real matrices.  Worldsheet parity action on the states 
$\left|a\right\rangle$ can be seen as an action on the coefficient $ \Lambda^{a}_{i\:j}$. Choosing them either symmetric or 
antisymmetric one has respectively $\frac{1}{2}(N^{2} + N)$ and  $\frac{1}{2}(N^{2} - N)$ of them. 
Since massless states transform with a minus under worldsheet parity, in order to create unoriented states one needs to couple these 
to Chan-Paton states $\left|a\right\rangle $ with antisymmetric coefficients. So a double minus gives a plus. Then the gauge field 
background, associated with those vertex operators, will be with values in the Lie Algebra of $N \times N$ anti-symmetric real matrices,
that is $SO(N)$.
From the spacetime effective action, with gauge field and matter in the adjoint, one has that the coupling of an  
$\left|a\right\rangle $ state with a generic  background $A = A_{b}\Lambda^{b}$ is of the kind
$\left[\Lambda^{a},A_{b}\Lambda^{b}\right]$.
If the background is diagonal with elements $a_{1} ... a_{N}$ and we consider the state $\left|a\right\rangle = \left|i,j\right\rangle$
this coupling gives an eigenvalue $+a_{i} - a_{j}$: the state $\left|i,j\right\rangle $  will shift the spacetime momenta as 
$p \rightarrow p + a_{i} - a_{j} $. This effect is  more precisely described changing the string action with the addition of a 
gauge field background, which will manifest itself inserting in the path integral a Wilson loop of the kind
\begin{equation} \label{1af}
\prod_{k}{\Tr} e^{i\int_{\partial\Sigma_{k}}A_{\mu}\dot X^{\mu}}
\end{equation}
where the sum is other all the connected components of the boundary. If one creates an open string state with a vertex operator on a 
boundary with non trivial homology, the left $i$ Chan-Paton sweeps in space giving an Aharonov-Bohm phase (\ref{1af}) $a_{i}$. 
The right $j$ Chan-Paton moves in the opposite direction on the same boundary and couples with a minus. For loop states with one 
Chan-Paton on a boundary and the second on another the situation is the same, always with one Chan-Paton moving along the orientation 
of the boundary and the other in the opposite \footnote{Notice that the state sweeping the loop should be consistent with the one 
created by a boundary vertex operator plus some string interaction}.

Now for any (constant) $SO(N)$  background one can always act with a rigid gauge transformation to put it in the form
\begin{equation} \label{1ag}
\bigoplus_i a_i\left( \begin{array}{cc} 0 & 1 \\ -1 & 0
\end{array}\right)
\end{equation}
This is still $SO(N)$ so the worldsheet parity still acts simply exchanging the $i - j$ factors. If in addition one wants to diagonalize 
it one needs to act with a gauge transformation that will change the  $SO(N)$ form and so will have effects also on the shape of 
$\Omega$. In fact we can rewrite $\left|i,j\right\rangle =  U_{i\:k'}\left|k',l'\right\rangle U^{-1}_{l' \:j}$ so that
\[
\left|a\right\rangle =\Lambda^{a}_{i\:j}U_{i\:k'}\left|k',l'\right\rangle U^{-1}_{l' \:j} = U^{T}_{k'\:i}\Lambda^{a}_{i\:j}(U^{\dagger})^{T}_{j\:l'}\left|k',l'\right\rangle
\]
This in order to transform $\Lambda^{a} $ so to diagonalize our background. But, acting in this way, the base $\left|i,j\right\rangle$ 
has changed and then also the worldsheet parity (\ref{1aa}) will be different. In particular
\[
\gamma ( = I )\rightarrow  U^{T}\gamma U ( = U^{T}U )
\]
When (\ref{1ag}) is reduced to the simple two dimensional case the matrix $U^{T}$ which diagonalizes $A$ and $\gamma$ ( in the primed base ) are
\[
U^{T} =  \left( \begin{array}{cc} i/\sqrt{2} & 1/\sqrt{2} \\ -i/\sqrt{2} & 1/\sqrt{2} \end{array}\right) \:\:\:\:\:\: \gamma =  \left( \begin{array}{cc} 0 & 1 \\ 1 & 0 \end{array}\right)
\]
The computation of the cylinder is straightforward. We have to sum over all states, and different Chan-Paton indices will modify the 
Hamiltonian with the usual momentum shift. Instead if we want to compute the M\"obius strip  we should look for diagonal states of 
$\Omega $. It is easy to see that, in the $\left|i',j'\right\rangle $ base, they are $\left|1,2\right\rangle  $ and 
$\left|2,1\right\rangle$, both with eigenvalue $+ 1$\footnote{ In the $ \left|a\right\rangle $ base there are four, one with 
eigenvalue $- 1$, that is the diagonalized background itself, and three with $+1$. }. Each diagonal term in the trace will contribute 
both with its eigenvalue and with its own Hamiltonian. In our case the two states will change the momenta respectively as 
$p \rightarrow p + a_{1} - a_{2} $ and $p \rightarrow p + a_{2} - a_{1} $ where, for our diagonalized $SO(2)$  background, 
$a_{1} = ia$ and $a_{2} = -ia$. Generalization to higher $N$ is straightforward. Then we end up with our amplitudes. 

The $Sp(N/2)$ situation is simpler. There the diagonal background is already an $Sp(N/2)$ algebra matrix if in the 
form\footnote{If one interchanges the positions of some diagonal elements, which can of course be done with a gauge transformation, 
that matrix is no longer $Sp(N/2)$.The cylinder is manifestly invariant under any such gauge transformation, but the M\"obius is not. 
In fact if one wants to compute the amplitude in the new background one should take care of the changing occurred to the worldsheet 
parity operator and find the new diagonal states with their gauge field couplings. Working properly the amplitude is of course 
invariant. } 
\[\diag\{a_1, \cdots, a_{N/2}, -a_1, \cdots, -a_{N/2}\}
\] 
 Therefore the worldsheet parity $\Omega$ is still the second of (\ref{1ae}). Diagonal states are now $\left|i,i + N/2\right\rangle  $ or 
$\left|i + N/2,i\right\rangle  $ for $i = 1 ... N/2$, note both with negative eigenvalues. The contribution to the Hamiltonian is 
again $p \rightarrow p \pm 2a_{i} $. 

\subsection{One loop amplitudes on $T^{2}$}

The relevant amplitudes for the mechanism of tadpole cancellation are
the cylinder, the M\"obius strip and the Klein bottle coupled to a constant gauge field. 
In the operator formalism, as usual for one loop amplitudes, we have
\begin{equation*}
\F_{cyl} = \int_{0}^{\infty}\frac{ds}{4s}{\Tr}_{o}\left( F(-1)^{F}e^{-2\pi s H}\right),\:\:\:\: 
{\cal F}_{m\ddot{o}b} = \int_{0}^{\infty}\frac{ds}{4s}{\Tr}_{o}\left(\P F(-1)^{F}e^{-2\pi s H}\right) 
\end{equation*}
\begin{equation} \label{1d}
\F_{kle} = \int_{0}^{\infty}\frac{ds}{4s}{\Tr}_{c}\left(\P F(-1)^Fe^{-2\pi s H}\right) 
\end{equation}
where $\P=\Omega\circ\sigma$ is the involution operator obtained by combining the worldsheet parity operator $\Omega$
and a target space involution $\sigma$, $F$ is the fermion number and $H$ is the Hamiltonian for worldsheet time translations. 
The trace is taken over all ( open or closed ) string states. In this section we consider D-branes wrapping the whole $T^2$
and take $\sigma$ to act trivially.
From the Hamiltonian $H$ of the $\sigma$-model with Wilson lines for gauge groups $SO(N)$ or $Sp(N/2)$, 
we have (setting $\alpha' = 1$):
\begin{equation} \label{1a}
\F_{cyl} = \sum_{n,m = -\infty}^{+\infty}\:\sum_{i,j = 1}^{N}\int_{0}^{\infty}\frac{ds}{4s}\: e^{-\frac{2\pi s}{\sigma_{2}t_{2}}\left|n -\, \sigma m - u_{i,j} \right|^2}
\end{equation}
\begin{equation} \label{1b}
\F_{m\ddot{o}b} = \pm \sum_{n,m = -\infty}^{+\infty}\:\sum_{i = 1}^{N}\int_{0}^{\infty}\frac{ds}{4s}\: e^{-\frac{2\pi s}{\sigma_{2}t_{2}}\left|n -\, \sigma m - 2u_{i} \right|^2}
\end{equation}
\begin{equation} \label{1c}
\F_{kle} = \sum_{n,m = -\infty}^{+\infty}\int_{0}^{\infty}\frac{ds}{4s}\: e^{-\frac{2\pi s}{2\sigma_{2}t_{2}}\left|n - \,\sigma m \right|^2}
\end{equation}
 
Here $u_{i,j} = u_{i} - u_{j}$ and $u_{i} = \phi_{i} + \sigma\theta_{i}$ with $\theta_{i}$ and $-\phi_{i}$ the $i$-th diagonal element 
of the Wilson lines 
\footnote{As reviewed in the previous section the unoriented theory selects either the $Sp(N/2)$ or the $SO(N)$ groups. 
In both cases one can diagonalize with a constant gauge transformation leading to N diagonal elements. 
These are purely imaginary for $SO(N)$ and real for $Sp(N/2)$, half of them being independent numbers $a_{1},...,a_{N/2}$ 
and the other half $-a_{1},...,-a_{N/2}$.} 
along the two 1-cycles of the torus with complex structure 
$\sigma = \sigma_{1} + i \sigma_{2} = \frac{R_{2}e^{i\rho}}{R_{1}}$ 
and area $t_{2} = R_{1}R_{2}\sin(\rho)$. 
This means that if one parametrizes the target space torus with $z = R_{1}x_{1} + R_{2}e^{i\rho}x_{2}$, then the gauge field 
reads $A_{i} = \theta_{i}dx_{1} - \phi_{i}dx_{2}$. 
The topological amplitudes get contribution from classical momenta only, due to 
a complete cancellation between the quantum bosonic and fermionic traces. 
The shift in the classical momenta by the Wilson lines $u_{i}$ is the only effect of the coupling to the gauge fields. 
Note that the different coupling between the cylinder and the M\"obius is due to the selection of diagonal $\P$ 
states for the M\"obius. The $\pm$ in front of the M\"obius corresponds to the $SO(N)$ and $Sp(N/2)$ theories respectively
coming from the eigenvalues of the Chan-Paton states in the trace under $\P$.

These amplitudes suffer of two kinds of divergences: the first one is from the $ s\rightarrow 0 $ part of the integral 
and will be removed by tadpole cancellation. The second one comes from the series which turns out to diverge for
vanishing Wilson lines \cite{RS}. In the superstring this second divergence is due to extra massless modes generated by gauge 
symmetry enhancement.
We will start with tadpole cancellation and deal later with the second divergence.

In order to analyze
the behavior at $ s\rightarrow 0 $, we 
Poisson re sum the $n,m$ sums in order to get an exponential going like $e^{-1/s}$. The result is:
\begin{equation} \label{1e}
\F_{cyl} = \sum_{m,n = -\infty}^{+\infty}\:\sum_{i,j = 1}^{N}\int_{0}^{\infty}ds\frac{t_{2}}{8s^2}\: 
e^{-\frac{\pi t_{2}}{2s\sigma_{2}}\left|n + \sigma m\right|^2}e^{2\pi i\left( m\phi_{i,j} - \,n\theta_{i,j}\right)}
\end{equation}
\begin{equation} \label{1f}
\F_{m\ddot{o}b} = \pm\sum_{m,n = -\infty}^{+\infty}\:\sum_{i = 1}^{N}\int_{0}^{\infty}ds\frac{t_{2}}{8s^2}\: 
e^{-\frac{\pi t_{2}}{2s\sigma_{2}}\left|n + \sigma m\right|^2}e^{2\pi i\left( 2m\phi_{i} - \,2n\theta_{i}\right)}
\end{equation}
\begin{equation} \label{1g}
\F_{kle} = \sum_{m,n = -\infty}^{+\infty}\:\int_{0}^{\infty}ds\frac{t_{2}}{4s^2}\: 
e^{-\frac{\pi t_{2}}{s\sigma_{2}}\left|n + \sigma m\right|^2}
\end{equation}
In order to extract the tadpole divergent part, let us
perform the change of variables $\frac{\pi}{s} \rightarrow s$,  $\frac{\pi}{4s} \rightarrow s$, and $\frac{\pi}{2s} \rightarrow s$
respectively for cylinder, M\"obius and Klein bottle 
\footnote{This is in order to normalize the three surfaces to have the same circumference and length 
(respectively $2\pi$ and $s$). They are
parametrized such that the M\"obius and the Klein are cylinders with one and two boundaries substituted by crosscaps respectively.}.
In the three cases the divergent parts come from the $n = m = 0$ term and adding the three contributions we get
\[
\int_{0}^{\infty}ds\frac{t_{2}}{8\pi}\:\left(N^2 \:\:( cylinder )\:\:\ \pm 4N \:\: ( M\ddot{o}bius ) \:\: + 4 \:\:( Klein )\right).
\]
The divergence is canceled by choosing $N = 2$ and requiring $Sp(N/2)$ gauge group
\footnote{In the case of target space $T^{2d}$ one finds $N=2^d$.}.

Once this divergence is removed, the M\"obius strip with non-zero Wilson lines is finite. In particular starting from ( \ref{1f} ) without the $n = m = 0$ part and changing variable to 
$s' = \frac{\pi t_{2}}{2s\sigma_{2}}\left|n + \sigma m\right|^2 $ you have, ( see \cite{RS} )

\[
{\cal F}_{m\ddot{o}b} = -\sum_{i = 1}^{N}\frac{t_{2}}{8}\sum_{m^2 + n^2 \neq 0 }e^{2\pi i\left( m2\phi_{i} - \,n2\theta_{i}\right)}\frac{2 \sigma_{2}}{\pi t_{2}}\left|n + \sigma m\right|^{-2}\int_{0}^{\infty} ds' e^{-s'} = 
\]
\[
= -\sum_{i = 1}^{N}\left[\sum_{n = 1 (m = 0)}^{\infty}\frac{\sigma_{2}}{4\pi}\frac{2}{n^2}cos(4\pi i n \theta_{i}) + \frac{1}{8 \pi}\sum_{m \neq 0}\: \sum_{n = -\infty}^{+ \infty}\frac{1}{\left|m\right|}e^{2\pi i m 2\phi_{i}}e^{-2\pi i n 2\theta_{i}}\frac{2 \left|m \right| \sigma_{2}}{\left|n + \sigma m\right|^{2}} \right]
\]

The result is known  and written in terms of the standard modular functions $\theta_1$ and $\eta$ reads

\begin{equation}
{\cal F}_{m\ddot{o}b} = +\frac{1}{2}log\prod_{i = 1}^{N}\left|e^{\pi i (2\theta_{i})^{2}\sigma}\theta_{1}
\left(2u_{i}|\sigma \right)\eta\left( \sigma\right)^{-1}\right|.
\label{1moebtorus}
\end{equation}

As it is evident from (\ref{1moebtorus}), a further divergence arises at vanishing Wilson lines, where $\theta_1$ vanishes.
In order to define a finite amplitude, notice that
for small value of one of the $u_{i}$'s we can expand to first order inside the logarithm getting
\bea
\F_{m\ddot{o}b} = ... + \frac{1}{2}log\left|e^{\pi i (2\theta_{i})^{2}\sigma}
\theta_{1}\left(2u_{i}|\sigma \right)\eta\left( \sigma\right)^{-1}\right| + ... \approx
\\
\approx ... + \frac{1}{2}log\left| 0 - 2\pi i \eta
\left( \sigma\right)^{2}\sqrt{\sigma_{2}}\frac{2u_{i}}{\sqrt{\sigma_{2}}} + ... \right| + ... 
\label{1cacca}\eea
Notice that both $\eta\left(\sigma\right)^{2}\sqrt{\sigma_{2}}$ and $\frac{2u_{i}}{\sqrt{\sigma_{2}}}$ are separately 
modular invariant under the $SL(2,{\mathbb Z})$ transformations
\footnote{Recall that $u_{i} = \phi_{i} + \sigma\theta_{i}$ and $- \phi$ and $\theta$ are the gauge fields along the two cycles.}
:
\[
\sigma \rightarrow \frac{c + d\sigma}{a + b\sigma} \:\:\:\:\:\:\:
\theta \rightarrow a\theta - b \phi \:\:\:\:\:\:\:
\phi \rightarrow -c\theta +d\phi 
\]
From (\ref{1cacca}), it is clear that the remaining 
finite part is the $\eta\left(\sigma\right)^{2}\sqrt{\sigma_{2}}$ term in the logarithm. One can
in fact also compute it for vanishing Wilson lines starting from (\ref{1b}) by first regulating the integral as
\begin{equation} \label{1h}
\int_{\epsilon}^{\infty}\frac{dt}{t}e^{-kt} = C -log(k\epsilon) + O(\epsilon) \approx -log(k) + C
\end{equation}
and discarding the $m = n = 0$ term, which take care of the tadpole divergence. Then by using zeta-function regularization 
to deal with the infinite product over the $k$-factors in the logarithm one gets\footnote{We use the formula 
$\sin(\pi z) = \pi z \prod_{n = 1}^{\infty}\left(1 - \frac{z^{2}}{n^{2}} \right)$.}
\begin{eqnarray}
\F_{m\ddot{o}b} =  +\frac{1}{4}log\prod_{i= 1}^{N}\frac{\sigma_{2}t_{2}}{2\pi 4 u_{i}\bar{u}_{i}}
\left| e^{i\pi 2u_{i} } - e^{-i \pi 2u_{i}} \right|^{2}e^{2 \pi i \sigma /12}e^{-2 \pi i \bar{\sigma} /12}\times\nonumber
\\
\times\prod_{m = 1 }^{\infty}
\left| \left(1 - e^{2 \pi i (m\sigma + 2u_{i})}\right)\left(1 - e^{2 \pi i (m\sigma - 2u_{i})}\right)\right|^{2}+C.
\label{1zio}
\end{eqnarray}
This is well behaved for $u_{i} \rightarrow 0$ giving, for each vanishing Wilson line element, a term
\begin{equation} \label{1i}
 \frac{1}{2}log\left(\frac{\sqrt{\sigma_{2} t_{2}}}{\sqrt{2}}\left|\eta(\sigma)\right|^{2}\right) +\frac{1}{4}log(4\pi) + C.
\end{equation}
The constant $C$ is arbitrary and can be chosen to reabsorb the term $\frac{1}{4}\log(4\pi)$. 
The extra dependence in (\ref{1zio}) on the K\"ahler modulus $t_2$ is indeed separated in an overall additional 
term which decouples from the one-point amplitudes
$\partial_\sigma\F$.
Using this regularization scheme also for the other amplitudes, that is deleting the tadpole term and, in case of vanishing Wilson lines, regulating the corresponding divergent series, we finally have:
\[
\F_{cyl} = - \sum_{i \neq j = 1}^{N}\Theta(|u_{i,j}|^{2})\frac{1}{2}log\left|e^{\pi i (\theta_{i,j})^{2}\sigma}\theta_{1}\left(u_{i,j}|\sigma \right)\eta\left( \sigma\right)^{-1}\right| - 
\]
\begin{equation}\label{1j}
- \left( N + \sum_{i \neq j = 1}^{N}(1 - \Theta(|u_{i,j}|^{2}))\right) \frac{1}{2}log\left(\frac{\sqrt{\sigma_{2} t_{2}}}{\sqrt{2}}\left|\eta(\sigma)\right|^{2} \right) 
\end{equation}
\[
\F_{m\ddot{o}b} = + \sum_{i = 1}^{N}\Theta(|u_{i}|^{2})\frac{1}{2}log\left|e^{\pi i (2\theta_{i})^{2}\sigma}\theta_{1}\left(2u_{i}|\sigma \right)\eta\left( \sigma\right)^{-1}\right| + 
\]
\begin{equation}\label{1k}
+ \left( \sum_{i = 1}^{N}(1 - \Theta(|u_{i}|^{2}))\right) \frac{1}{2}log\left(\frac{\sqrt{\sigma_{2} t_{2}}}{\sqrt{2}}\left|\eta(\sigma)\right|^{2} \right) 
\end{equation}
\begin{equation}\label{1l}
\F_{kle} = -\frac{1}{2}log\left(\sqrt{\sigma_{2} t_{2}}\left|\eta(\sigma)\right|^{2} \right) 
\end{equation}
where $\Theta(x)$ is the step function, zero for $x\leq 0$ and one for $x>0$.

Let us now make a couple of observations on the above results.
First, notice that all the above free energies satisfy at generic values of the Wilson line a standard
holomorphic anomaly equation in the form
$
\partial_\sigma\partial_{\bar\sigma}\F \sim \frac{1}{(\sigma-\bar\sigma)^2}
$
with a proportionality constant counting the number of states in the appropriate vacuum bundle. More in detail let us derive with respect to the complex structure moduli.   Naively \footnote{ from the simple rule $\partial_{z}\partial_{\bar{z}}log\left(\left|f(z)\right|^{2}\right) = 0 \:\:for\:\: f(0) \neq 0$  with $f(z)$ a holomorphic function} we would have for the cylinder
\[
\partial_{\sigma}\partial_{\bar{\sigma}}Z_{cyl} \approx - \frac{1}{4}N\frac{1}{(\sigma - \bar{\sigma})^{2}} - \sum_{\stackrel{i,j = 1 }{i \neq j}}^{N}(1 - \theta(|u_{i,j}|))\frac{1}{4}\frac{1}{(\sigma - \bar{\sigma})^{2}} 
\]
Note that the only non holomorphic contribution from $e^{-\pi (\theta_{i,j})^2 \sigma_{2}}$ does not contribute for closed string moduli. However $u$ contains itself a $\sigma$ contribution so we also need to add the composite derivatives
\[
\partial_{\sigma} \rightarrow \partial_{\sigma} + \frac{\partial u_{k}}{\partial \sigma}\partial_{u_{k}}
\]
Contribution from these terms will give delta functions ( or worst ) in $u_{i,j}$, either because of the derivations of the step functions or because of the holomorphicity but in $u_{i,j} = 0$ of $\theta_{1}\left(u_{i,j}|\sigma \right)$, or both. We interpret these terms as discontinuities in the moduli space of the theory, passing from a purely closed moduli space ( with $u = 0$ ) to an open plus closed space. So it is perfectly resonable that the moduli space metric has in these points some divergences. Looking only at the smooth contributions for the M\"obius we get similarly:
\[
\partial_{\sigma}\partial_{\bar{\sigma}}Z_{m\ddot{o}b} \approx \sum_{i= 1}^{N}(1 - \theta(|u_{i}|))\frac{1}{4}\frac{1}{(\sigma - \bar{\sigma})^{2}} 
\]

And for the Klein

\[
\partial_{\sigma}\partial_{\bar{\sigma}}Z_{kle} = - \frac{1}{4}\frac{1}{(\sigma - \bar{\sigma})^{2}}. 
\]

In particular, at vanishing Wilson lines, we recover the results stated in \cite{Wal3}.

A more accurate discussion on the holomorphic anomaly equation for general target spaces is deferred to the next section.
The second comment  concerns the interpretation of the amplitudes we just calculated in terms of the analytic
Ray-Singer torsion \cite{RS}.
This is defined as \cite{pestunwitten}
\begin{equation}
log T(V)=\frac{1}{2}\sum_{q=0}^d (-1)^{q+1}q log ~det' \Delta_{V\otimes\Lambda^q \bar{T}^*X}
\label{1torsion}
\end{equation}
where $d=dim_{\C}X$.
On an elliptic curve with complex structure $\sigma$ the analytic torsion of a flat line bundle ${\cal L}$ 
with constant connection $u$ is
given by (as it can be found in Theorem 4.1 in \cite{RS})
\bea
\T\left({\cal L}\right)= 
\begin{cases} 
\quad \left\vert e^{-\pi \frac{\left(Im u\right)^2}{Im \sigma}}
\frac{\theta_1(u|\sigma)}{\eta(\sigma)}\right\vert                                            & \mbox{if } u\not=0 \\
\quad                                                                                              & \quad \\ 
\quad \sqrt{Im\sigma}\vert\eta(\sigma)\vert^2                                                     & \mbox{if } u=0 
\end{cases}
\eea
where the second case corresponds to the trivial line bundle ${\cal O}$.

On an elliptic curve equipped with a flat vector bundle $E=\oplus_i {\cal L}_i$, 
an extension of the formula for the Ray-Singer torsion 
implies that
\begin{eqnarray}
\F_{cyl}&=&-\frac{1}{2}\sum_{i,j}log \T\left({\cal L}_i\otimes{\cal L}^*_j\right) =
-\frac{1}{2}log \T\left(E\otimes E^*\right)
\nonumber\\
\F_{m\ddot{o}b}&=&+\frac{1}{2}\sum_{i}log \T\left({\cal L}_i^2\right)=
+\frac{1}{2}log \T\left(diag(E\otimes E)\right)
\nonumber\\ 
\F_{kle}&=&-\frac{1}{2}log \T\left({\cal O}\right).
\label{1rst2}
\end{eqnarray}
The possibility to rewrite one-loop topological string amplitudes for the  B-model in terms of the analytic Ray-Singer torsion 
on the target space also for the unoriented open sector will be discussed in more detail in section \ref{1storsion}.

Let us notice that the chamber structure in the amplitudes (\ref{1rst2}) reflects exactly the multiplicative properties 
of the Ray-Singer torsion under vector bundle sums $log \T(V_1\oplus V_2)=log \T(V_1)+log \T(V_2)$ in the specific case of 
the torus.
In fact, the limit of vanishing Wilson line corresponds to the gauge bundle $E=E'\oplus {\cal O}$ and therefore one 
finds $log \T(E' \oplus {\cal O})=log \T(E')+log \T({\cal O})$.

\section{Unoriented topological string amplitudes at one loop}
\label{1hae}

\subsection{Holomorphic anomaly equations}
\label{1hanom}

In last section we considered a B-model topological string on a torus. Now we will generalize the computation to a generic Calabi-Yau 
3-fold. Namely, we will follow the standard BCOV's computation \cite{BCOV} to derive holomorphic anomaly equations for the amplitudes 
of the cylinder, the M\"obius strip, and the Klein bottle. 

Firstly, we compute the cylinder amplitude $\F_{cyl}$. We fix the conformal Killing symmetry and the A- or B-twist on the cylinder by inserting a derivative with respect to the right moduli, that is, K\"ahler moduli for A-model and complex structure moduli for B-model 
of Calabi-Yau moduli space. We get the anomaly equation for the unoriented string amplitude

\begin{equation}
 \frac{\partial}{\partial \bar t^{\bar i}}\frac{\partial}{\partial t^{j}}\F_{cyl}=\frac14\int^\infty_0 ds
\left\langle\int d^2z \left\{(Q^++\bar Q^+),\bar{O}^{[1]}_{\bar i}\right\}\int_l  (G^-+\bar G^-)\int_{l'}O_j^{[1]}\right\rangle,
\end{equation}

where $\int_{l}$ is some space worldsheet integral at fixed time and

\begin{equation}
  \label{1eq:BRST}
  Q_{BRST}=Q^++\bar Q^+,
\end{equation}

and

\begin{equation}
  \bar{O}_{\bar i}^{[1]}:=\frac12[(Q^+-\bar Q^+) ,\bar{O}_{\bar i}], \quad\quad O_j^{[1]}:=\frac12[(Q^--\bar Q^-),O_j].
\end{equation}

The degeneration gives rise to two contributions: the cylinder can reduce to a long narrow tube, and it is called the closed channel, or it can become a short and large diameter one, that is a long thin strip closed on itself; the open channel. This last one is \cite{BCOV}

\begin{equation}
  \label{1eq:open_channel}
 \frac14\bar\partial_{\bar i}\partial_j {\Tr}_{open}(-1)^Flog ~ g_{tt^*},
\end{equation}
where the trace is taken on the open string ground states, $g_{tt^*}$ is the $tt^*$ metric for the open string. 

\begin{figure}[h]
  \centering
  \subfigure[disk two-point function]{ \label{1fig:disc_two_point_solo}
  \input{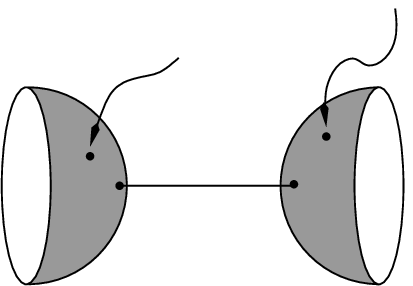}
                   }
\hspace{0.05\textwidth}
  \subfigure[tadpole for the cylinder degeneration]{ \label{1fig:cylinder_tadpole_solo}
 \input{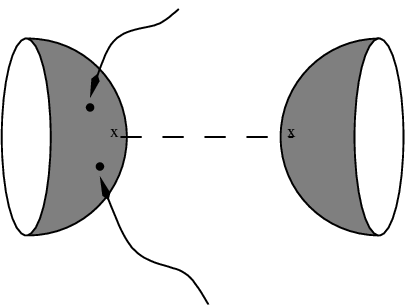}
                   }
\caption[cylinder_tadpole]{the degeneration of a cylinder}
\label{1fig:cylinder_solo}
\end{figure}

For the closed channel, there are two cases.

i) The two operator insertions $\bar{O}_{\bar i}^{[1]}$ and $O_j^{[1]}$ are on different sides. It contributes to the equation by
\begin{equation}
  \label{1eq:closed_twopoint}
  -\bar{\cal D}_{\bar i\bar k}{\cal D}_{jk}g^{\bar kk},
\end{equation}
where $g^{i\bar j}$ is the $tt^*$ metric for the closed string and ${\cal D}_{ij}$ is the disk two-point function (figure \ref{1fig:disc_two_point_solo}). The change in notation from the previous chapter, as we are now calling ${\cal D}$ instead of $\Delta$ the disk two point function, is because for an unoriented theory $\Delta$ will be the disk plus the crosscap, generalizing our old $\Delta$.

ii) The two operator insertions are on the same side. It is a tadpole multiplied by a disk three-point function, where one operator insertion belongs to the wrong moduli, namely, complex structure moduli in A-model and K\"ahler moduli in B-model. In figure \ref{1fig:cylinder_tadpole_solo}, we denote them as  $p$ and $\bar q$, the metric in between is $g^{p\bar q}$.

Next let us consider the amplitude's for a M\"obius strip
\begin{equation}
  \F_{m\ddot{o}b}=\int^\infty_0\frac{ds}{4s}{\Tr}[\P (-1)^FFe^{-2\pi sH}],
\end{equation}
where $\P$ is the involution operator. 

\begin{figure}[h]
  \centering
  \includegraphics[width=0.2\textwidth]{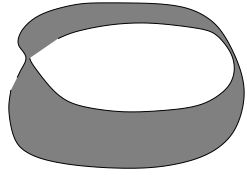}
  \caption{the pinching of a M\"obius strip}
  \label{1fig:pinching_moebius}
\end{figure}

The holomorphic anomaly equation is then
\begin{equation}
\bar\partial_{\bar i}\partial_j \F_{m\ddot{o}b}=\frac14\int^\infty_0 ds\left\langle \P \int d^2z \left\{(Q^++\bar{Q}^{+}), \bar{O}^{[1]}_{\bar i}\right\}\int_l (G^-+\bar{G}^{-})\int_{l'}O_j^{[1]} \right\rangle.
\end{equation}
Now the degeneracy has two types. One is the pinching of the strip, it gives rise to a contribution
\begin{equation}
  \label{1eq:pinching_moebius}
  \frac14\bar\partial_{\bar i}\partial_j{\Tr}_{open}(-1)^F{\cal P}  log ~ g_{tt^*}.
\end{equation}
The only difference between the pinching of a cylinder and of a M\"obius strip (figure \ref{1fig:pinching_moebius}), is the insertion 
of the involution operator $\P$ acting on the remaining strip amplitude. 

\begin{figure}[h]
  \centering
   \subfigure[crosscap two-point function and disk two-point function]{\label{1fig:left_crosscap_solo}
   \input{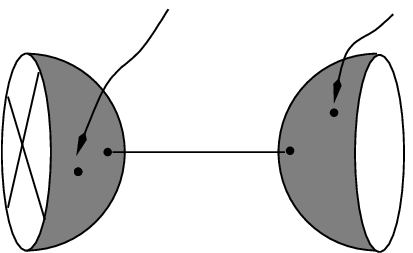}
             }
\hspace{0.1\textwidth}
  \subfigure[tadpole for the M\"obius degeneration]{\label{1fig:left_crosscap_tadpole_solo}
  \input{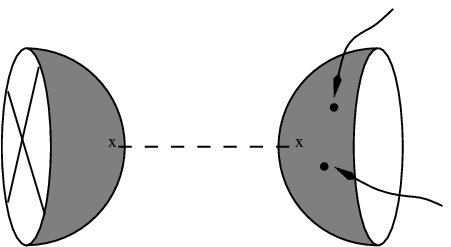}
             }
\caption{the degeneration of a M\"obius strip}
\label{1fig:moebius_solo}
\end{figure}

The remaining degeneration amounts to remove the boundary from the M\"obius strip. There are two cases.

i) The two operator insertions are on the different sides (figure \ref{1fig:left_crosscap_solo}). It gives rise to a disk two-point function multiplied by a crosscap two-point function
\begin{equation}
  \label{1eq:disk_crosscap_twopoint}
  -(\bar {\cal C}_{\bar i\bar k}{\cal D}_{jk}+{\cal C}_{jk}\bar {\cal D}_{\bar i\bar k}) g^{k\bar k}.
\end{equation}
ii) The two operator insertions are on the same side (figure \ref{1fig:left_crosscap_tadpole_solo}). It is a tadpole multiplied by a crosscap three-point function or a crosscap tadpole multiplied by a disk three-point function with one wrong modulus.

\begin{figure}[h]
  \centering
  \subfigure[two crosscap two-point functions]{\label{1fig:klein_twopoint}
  \input{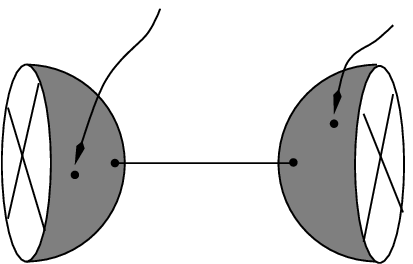}
            }
\hspace{0.1\textwidth}
  \subfigure[tadpole for the Klein degeneration]{\label{1fig:kleintadpole}
  \input{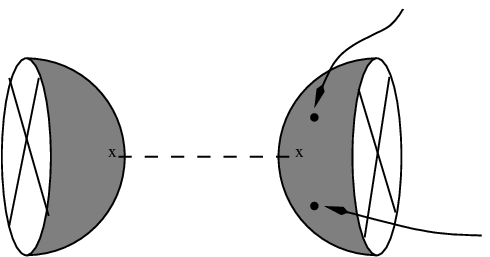}
            } 
  \caption{one degeneration of a Klein bottle}
  \label{1fig:klein_solo}
\end{figure}

Finally, for the Klein bottle we have
\begin{equation}
  \F_{kle}=\int^\infty_0\frac{ds}{4s}{\Tr}[\P (-1)^F F e^{-2\pi sH}].
\end{equation}
There are two degenerations. Firstly, we consider the degeneration that splits the Klein bottle to two crosscaps. Again we have two cases.

i) The two operator insertions are on different sides (figure \ref{1fig:klein_twopoint}).  It gives rise to two crosscap two-point functions
\begin{equation}
  \label{1eq:klein_twopoint}
  -{\cal C}_{ik}\bar {\cal C}_{\bar j\bar k}g^{k\bar k}
\end{equation}

ii) The two operator insertions are on the same side (figure \ref{1fig:kleintadpole}). It gives rise to  a crosscap tadpole multiplied by a crosscap three-point function with one wrong operator insertion.

Secondly, let us consider the complex double of the Klein bottle. Since this is a torus, the holomorphic anomaly equation is inherited from the torus. The only difference is that instead of a Yukawa coupling, we obtain an involution operator acting on the chiral/twisted chiral rings. The doubling torus degeneration gives rise, keeping into account a further factor $1/2$ from left/right projection, to
\begin{equation}
  \label{1eq:Klein}
  \frac18 {\Tr}_{closed} [\P\bar C_{\bar i} C_j].
\end{equation}
This term corresponds to 
\begin{equation}
  \label{1eq:quillen_klein}
 \frac18 \bar\partial_{\bar i}\partial_j {\Tr}_{closed}\P log ~ g,
\end{equation}
where $g$ is the $tt^*$ metric for the closed string.

\subsection[The derivative of the string amplitudes with resp. to the wrong moduli]{The derivative of the string amplitudes with respect to the wrong moduli}

In previous subsection we discussed about the anti-holomorphic dependence of one-loop open string amplitudes of the right moduli $\bar t^{\bar i}$. 
We can also calculate the derivative $\partial_i \F$ with respect to the wrong moduli $y^p$'s. 

Now we will study the different amplitudes separately. 

Firstly, we can consider what is the wrong moduli dependence of $\partial_i\F_{cyl}$.
\begin{equation}
\frac{\partial}{\partial t^{i}}\frac{\partial}{\partial y^{ p }}\F_{cyl}=\frac14\int^\infty_0 ds
\left\langle\int_l  (G^-+\bar{G}^{-})\int_{l'} O^{[1]}_{i}\int d^2z\left\{(Q^++\bar{Q}^{+}),  
[Q^-,{\cal O}_{p}]\right\}\right\rangle
\end{equation}

where we use the same notation $O_{i}^{[1]}=\frac12[(Q^--\bar{Q}^{-}), O_{i}]$. We can check that this operator carries charge $1$.  
We define ${\cal O}^{[1]}_p=[Q^-, {\cal O}_p]$, which has charge $-1$. 
Then we will perform a similar analysis as in the previous subsection. 

1) For the degeneration as the pinching of the two boundaries, we obtain

\begin{eqnarray}
  \label{1eq:wrong_pinching}
&& \eta^{\alpha\beta}\left.\left\langle \Theta_\alpha(-\infty) \int_{l'}O^{[1]}_i\int d^2z{\cal O}^{[1]}_p (z)
\Theta_\beta(+\infty)\right\rangle\right|_{s\rightarrow\infty}\\
&=&\left.\frac12\eta^{\alpha\beta}\left[\langle\alpha |\int_{l'}(Q^--\bar{Q}^{-})
  O_i\int d^2z[Q^-,{\cal O}_p]|\beta\rangle-\langle\alpha |\int_{l'}
  O_i (Q^--\bar{Q}^{-})\int d^2z[Q^-,{\cal O}_p]|\beta\rangle \right]\right|_{s\rightarrow\infty}\nonumber
\end{eqnarray}

where $\alpha, \beta$ are open string ground states and $\eta^{\alpha\beta}$ is the open string topological metric. 
This amplitude is independent of the time ($s$) position of the line $l'$, so we can put it in the center of the infinite strip. 
Thus the first piece of (\ref{1eq:wrong_pinching})  is zero, because the $|\alpha\rangle$ state is projected to zero energy state 
by $e^{-2\pi sH}$ for $s\rightarrow\infty$, and so annihilated by $Q^--\bar{Q}^{-}$. 
The second piece is also zero, because now the $|\beta\rangle$ state is annihilated by $Q^--\bar{Q}^{-}$ for the same reason. 
Notice that the position of $[Q^-, {\cal O}_p]$ does not matter, since it anti-commutes with $Q^--\bar{Q}^{-}$. 
Comparing with the right moduli case (\ref{1eq:open_channel}),  we obtain 

\begin{equation}
  \label{1eq:wrong_quillen}
  \partial_i\partial_p {\Tr}_{open}(-1)^Flog ~g_{tt^*}=0. 
\end{equation}

\begin{figure}[h]
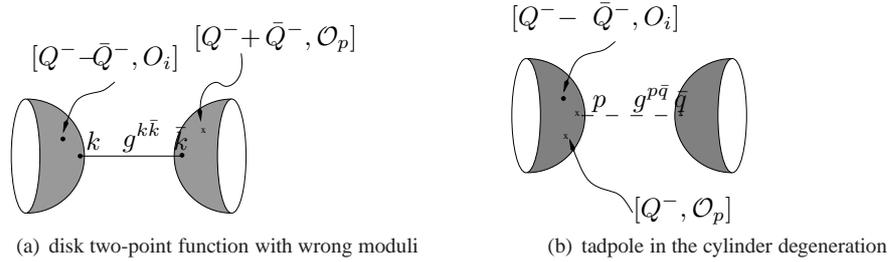

  \centering
  \subfigure[disk two-point function with wrong moduli]{\label{1fig:disc_two_point}
  \input{disc_two_point.tex}
   }
\hspace{0.05\textwidth}
  \subfigure[tadpole in the cylinder degeneration]{\label{1fig:cylinder_tadpole}  
  \input{cylinder_tadpole.tex}
  }
\label{1fig:cylinder_wrong_moduli}
\caption{one degeneration of a cylinder}
\end{figure}

2) The second degeneration is the removing of a boundary from the cylinder. As before, there are two cases. 

i) The two operators insertions are on different sides (figure \ref{1fig:disc_two_point}). 
Since $O_{i}^{[1]}$ and ${\cal O}_{p}^{[1]}$ have charges  $1$ and $-1$ respectively, 
in order to get charge $3$ or $-3$ on the disk we need to project the ground states to $(1, 1)$ and $(-1, -1)$ respectively. 
On one disk which has the wrong type of operator insertion ${\cal O}_p^{[1]}$, we can turn ${\cal O}_p^{[1]}$ into $[Q^-+\bar{Q}^{-}, {\cal O}_p]$. We know  that $G^-$ and $\bar{G}^{-}$ annihilate $(a, a)$ rings,  and $G^-+\bar{G}^{-}$ vanishes on the boundary. Therefore, this diagram does not contribute to the holomorphic anomaly equation. 

ii) The two operators insertions are on the same side (figure \ref{1fig:cylinder_tadpole}). Again we obtain a tadpole multiplied 
by a disk three-point function. 

Secondly, we consider the M\"obius strip. It contains two cases: 1) The pinching of the boundary (see figure \ref{1fig:pinching_moebius})
\begin{equation}
  \label{1eq:pinching_moebius_wrong}
\eta^{\alpha\beta}\left.\left\langle  \Theta_\alpha(-\infty) \P \int_{l'}O^{[1]}_i\int d^2z{\cal O}^{[1]}_p (z)
\Theta_\beta(+\infty)\right\rangle\right|_{s\rightarrow\infty}.
\end{equation}
According to the similar argument as the case of the cylinder (\ref{1eq:wrong_pinching}), we get zero.

\begin{figure}[h]
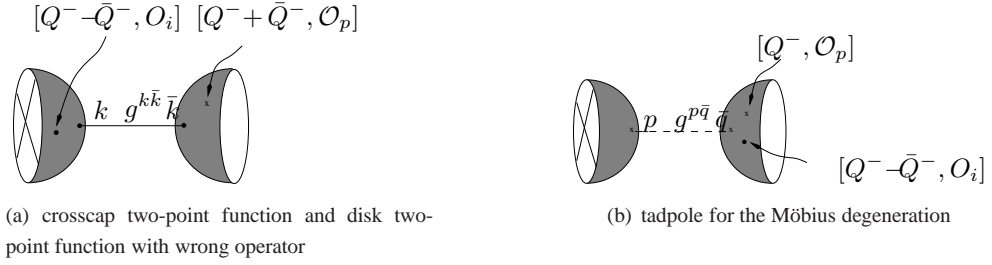

  \centering
  \subfigure[crosscap two-point function and disk two-point function with wrong operator]{\label{1fig:left_crosscap}
  \input{left_crosscap.tex}
  }
\hspace{0.1\textwidth}
  \subfigure[tadpole for the M\"obius degeneration]{\label{1fig:left_crosscap_tadpole}
  \input{left_crosscap_tadpole.tex}
  }
\label{1fig:moebius_wrong_moduli}
\caption{one degeneration of a M\"obius strip}
\end{figure}

2) The removing of the boundary from the M\"obius strip.

i) One operator insertion is near the boundary, and the other is away from the boundary 
(figure \ref{1fig:left_crosscap}).  If ${\cal O}_p^{[1]}$ is near the boundary, the degeneration for that disk will be a $(a, a)$ ring
inserted on the disk. From the same argument as for the cylinder, the disk two-point function is zero. If ${\cal O}_p^{[1]}$ is away 
from the boundary, namely, it is inserted on the crosscap, then that function is also zero. 

ii) The two operators insertions are on the same side (figure \ref{1fig:left_crosscap_tadpole}). We obtain a tadpole multiplied by 
crosscap three-point insertions, or a crosscap multiplied by disk three-point insertions. 

\begin{figure}[h]
  \centering
  \input{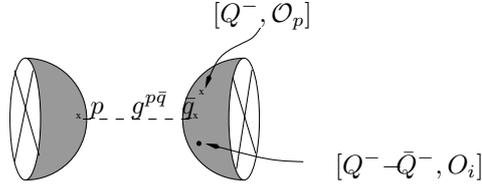}
\caption{one degeneration of a Klein bottle}
  \label{1fig:klein_tadpole}
\end{figure}

Finally, for the Klein bottle, there is only one contribution. That is when the two insertions are on one side, we get the following 
diagram (figure \ref{1fig:klein_tadpole}).

\subsection{Tadpole cancellation at one-loop}
\label{1tadpole_oneloop}

\begin{figure}[h]
  \centering
  \includegraphics[width=0.4\textwidth]{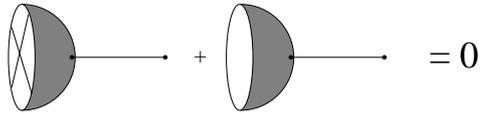}
  \caption{tadpole cancellation for one-loop}
  \label{1fig:tadpole_cancellation}
\end{figure}

When we add up the holomorphic anomaly equations for the cylinder, M\"obius strip, and Klein bottle,
requiring tadpole cancellation 
(figure \ref{1fig:tadpole_cancellation}), we get 
\begin{eqnarray}
  \label{1eq:hol_anomaly}
  \bar\partial_{\bar i}\partial_j [\F_{cyl}+\F_{m\ddot{o}b}+\F_{kle}]&=&\frac18 \bar\partial_{\bar i}\partial_j {\Tr}_{closed} 
\left[\P log ~ g  \right]-\bar\Delta_{\bar i\bar k}\Delta_{jk}g^{k\bar k}, \nonumber\\
&&+\frac{1}{4}\bar\partial_{\bar i}\partial_j {\Tr}_{open} \left[(-1)^F (1+\P)log ~ g_{tt^*}\right]\\
  \partial_i\partial_p [\F_{cyl}+\F_{m\ddot{o}b}+\F_{kle}]&=&0, 
\end{eqnarray}where $\Delta_{ij}={\cal D}_{ij}+{\cal C}_{ij}$ is the sum of the disk and the crosscap two-point function. 

Eq.(\ref{1eq:hol_anomaly}) reproduces the results stated in \cite{Wal3} and extend
them to the presence of non-trivial open string moduli.  

\section{Unoriented one-loop amplitudes as analytic torsions}
\label{1storsion}

In this section we discuss B-model unorientable one-loop amplitudes for generic Calabi-Yau threefolds
and provide a geometrical interpretation of them in terms of holomorphic torsions
of appropriate vector bundles.
 
Let us consider the Klein bottle amplitude first. As we have already seen,
this is given by the insertion of the involution operator ${\cal P}$ in the unoriented closed string trace
as
\begin{equation}
\label{1klein0}
{\cal F}_{kle}= \int_0^\infty \frac{ds}{4s} {\Tr}_{{\cal H}_c} (-1)^F F {\cal P} e^{-2\pi sH_c}
\end{equation}
which we can compute as follows. We recall from (\ref{0sezcit}) that
the closed topological string Hilbert space is given by $H_{\bar{\partial}}^{(0,q)}\otimes\wedge^{m}T^{p,0}X$ so
$
{\cal H}_c
\in
\Lambda^\bullet TX \otimes \Lambda^\bullet \bar T^*X=
\bigoplus_{p,q}\Lambda^p TX \otimes \Lambda^q \bar T^*X
$, where $TX$ and $\bar T^* X$ denote the holomorphic tangent bundle and anti-holomorphic
cotangent bundle respectively. 
At the level of the worldsheet superconformal field theory
these spaces are generated by the zero modes of the $\eta^{\bar I}$ and $\theta_{I}=\theta^{\bar I}g_{\bar I I}$
fermions respectively. The parity ${\cal P}$ acts as ${\cal P}\eta^{\bar I}=\eta^{\bar I}$ and 
${\cal P}\theta_{I}=-\theta_{I}$ \cite{horibrunner}. It is thus clear that the projection operator acts as ${\cal P}=(-1)^p$
on the closed string Hilbert space. 
By inserting in (\ref{1klein0}) the expressions for the total fermion number $F=F_L+F_R=q-p$, a factor of $\frac{1}{2}$
which takes care of left/right identification and the closed string Hamiltonian in terms of the Laplacian $H=\Delta_{p,q}$, we get 
$$
{\cal F}_{kle}=\frac{1}{8}\sum_{p,q}(-1)^{q} q~ log\left( {\det}'\Delta_{p,q}\right)
=-\frac{1}{4}\sum_p log \T \left(\Lambda^p  TX\right)=
-\frac{1}{4}log \T\left(\Lambda^\bullet  TX\right)
$$
in terms of the analytic Ray-Singer torsion $\T(V)$ of the bundle $V=\Lambda^\bullet  TX \sim \Lambda^\bullet T^{*}X$ ( using the well known isomorphism given by contraction with the holomorphic three form on the Calabi Yau )

The cylinder amplitude is given by
$$
{\cal F}_{cyl}= \int_0^\infty \frac{ds}{4s} {\Tr}_{{\cal H}_o} (-1)^F F e^{-2\pi sH_o}
$$
where the assignment of the Chan-Paton factors 
selects ${\cal H}_o\in\oplus_q\Lambda^q \bar{T}^*X \otimes E\otimes E^*$ and the Hamiltonian 
$H_o=\Delta_{E\otimes E^*,q}$ is the corresponding Laplacian.
The result is (as already found in \cite{BCOV})
$$
{\cal F}_{cyl}=-\frac{1}{2}log\T\left(E\otimes E^*\right).
$$

The last term to compute is the M\"obius strip amplitude that is
$$
{\cal F}_{m\ddot{o}b}= \int_0^\infty \frac{ds}{4s} {\Tr}_{{\cal H}_o} (-1)^F F {\cal P} e^{-2\pi sH_o}
$$
The only issue to discuss here is how to compute the trace with the ${\cal P}$ insertion. As explained in (\ref{1chan}) the trace over $\P$, for a $Sp(N/2)$ bundle, selects the $\diag(E\otimes E)$ states with $-1$ eigenvalue. 
This is the only non trivial action of the ${\cal P}$ operator 
on the Hilbert space. Indeed, the boundary conditions project away the $\theta_{\bar I}$'s and we are left with 
the $\eta^I$'s only, on which ${\cal P}$ acts as the identity.
Thus we get 
$$
{\cal F}_{m\ddot{o}b}=+\frac{1}{2}log\T\left({\rm diag}(E\otimes E\right)).
$$

Notice that the above conclusions agree with the explicit calculations of section \ref{1tad1}, once restricted to 
the $T^2$ target space.

\subsection{Wrong moduli independence and anomaly cancellation}
\label{1torsion_bismut}

In this section we show that the decoupling of wrong moduli in the unoriented 
open topological string on a Calabi-Yau threefold $X$ is equivalent to the usual D-brane/O-planes
anomaly cancellation. This is performed for the B-model with a system of $N$ spacefilling D-branes.
These are described by
a Chan-Paton gauge bundle $E$ over $X$ with structure group $U(N)$. 
As it is well known however, in order to implement the orientifold projection, $E\sim E^*$
has to be real therefore reducing the structure group to $SO(N)$ or $Sp(N/2)$ 
if the fundamental representation is real or pseudo-real respectively.

Let us now calculate the variation of the unoriented topological string free energy at one loop
under variations of the K\"ahler moduli.
In order to do it, we use the Bismut formula \cite{bismut} for the variation of the Ray-Singer torsion 
under a change of the base and fiber metrics $(g,h)\to (g+\delta g, h+\delta h)$
\begin{equation}
\left.\frac{1}{2\pi}\frac{\partial}{\partial t}\right\vert_{t=0} log \T(V)
=
\left.\frac{1}{2}\int_X \frac{\partial}{\partial t}\right\vert_{t=0}
\left[
Td\left(\frac{1}{2\pi}\left(iR + t g^{-1}\delta g\right)\right)
Ch\left(\frac{1}{2\pi}\left(iF + t h^{-1}\delta h\right)\right)
\right]_{8}
\label{1bismut}
\end{equation}

By applying the Bismut formula to the whole unoriented string free energy
${\cal F}^u_{\chi=0}=
{\cal F}_{cyl}+{\cal F}_{m\ddot{o}b}+
{\cal F}_{kle}$
and specializing to the variations of the K\"ahler form only (that is at a fixed metric on the Chan-Paton holomorphic vector bundle) 
we get\footnote{Here and in the following calculations we insert for convenience a formal parameter $n_o$ which keeps track
of the number of crosscaps. It will be eventually put to $1$.} 
\begin{equation}
\sim\int_X\left\{ \left[(Ch(E))^2 - n_o Ch({\rm diag}(E\otimes E))\right]
\frac{\partial}{\partial t}[ Td(TX)]_{t=0} + \frac{1}{2}n_o^2 
\frac{\partial}{\partial t} \left[Td(TX)
Ch(\Lambda^\bullet T^*)\right]_{t=0}\right\}
\label{1variation}
\end{equation}
We will use $ch_k(2E)=2^kch_k(E)$ and $ch_k(E^*)=(-1)^kch_k(E)$, so that for $E=E^*$, $ch_k(E)=0$ for $k$ odd. 
From the definitions\footnote{See the book \cite{hirzebruch} for the notation.}
\begin{equation}
Td(TX)=\prod_a \frac{\gamma_a}{1-e^{-\gamma_a}}
\end{equation}
\begin{equation}
Ch(\Lambda^\bullet T^*X)=\prod_a (1+e^{-\gamma_a})
\end{equation}
we rewrite $Td(TX)=e^{c_1(T)/2}\hat A(TX)$ and $Td(TX)Ch(\Lambda^\bullet T^*X)=2^3 L(TX)$.
Using the standard expansions 
\begin{eqnarray}
\hat A&=& 1 - \frac{2}{3}p_1 2^{-4} +\frac{2}{45}\left(-4p_2+7p_1^2\right) 2^{-8} + \ldots
\\ \nonumber
L&=& 1 + \frac{1}{3}p_1 2^{-2} +\frac{1}{45}\left(7p_2-p_1^2\right) 2^{-4} + \ldots
\label{1expans}
\end{eqnarray}
in (\ref{1variation}) we 
calculate the variations of the cohomology classes above and obtain
\begin{equation}
\delta {\cal F}^u_{\chi=0} = \left.\int_X \sum_{i=1}^2 C_i \frac{\partial}{\partial t}p_i\right|_{t=0}
\label{1variation2}
\end{equation}
with 
\begin{eqnarray}
C_1&=&-\frac{2^{-3}}{3}J_4(E)+\frac{2^{-6}\cdot 7}{45}p_1J_0(E)-\frac{2^{-1}}{45}n_o^2p_1\label{1coeff0}
\\
C_2&=&-\frac{2^{-5}}{45}J_0(E)+\frac{2^{-2}\cdot 7}{45}n_o^2
\label{1coeff}
\end{eqnarray}
where $J(E)=(Ch(E))^2-n_oCh({\rm diag}(E\otimes E))=J_0(E)+J_4(E)+\ldots$.
One verifies that, setting $n_o=1$, the vanishing of the coefficients (\ref{1coeff0}) and (\ref{1coeff}) is realized by
\bea 
ch_0(E) =8
\\ \nonumber
ch_2(E)= \frac{1}{4} p_1
\label{1queste}
\eea
that can be rewritten in the more familiar form
\be
\sqrt{\hat A(TX)}Ch(E)-2^3\sqrt{\hat L(TX)}=0
\label{1tadpo}\ee
that is\footnote{We denoted $\hat L=\prod_i\frac{\gamma_i/4}{th\left(\gamma_i/4\right)}$.}
the tadpole/anomaly cancellation condition for a system of spacefilling D-branes/O-planes
on a Calabi-Yau threefold \cite{anomacanc}.

\subsection{Quillen formula and holomorphic anomaly}
\label{1Quillen}

In this subsection we compute the holomorphic anomaly equations of Section \ref{1hae} from the expressions 
of the free energies in terms of Ray-Singer analytic torsion.

In order to do this, we apply the Quillen formula for torsions 
\begin{equation}
\partial\bar\partial log[T(V)]= \partial\bar\partial \sum_q \frac{(-1)^{q+1}}{2} log[\det g^{(q)}_V] -
\pi i \int_X \left[Td(TX)Ch(V)\right]_{(4,4)}
\label{1Q}
\end{equation}
 where $\det g^{(q)}_V$ is the volume element in the kernel of $\bar\partial_V$ on $\Lambda^q T^* X\otimes V$ 
and $V$ is the relevant vector bundle for each contribution (that is $V_{cyl}=E\otimes E$, etc. see the beginning of the section) corresponding 
to the first and the last terms in the r.h.s. of formula
(\ref{1eq:hol_anomaly}), while its second term matches the second term above.

In the notation of the previous subsection we get, up to the $\partial\bar\partial$-volume terms and setting $n_o=1$
\begin{eqnarray}
\partial\bar\partial {\cal F}^u_{\chi=0}= 
-\frac{\pi i}{2} \int_X \left[Td(TX)J(E)+\frac{1}{2}Td(TX)Ch(\Lambda^\bullet T^*X)  \right]_{(4,4)}  \\
+\frac{1}{4}\partial\bar\partial\left[\sum_q(-1)^q
\left(log\left[\det g^{(q)}_{E\otimes E^*}\right]]-log\left[\det g^{(q)}_{diag(E\otimes E)}\right]
+\frac{1}{2}log\left[\det g^{(q)}_{\Lambda^\bullet T^*X}\right]
\right)\right] \nonumber 
\label{1Q1}
\end{eqnarray}
which we can calculate using the expansions (\ref{1expans}) for the vector bundle $E$
satisfying (\ref{1queste}).
The first line of the previous equation is
\begin{eqnarray}
\pi i \int_X \left(
\left(ch_2(E)\right)^2 + \hat A_4\cdot 12 ch_2(E)+ 7\cdot 8\hat A_8+4L_8
\right)=0.
\label{1Q2}
\end{eqnarray}
and vanishes. This result means that, once tadpoles are canceled, 
the $\bar\Delta\Delta$-term in (\ref{1eq:hol_anomaly}) vanishes 
{\it for spacefilling branes/orientifolds}.
This is in agreement with the result for $T^2$ target found in section \ref{1tad1}.


\section{Tadpole cancellation at all loops}
\label{1all}

\subsection[Compactification of the moduli space...]{Compactification of the moduli space of Riemann surfaces with boundaries}
\label{1comp}

\begin{figure}[h]
  \centering
  \includegraphics[width=0.5\textwidth]{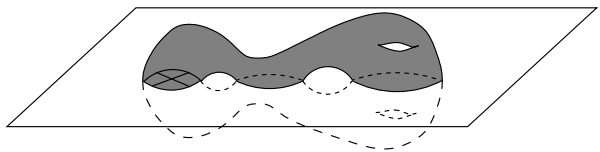}
  \caption{one example of an anti-holomorphic involution on $\Sigma$}
  \label{1fig:antiholomorphic_involution}
\end{figure}

The moduli space of Klein surfaces with boundaries $\Sigma$ can be usefully described by referring 
to the notion of complex double $\left(\Sigma_{\mathbb C},\Omega\right)$, that 
is a compact orientable connected Riemann surface with an anti-holomorphic involution $\Omega$
(see figure \ref{1fig:antiholomorphic_involution}).

The topological type of $\Sigma=\Sigma_{\mathbb C}/\Omega$ is classified by the fixed locus $\Sigma_{\mathbb R}$
of the involution \cite{natanzon}. 
If $\Sigma_{\mathbb R}=\emptyset$, then $\Sigma$ is non orientable and without boundaries, while if
$\Sigma_{\mathbb R}$ is not empty, then $\Sigma$ has boundaries. In the latter case, 
$\Sigma$ is orientable
if $\Sigma_{\mathbb C}\setminus\Sigma_{\mathbb R}$  is not connected and non orientable otherwise.

We recall that on a local chart $z\in{\complex}^*$, the anti-holomorphic involution acts as
$\Omega_\pm(z)=\pm\frac{1}{\bar z}$. The involution $\Omega_+$
has a non empty fixed set with the topology 
of a circle, which after the quotient becomes a boundary component. 
The involution $\Omega_-$ doesn't admit any fixed point and leads to a crosscap.

\begin{figure}[h]
  \centering
  \includegraphics[width=0.8\textwidth]{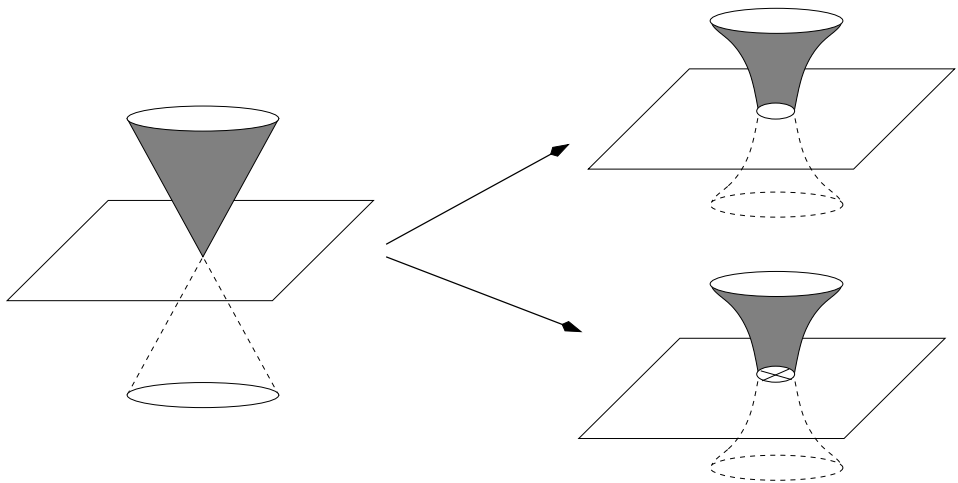}
  \caption{the complete resolution of a cone singularity in the doubling space}
  \label{1fig:cone}
\end{figure}

The compactification of the moduli space of open Klein surfaces can be studied from the point of 
view of the complex double \cite{KL}. 
In this context, the boundary is given as usual by nodal curves, but with respect to the closed orientable 
case there are new features appearing due to the quotient.
In particular, nodes belonging to $\Sigma_{\mathbb R}$ can be smoothed either as boundaries or as crosscaps
(see figure \ref{1fig:cone}).
Thus the moduli spaces of oriented and non orientable surfaces intersect at these boundary components
of {\it complex} codimension one.

\begin{figure}[h]
  \centering
  \includegraphics[width=0.8\textwidth]{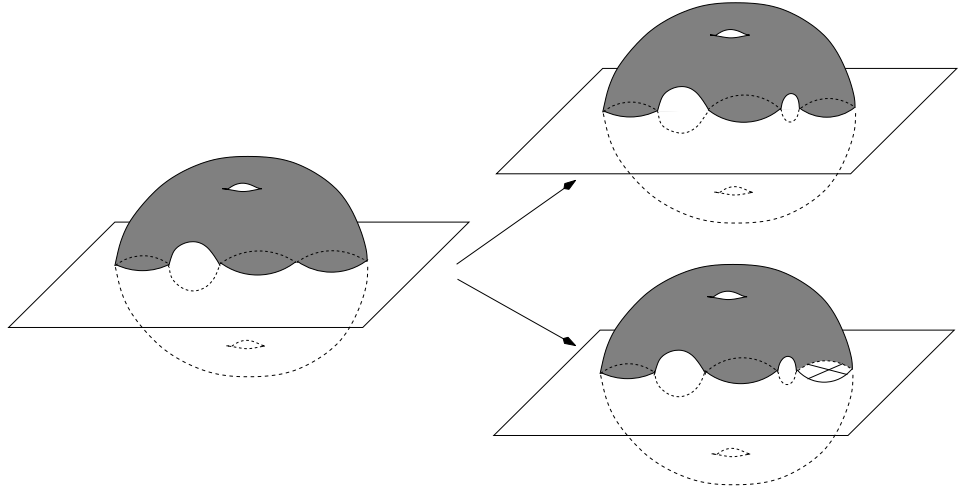}
  \caption{the resolution of a pinching point in the doubling space}
  \label{1fig:pinching_doubling}
\end{figure}

Actually there are also boundary components of {\it real} codimension one
which are obtained when the degenerating 1-cycle of the complex double intersects 
$\Sigma_{\mathbb R}$ at points.
In this cases, one obtains the boundary open string degenerations as described in (\ref{0trallalla}).
The resolution of the real boundary nodes can be performed either as straight strips 
or as twisted ones. For example, when we have  colliding boundaries, their singularity can be resolved
either as splitting in two boundary components or as splitting in a single boundary and a crosscap
(see figure \ref{1fig:pinching_doubling}). 
Thus the moduli space of oriented and unoriented surfaces intersect also along these components.
For a more detailed and systematic description, see \cite{chiuchu}.

More precisely, as discussed also in \cite{blau}, the moduli space of the quotient surface $\Sigma$ is obtained by 
considering the relative Teichm\"uller space $T(\Sigma_\complex,\Omega)$, that is the $\Omega$-invariant locus of 
$T(\Sigma_\complex)$, modding the large diffeomorphisms $\Gamma(\Sigma_\complex,\Omega)$ 
which commute with the involution $\Omega$ 
\be
{\cal M}_\Sigma=T(\Sigma_\complex,\Omega)/\Gamma(\Sigma_\complex,\Omega)
\label{1moduli}\ee

\begin{figure}[h]
  \centering
  \psfrag{tau}{$\tau$}
 \includegraphics[width=0.3\textwidth]{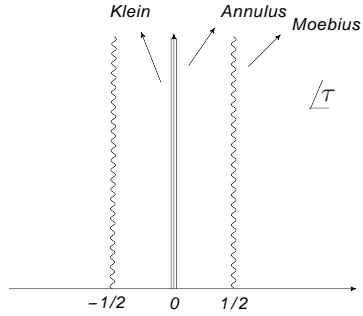}
  \caption{different involutions represent in the moduli space of Torus}
  \label{1fig:involution_F1}
\end{figure}

Let us consider as an example the case of null Euler characteristic.
In this case the complex double is a torus and the annulus, M\"obius strip and Klein bottle
can be obtained by quotienting different anti-holomorphic involutions. 
The conformal families of tori admitting such involutions are 
Lagrangian submanifolds in the Teichm\"uller space of the covering torus modded by\footnote{The other generator $S$ of 
$\Gamma(T^2)=PSL(2,\mathbb{Z})$ is not quotiented because it does not commute with the involutions.} 
the translations $\tau\to\tau+1$
$\left\{\tau\in\complex | Im(\tau)>0, -\frac{1}{2}\leq Re(\tau)\leq \frac{1}{2}\right\}$.
These are vertical straight lines at $Re(\tau)=0$ for the annulus and the Klein bottle while
at $Re(\tau)=\pm\frac{1}{2}$ for the M\"obius strip
\footnote{Notice that the annulus and the Klein bottle are distinguished by different anti-holomorphic involutions.}
(see figure \ref{1fig:involution_F1}).
Notice that all vertical lines meet at $\tau=i\infty$ which is the intersection point of the different moduli spaces.

At a more general level, one should similarly discuss the moduli space of holomorphic 
maps from the worldsheet $\Sigma$ to the Calabi-Yau space $X$ with involution $\sigma$ in terms of 
equivariant maps $(\Sigma_{\mathbb C},\Omega) \to (X, \sigma)$ \cite{KL}. The above discussion suggests that 
the proper definition of open topological strings can be obtained by summing over all possible nonequivalent 
involutions of $\Sigma_{\mathbb C}$.  In particular one should include the contribution of non-orientable surfaces
in order to have a natural definition of the compactification of the space of stable maps.
 Actually, once the perturbative expansion of the string amplitudes
is set in terms of the Euler characteristic of the worldsheet, we have to
sum over all possible contributions ${\cal F}_\chi=\sum_{g,h,c|\chi=2-2g-h-c}{\cal F}_{g,h,c}$
at given genus $g\geq 0$, boundary number $h\geq 0$ and crosscaps number $0\leq c\leq 2$.

At fixed Euler characteristic, the set of Riemann surfaces admitting an anti-holomorphic involution
is a Lagrangian submanifold $L_\Omega$ of the Teichm\"uller space of the complex double $T_{\Sigma_{\mathbb C}}$
as in formula (\ref{1moduli}).
Actually it might happen that the same Lagrangian submanifold corresponds to Riemann surfaces admitting
nonequivalent involutions which have to be counted independently, as for the example of the annulus and
Klein bottle that we just discussed.
The complete amplitude is then given schematically as
$$
{\cal F}^u_\chi= \sum_\Omega
\int_{T_{\Sigma_{\mathbb C}}} \delta(L_\Omega)
\int_{\{O:\Sigma\to X, O\circ\Omega=\sigma\circ{O}\}}
|\mu G^-|^{3\chi}
$$
which provides a path integral representation for open/unoriented topological string amplitudes.

The above is the counterpart in topological string of the well-known fact that in open superstring theory 
unoriented sectors are crucial in order to obtain a consistent (i.e. tadpole and anomaly free) theory at all loops
\cite{BS}.
Evidence of these requirements has been found from a computational point of view in \cite{Wal3}
where the contribution of unoriented surfaces has been observed to be necessary to obtain integer
BPS counting formulas for A-model open invariants on some explicit examples.
Let us remark that this picture applies to any compact or non compact Calabi-Yau threefold in principle. 
It might happen, however, that in the non compact case for some specific D-brane geometries tadpole cancellation 
can be ensured by choosing suitable boundary conditions at infinity 
so that the orientable theory is consistent by itself as in the case of \cite{KL,SV}.

\subsection{Local tadpole cancellation and holomorphic anomaly}
\label{1local}

As we have seen in the Section \ref{1comp}, non-orientable Riemann surfaces should be included in 
order to provide a consistent compactification of the moduli space of open strings.
It was found in \cite{new} that a dependence on wrong moduli appears when one considers
holomorphic anomaly equations for orientable Riemann surfaces with boundaries.
However, it follows from the discussion of Section \ref{1comp} that whenever we consider a closed string degeneration
in which one of the boundaries shrinks ( that is it is pushed far away ), there is always a corresponding component in the boundary
of the complete moduli space where one crosscap is sent to infinity as well.  
Therefore, we always have this type of degeneration 
\begin{equation}
  \label{1eq:tadpole_crosscap}
  {\cal A}^p(\langle\omega_p|B\rangle +\langle\omega_p|C\rangle),
\end{equation}
where $|B\rangle$ and $|C\rangle$ are the boundary and crosscap state respectively, 
$\omega_p$ is the operator inserted in the degenerated point which corresponds to a wrong modulus, 
${\cal A}^p$ is the amplitude of the remaining Riemann surface with a wrong moduli operator insertion.  
Tadpole cancellation implies
\begin{equation}
  \label{1eq:tadpole_cancellation}
  \langle\omega_p|B\rangle +\langle\omega_p|C\rangle=0.
\end{equation}
which ensures the cancellation of the anomaly of \cite{new} at all genera.
This cancellation has a simple geometrical interpretation in the A-model:
in this case, we can have D6-branes and O6-planes wrapping 
3-cycles of the Calabi-Yau 3-fold $X$, and the condition (\ref{1eq:tadpole_cancellation}) reads 
\begin{equation}
  \label{1eq:tadpole_geometry}
  \langle \omega_p|B\rangle+\langle\omega_p|C\rangle=\partial_{y^p}\left(\int_L \Omega^{(3,0)}+ \int_{X^\sigma}\Omega^{(3,0)}\right)=0,
\end{equation}where $L$ is a Lagrangian 3-cycle, $\Omega^{(3,0)}$ is the holomorphic 3-form, and $X^\sigma$ is the fixed point set of the involution $\sigma: X\rightarrow X$. From 
(\ref{1eq:tadpole_geometry}) we can interpret the local cancellation of the wrong moduli dependence (\ref{1eq:tadpole_crosscap})
as a stability condition for the vacuum against wrong moduli deformations.

    \chapter{Target space theories}
\label{sec:cap2}
\noindent

\section{Introduction}

A proper target space formulation of open plus closed topological strings is important for several reasons,
the most compelling in our opinion being a better understanding of open/closed string duality which, once an off shell
formulation of the theory is given, should become manifest.
Actually, this is the main subject of the next chapter. For now we will limit first to a review of the known target space theories for open and closed strings, and then to a derivation ( plus some checks ) of the complete open and closed target space effective field theory.

The topological open string target space formulation has been actually obtained long ago in \cite{W1}
where it was shown to be given by the Chern-Simons theory for the A-model and its holomorphic version
for the B-model. These are formulated for a fixed on shell background geometry, in particular for the B-model
the holomorphic Chern-Simons is formulated with respect to an integrable complex structure on the Calabi-Yau
target.

Since the aim of this paper is to study a string field theory formulation of topological open plus closed
strings on equal footing, we will extend this framework to non-integrable structures. In particular the closed string field theory of the B-model
derived in \cite{BCOV}, the so called Kodaira-Spencer theory of gravity, is nothing but a consistent theory for generic non integrable complex structures which, only on shell, reduce to a true, integrable one. So the idea is to write an action containing the holomorphic Chern-Simons functional written with respect to a generic non integrable complex structure, weighted by the Kodaira-Spencer action. The result will mainly deal with the coupling between closed and open strings, and this will be analyzed in detail. In particular we will discuss the relation among our result and some statements already done by Witten in \cite{W1}.

As we have seen a distinctive feature of topological strings is that the non-holomorphic dependence of its amplitudes can be
recursively computed by means of the holomorphic anomaly equations. 
It turned out that the target space formulation of the closed string in terms of the Kodaira-Spencer gravity
is very effective in reproducing these recurrence relations from a Feynman diagram's expansion.
This also provides a target space interpretation of the various coefficients appearing in the HAE.
These latter have been more recently extended to open strings in \cite{Wal1} and \cite{BT}.
These were further studied in \cite{Alim}. We will extend this analysis to all the building block tree level amplitudes appearing in (\ref{0trallalla}), both as a check for our theory and as a prediction for their value in terms of target space quantities.

This and the next chapter are based on \cite{Bonelli:2010cu}.

\section{Open and closed effective field theories}

It is well known from \cite{BCOV} that the effective space-time theory corresponding to the B-model for closed strings is given by the Kodaira-Spencer 
theory of gravity:

\begin{equation}\label{2d}
\lambda^{2}S_{KS} = \int_{X}\frac{1}{2}A'\frac{1}{\partial}\bar{\partial}A' - \frac{1}{3}[(A + x)(A + x)]'(A + x)'
\end{equation}

where $\lambda$ is the string coupling and
$A$ and $x$ are $(0,1)$ forms with values in the $(1,0)$ vector field that is, in coordinates, 
$A = A_{\bar{i}}^{j}dz^{\bar{i}}\frac{\partial}{\partial z^{j}}$ and similarly for $x$. 
In (\ref{2d})
$A'=i_A\Omega_0=3(\Omega_0)_{ijk}A^i_{\bar i}dz^j dz^k dz^{\bar i}$ and similarly for $x'$ where
$\Omega_{0}$ is
the holomorphic three form on the Calabi-Yau target space $X$\footnote{Factors may change depending on the conventions; 
we will use the ones of \cite{Tod} and \cite{KNS}.}.
$A + x$ is defined to be a ( non integrable ) deformation of the complex structure of $X$ split into an infinitesimal part, $x$, required to be a proper integrable deformation, and a generic finite one, $A$. 
The full deformation, $A + x$, is parametrized by the shift 
$\bar{\partial}_{\bar{i}} \rightarrow \bar{\partial}_{\bar{i}} - (x_{\bar{i}}^{j} + A_{\bar{i}}^{j})\partial_{j} $. 
By definition the coefficients of forms with barred indices transform in the same way : 
$w_{\bar{i}} \rightarrow w_{\bar{i}} - (x_{\bar{i}}^{j} + A_{\bar{i}}^{j})w_{j} $. 
In addition  $dz^{j} \rightarrow dz^{j} + ( x^{j}_{\bar{i}} + A^{j}_{\bar{i}} )dz^{\bar{i}}$, while
 $\partial$ and $d\bar{z}$ are fixed (their shift would refer to the antitopological theory). 
In this way real objects as the de Rham differential $d$ or a real form $w_{i}dz^{i} + w_{\bar{i}}dz^{\bar{i}}$ remains unchanged. 
The condition of integrability of the modified complex structure is   

\[
0 = (\bar{\partial} - x - A)(\bar{\partial} - x - A) = - \bar{\partial}(A + x) + \frac{1}{2}[A + x,A + x] = 0
\]

Because $x$ does not appear in the kinetic term of (\ref{2d}) it is interpreted as a background parameter, valued in 
$H^{0,1}_{\bar{\partial}}\otimes T^{(1,0)}X$, and its property to correspond to an integrable deformation of the complex structure, $\bar{\partial}x = 0$, is exactly the requirement to drop from the kinetic term. Then the condition of integrability for $A + x$ translates into an equation for the only $A$ 
\be 
\bar{\partial}A'  = \partial((A + x)\wedge ( A + x ))' .
\label{2kseom}\ee

which is nothing but the equation of motion of (\ref{2d}). This is not surprising as (\ref{2d}) was explicitly built to fulfill this requirement: integrability condition $\leftrightarrow$ equation of motion.

Being the kinetic term non local, at least for the fields we are using, $A$ is required to satisfy the so called Tian's gauge, $\partial A' = 0$, 
in order to make it well defined. In summary we can say we are dealing with a theory of finite non integrable complex structures written with respect to a fixed integrable one ( the one with respect to which (\ref{2d}) is written ) and an infinitesimal deformation required to be integrable as well. The equation of motion for the fields representing these deformations correspond to the requirement of integrability, so that $x$ is correctly translated into a static background field while $A$ is off shell.  

The symmetries of (\ref{2d}) turn out to be the $\Omega_0$ preserving reparametrization of the complex coordinates
\[
z^{i} \rightarrow z^{i} + \chi^{i}(z,\bar{z}) \;\;\;\; z^{\bar{i}} \rightarrow z^{\bar{i}} 
\]

while the condition of being $\Omega_{0}$ preserving reads $\partial\chi' = 0$. 
According to $\bar{\partial} \rightarrow \bar{\partial}  - (A + x)$, we can apply to both sides the above $\Omega_{0}$-preserving coordinate transformations and define the variation of $A$ to include all the variations of the right hand side. This because $x$ is kept fixed being a background; so 

\begin{equation}\label{2e}
\delta A =  -\bar{\partial}\chi - {\cal L}_{\chi}(A + x) = -\bar{\partial}\chi - [\chi,(A + x)]
\end{equation}

Reinterpreting $\chi$ as a ghost field, 
this transformation can be promoted to a nilpotent BRST if
\begin{equation}
\delta \chi = -\frac{1}{2}{\cal L}_{\chi}\chi = -\chi^{i}\partial_{i}\chi.
\end{equation}

The open effective theory has been analyzed by Witten in \cite{W1} and for the B-model it is 
given by the holomorphic Chern-Simons action
\begin{equation}\label{2a}
\lambda S_{HCS} = \int_{X}\Omega_{0} Tr(\frac{1}{2}B^{0,1}\bar{\partial}B^{0,1} + \frac{1}{3}B^{0,1}B^{0,1}B^{0,1})
\end{equation}
with $B^{0,1}$ a Lie algebra valued $(0,1)$-form. 

Indeed (\ref{2a}) is globally ill defined; the precise definition of the model has been presented in \cite{Tho}.
From the Chern-Weil theorem we know that only the difference of two invariant polynomials with respect to two different connections 
$\hat{B}$ and $B_{0}$ (dropping for the moment the label $(0,1)$) is an exact form. So using the reference connection $B_{0}$ we can write
\begin{eqnarray}\label{2m}
-\int_{K_{4}}\frac{\Theta}{2} Tr(\hat{F}^{2} - F_{0}^{2}) &=& -\int_{K_{4}}\Theta Tr \;\bar{\partial}(\frac{1}{2}\hat{B}\bar{\partial}\hat{B} + \frac{1}{3}\hat{B}^{3} - \frac{1}{2}B_{0}\bar{\partial}B_{0} - \frac{1}{3}B_{0}^{3}) = \nonumber \\
&=& \int_{X}\Omega_{0} Tr(\frac{1}{2}\hat{B}\bar{\partial}\hat{B} + \frac{1}{3}\hat{B}^{3} - \frac{1}{2}B_{0}\bar{\partial}B_{0} - \frac{1}{3}B_{0}^{3})
\end{eqnarray}
where $K_{4}$ is a fourfold containing $X$ as a divisor while $\Theta$ is a connection of the associated line bundle $\call_X$ so that
$\bar{\partial}\Theta = \Omega_{0}\delta(X)$. 
We expand $\hat B$ with respect to the reference connection as
\[
\hat{B} = B + B_{0} 
\]
so that
(\ref{2m}) provides the globally well defined action  
\begin{equation}\label{2b}
\lambda S_{HCS} = \int_{X}\Omega_{0} Tr(\frac{1}{2}B^{0,1}\bar{\partial}_{B_{0}^{0,1}}B^{0,1} + \frac{1}{3}(B^{0,1})^{3} + F^{0,2}_{0}B^{0,1})
\end{equation}

with $\bar{\partial}_{B_{0}^{0,1}}\varphi \equiv \bar{\partial}\varphi + [B_{0}^{0,1},\varphi]_{\pm}$ with $\pm$ 
depending on the grade of the form $\varphi$. $B_{0}$ is the open counterpart of what was $x$ for the closed theory: it is the open string background and as such
it obeys the holomorphicity condition ( equations of motion ) $F_{0}^{0,2} = 0$. 

The symmetries of (\ref{2b}) -- at fixed background $B_0$ -- are given by

\begin{equation}\label{2gt}
 \delta B^{0,1} = \bar{\partial}_{B_{0}^{0,1}}\epsilon + [B^{0,1},\epsilon].
\end{equation}

\section{Open-Closed effective field theory}

Now we want to explicitly couple the open theory to the closed field that is we want to deform the complex structure of $X$, 
over which the theory is defined, using the fields $A$ and $x$. 
Of course the closed field $A$ is in general not on shell so the new complex structure (better call it almost complex structure) 
is generically not integrable. 
In addition we want to write the new action with respect to the undeformed complex structure in order to keep the closed field explicit. 
Actually, under the deformation $\Omega_{0}$ is mapped to \cite{Tod}
\begin{equation}\label{2x}
\Omega = \Omega_{0} + (A + x)' - [(A + x)(A + x)]' - [(A + x)(A + x)(A + x)]'
\end{equation}
which is a $(\tilde{3},\tilde{0})$ form with respect to the new complex structure (from now on always indicated with a tilde) 
while with respect to the old one
decomposes in forms of total degree 3, namely $(p,q)$ forms with $p+q=3$.
We can now deform also the remaining $(0,3)$ part of the action, $L_{CS}^{0,3}$, with 
$ L_{CS}^{0,3} \equiv Tr(\frac{1}{2}B^{0,1} \bar{\partial}_{B_{0}^{0,1}}B^{0,1} + \frac{1}{3}(B^{0,1})^{3} + F^{0,2}_{0}B^{0,1}) $,  
into a $(\tilde{0},\tilde{3})$ form. 

In order to keep into account the deformation of the complex structure of the full action the simplest way is to use a real form for 
the Chern-Simons term, rewriting
\be
\int_{X}\Omega^{\tilde{3},\tilde{0}} L_{HCS}^{\tilde{0},\tilde{3}} = \int_{X}\Omega^{\tilde{3},\tilde{0}}
L_{CS}=\int_{X}\Omega^{\tilde{3},\tilde{0}}Tr\left(\frac{1}{2}Bd_{B_{0}}B + \frac{1}{3}B^{3} + F_{0}B\right)
\label{2real}\ee
where $B$ is a real Lie algebra valued 1-form on $X$. Indeed,
being $\Omega$ a $(\tilde{3},\tilde{0})$ form, the added piece is zero.
 However, from the path integral quantization viewpoint, we have to define a suitable measure for the new field component 
$B^{\tilde{1},\tilde{0}}$. We will discuss this issue in the next section by using the Batalin-Vilkovisky formalism\footnote{For a complete discussion on the Kodaira-Spencer gravity in antifield formalism see \cite{BCOV}}.

.
Let us notice that the real form $L_{CS}$ is completely independent from the closed field, while it is $\Omega$ which really takes care to project the 
action onto the new complex structure selecting the complementary form degree from $L_{CS}$.

Let us consider the symmetries of (\ref{2real}). As far as diffeomorphisms (\ref{2e}) are concerned, $\Omega$ in (\ref{2x}) transforms as ${\cal L}_{\chi}\Omega$
so that the whole action is invariant under the standard action on $B$, namely $\delta B=-\call_\chi B$.

The situation for the Chan-Paton gauge symmetry is more subtle. Indeed, being the field $A$ off shell, we do not have $d\Omega = 0$. In fact it can be shown \cite{Tod} that $d\Omega = 0$ is equivalent to the equations of motion 
for the Kodaira-Spencer action, $\bar{\partial}A' = \partial((A + x ) \wedge ( A + x ))'$. 
The real version of (\ref{2gt}) is

\begin{equation}
\delta B = d_{B_{0}}\epsilon + [B,\epsilon]  
\end{equation}

and if $d\Omega\neq 0$ under this transformation we have that the action (\ref{2real}) is no longer gauge invariant but its variation amounts to

\begin{equation}
\delta S_{HCS} = \frac{1}{\lambda }\int_{X}\Omega Tr\; d(\frac{1}{2}\epsilon d_{B_{0}}B + F_{0}\epsilon)
\end{equation}
We can save the day by adding the term $-\frac{1}{2}\Omega db$, where $b$ is a real 2-form field transforming as \cite{BW}:  
\begin{equation}
\delta b = Tr(\epsilon d_{B_{0}}B +2 F_{0}\epsilon) 
\end{equation}

We can either keep the field $b$ or integrate it out. In the first case transforming all the fields in the modified action leaves it gauge invariant. Instead it is clear that the field $b$ acts as a Lagrange multiplier whose integration enforces the Kodaira-Spencer equations for the closed field $A$ so $d\Omega = 0$ and the action is again gauge invariant. 
However the role of implementation of the associated delta function requires also a determinant factor such that
\begin{equation}\label{2fp}
\int {\cal D}A{\cal D}b e^{-\frac{1}{2}\int_{X}\Omega db} det_{FP}  = 1
\end{equation}
This determinant measure has to be included in the very definition of the theory and will be explicitly derived in the next section.

This isn't really the end of the story as $b$ has shift symmetries along its $(\tilde{2},\tilde{0})$ and $(\tilde{1},\tilde{1})$ components.
In addition we should specify the full nilpotent symmetries and the gauge fixing. This will be the subject of the next section. 

Summarizing, the classical action for open and closed B-model is
\begin{equation}\label{2f}
S_{tot}= \frac{1}{\lambda^{2}}\int_{X}\left(\frac{1}{2}A'\frac{1}{\partial}\bar{\partial}A' - \frac{1}{3}[(A + x)(A + x)]'(A + x)' \right)+
\end{equation}
\[
+ \frac{1}{\lambda}\int_{X} \Omega Tr(\frac{1}{2}Bd_{B_{0}}B + \frac{1}{3}B^{3} + F_{0}B) -\frac{1}{2}\Omega db 
\]

\section{On the BV quantization of Holomorphic Chern-Simons}

In this section we provide the BV action for the holomorphic Chern-Simons theory and a non singular gauge fixing fermion. 
For simplicity in this section we will drop the tilde in the notation for forms in the new complex structure. 
Still the coupling with the closed field is always present.

The classical action is
\be
\lambda S_o=\int_X \Omega^{(3,0)} \left[Tr\left(\frac{1}{2} B d_{B_0} B + \frac{1}{3} B^3 + B F_0\right)-\frac{1}{2}db\right]
\label{2hcsa}\ee
This is invariant under the infinitesimal gauge transformations
\bea
s B &=& d_{B_0}\epsilon +[B,\epsilon]+\psi^{(1,0)} \nonumber\\
s b &=& Tr\left(Bd_{B_0}\epsilon+2F_0\epsilon\right) + d\gamma +\eta^{(1,1)}+\eta^{(2,0)}
\label{2gsym}\eea
where $\epsilon$ is the usual gauge symmetry ghost while $\psi^{(1,0)}$, $\eta^{(2,0)}$ and $\eta^{(1,1)}$
are the ghosts for the shift symmetries.

By further defining
\bea
s\epsilon &=& -\epsilon^2 \nonumber\\
s\psi^{(1,0)} &=& \left[\epsilon,\psi^{(1,0)}\right] \nonumber\\
s\gamma^{(1,0)} &=& n^{(1,0)}- Tr\left(\epsilon \partial^{(1,0)}_{B_0}\epsilon\right)\nonumber\\
s\gamma^{(0,1)} &=& \partial^{(0,1)}m - Tr\left(\epsilon \partial^{(0,1)}_{B_0}\epsilon\right)\nonumber\\
s\eta^{(1,1)} &=& -Tr\left(\psi^{(1,0)}\partial^{(0,1)}_{B_0}\epsilon\right) -\partial^{(1,0)}\partial^{(0,1)} m
-\partial^{(0,1)} n^{(1,0)}\nonumber\\
s\eta^{(2,0)} &=& -Tr\left(\psi^{(1,0)}\partial^{(1,0)}_{B_0}\epsilon\right) -\partial^{(1,0)} n^{(1,0)}\nonumber\\
sn^{(1,0)}&=&0\nonumber\\
sm &=& 0
\label{2pseudobrs}\eea
we get a {\it pseudo}-BRST operator.
Actually the operator $s$ defined by (\ref{2gsym}) and (\ref{2pseudobrs}) is nilpotent only on shell.
Explicitly, one gets
\be
s^2 b^{(0,2)}=\left(\partial^{(0,1)}\right)^2 m
\label{2nnp}\ee
which is vanishing only on shell. 
(\ref{2nnp}) is proportional either to the closed field e.o.m. or, which is the same, to the ones of $b$.
On all other fields one gets $s^2=0$.

The BV recipe is in this case still simple, since one can check that second order in the antifields 
already closes in this case.
By labeling all the fields entering 
(\ref{2gsym}) and (\ref{2pseudobrs}) as $\phi^i$, we have therefore
\footnote{Here we use the $\lor$-operator as in \cite{BCOV} so that the $\lor$ of a $(3,p)$-form is a $(0,p)$-form.}
\be
S_{BV}= S_o + \int_X \sum_i \phi^*_i s\phi^i + c \int_X \left((b^*)^{(2,2)}\partial^{(1,0)}m\right)^{\lor} (b^*)^{(3,1)}
\label{2BV}\ee
where $c$ is a non zero numerical constant which will not be relevant for our calculations (see later).
One can explicitly show that $S_{BV}$ satisfies $\Delta S_{BV}=0$, where $\Delta$ is the BV-Laplacian
and $(S_{BV},S_{BV})=0$ the corresponding bracket.
In our conventions, all antifields have complementary form degree with respect to fields.

Let us notice that a parallel result has been obtained in \cite{camillo} by C.~Imbimbo for the A-model. 
Indeed, also in the case of the real Chern-Simons theory, the coupling with the gravitational background requires the 
use of the full BV formalism giving rise to quadratic terms in the anti-fields.

While gauge fixing, we need to add the anti-ghost multiplets for all gauge fixed parameters.
Actually we are going to gauge fix our theory only partially, that is we will keep the 
($\epsilon$-)gauge freedom relative to the Chan-Paton bundle.
By introducing the relevant anti-ghost multiplets, we define the gauge fixing fermion
\bea
\Psi=
\int_X \left\{
{\bar\psi}^{(1,3)}\left(d_{B_0}B + B^2 + F_0\right)^{(2,0)}
+{\bar\eta}^{(2,2)}b^{(1,1)}
+{\bar\eta}^{(1,3)}b^{(2,0)}
\right.\\\left.
+{\bar n}^{(2,3)}\gamma^{(1,0)}
+{\bar m}^{(3,3)} \left(\partial^{(0,1)}\right)^\dagger \gamma^{(0,1)}
+{\bar\gamma}^{(3,2)}\left[\left(\partial^{(0,1)}\right)^\dagger b^{(0,2)}+\partial^{(0,1)}p\right]
\right\}
\eea
by adding the anti-ghost (trivial) part of the BV action in the usual form.
We extend therefore the s-operator action, that is the BV-bracket with the part of the BV action linear in the anti-fields, 
to the anti-ghosts in the trivial way, namely for any anti-ghost $\bar\psi$ we have
$s\bar\psi=\Lambda_{\bar\psi}$ and $s\Lambda_{\bar\psi}=0$. The anti-ghost gauge freedom is fixed by the addition of the 
relevant further sector.

Finally we can compute the (partially) gauge fixed action by specifying all anti-fields as derivatives with respect to 
their relative fields of gauge fermion $\Psi$.
All in all, the (partially) gauge fixed action reads
\bea
 S_{g.f.}=  S_o + s\Psi + c\int_X\left(
{\bar\eta}^{(2,2)}\partial^{(1,0)}m\right)^{\lor}\left(\partial^{(0,1)}\right)^\dagger{\bar\gamma}^{(3,2)}
\label{2gfa}
\eea

Let us now perform the path-integral in the different sectors (by naming them by the relative anti-ghost as appearing in 
the gauge fermion).
\begin{itemize}
\item The ${\bar\psi}^{(1,3)}$ is seen to decouple since 
$$s\left\{d_{B_0}B + B^2 + F_0\right\}^{(2,0)}=\partial^{(1,0)}_{B_0}\psi^{(1,0)}+\left[B^{(1,0)},\psi^{(1,0)}\right]_+$$
Therefore we get the contribution 
$$
\int \cald[B^{(1,0)}]\delta\left(\partial^{(1,0)}_{B_0}B^{(1,0)} + B^{(1,0)}B^{(1,0)} + F_0^{(2,0)}\right)
{\rm det'}\left\{\partial^{(1,0)}_{B_0}+\left[B^{(1,0)},\cdot\right]_+\right\}
$$
which counts the volume of the space of holomorphic connections.

\item The two $\bar\eta$-sectors are just algebraic and give a constant contribution to the path-integral.
Notice that while integrating over $\bar\eta^{(2,2)}$ also the last term in (\ref{2gfa}) gets involved being reabsorbed in a shift of $\eta^{(1,1)}$.
This gauge fixing of course restricts the field $b$ to be a $(0,2)$-form only and set to zero $\eta^{(1,1)}$ and $\eta^{(2,0)}$.

\item The ${\bar n}^{(2,3)}$ sector is algebraic too and simply sets to zero $\gamma^{(1,0)}$ and its partner. 

\item The last part is the standard term for higher form BV quantization (see for example \cite{HT}).
The fermionic bilinear operator reduces to 
$$
\calb=
\left(\begin{matrix}-{\partial^{(0,1)}}^\dagger\partial^{(0,1)} & -\partial^{(0,1)} \\
                {\partial^{(0,1)}}^\dagger & 0 \end{matrix}\right)$$ 
mapping $\Omega^{(0,1)}(X)\oplus \Omega^{(0,0)}(X)$ to itself.
The bosonic bilinear operator is instead the anti-holomorphic Laplacian 
$\Delta^{(0,0)}={\partial^{(0,1)}}^\dagger\partial^{(0,1)}$ on the scalars $\Omega^{(0,0)}(X)$.
One therefore stays with the gauge fixed measure 
\be
\int \cald[Y] e^{-\frac{1}{2}\int_X Y \calc Y +\int_X J^tY}
\label{2circa}\ee
where
$Y=\left(p,\Lambda_{\bar\gamma},b^{(0,2)}\right)$, 
$$
\calc=
\left(
\begin{matrix}
0               & -\partial^{(0,1)} & 0           \\
\partial^{(0,1)}&      0            & {\partial^{(0,1)}}^\dagger\\
0               & -{\partial^{(0,1)}}^\dagger & 0
\end{matrix}
\right)
$$
and the source $J=(0,0,d\Omega)$ takes into account the classical action.
Eq.(\ref{2circa}) can be integrated being a Gaussian.

\end{itemize}

Therefore, all in all, we find that the quantum measure for the holomorphic Chern-Simons theory
is
\be
\frac{det'[\calb]}{det'[\Delta^{(0,0)}]\left(det'[\calc]\right)^{1/2}}
e^{J^t(\calc)^{-1}J}
\label{2qm}\ee
for a (generically non integrable) almost complex structure.
The determinant of the operator $\calc$ is easily obtained by noticing that
$$
\{\calc,\calc^\dagger\}=
\left(
\begin{matrix}
\Delta^{(0,0)}               & 0            & \left(\partial^{(0,1)}\right)^2 + \left({\partial^{(0,1)}}^\dagger\right)^2\\
0                            & 2 \Delta^{(3,2)}                 &                           0           \\
       \left(\partial^{(0,1)}\right)^2 + \left({\partial^{(0,1)}}^\dagger\right)^2                 &0            & \Delta^{(2,0)} 
\end{matrix}
\right)
$$

We want to compare our open theory, defined as coupled to the closed field $A$, with the standard holomorphic Chern-Simons, 
defined for an integrable complex structure. In particular the two theories should match once we put on shell the closed field. 
So the integral of all the additional fields should contribute as one. Notice that, if the complex structure is integrable, 
then $d\Omega=0$ and the source term is not contributing.
On top of it, since $\left(\partial^{(0,1)}\right)^2=0$, the bosonic operator block-diagonalizes.
Moreover, in this case, the determinant of the fermionic operator $\calb$ can be easily computed
\footnote{This can be done by writing the eigenvector equation for $\calb$ as $\calb \tbinom{a}{b}=\lambda\tbinom{a}{b}$
and then expanding the 1-form $a=\partial^{(0,1)}x+{\partial^{(0,1)}}^\dagger y$ in exact and co-exact parts. 
Then one finds that $b=\lambda x$ and that the eigenvalues of $\calb$ coincide with 
those of $\Delta^{(0,2)}$ for $x=0$ or with the square roots of those of $\Delta^{(0,0)}$ for $y=0$.}
to be equal to $det'\Delta^{(0,2)} \left(det'\Delta^{(0,0)}\right)^{1/2}$.

All in all, we find an overall 
\be
\frac{det'[\Delta^{(0,2)}] \left(det'[\Delta^{(0,0)}]\right)^{1/2}}
{det'[\Delta^{(0,0)}]\left\{\left(det'[\Delta^{(0,2)}]\right)^2
det'[\Delta^{(0,0)}]\right\}^{1/2}
}=\frac{1}{det'[\Delta^{(0,0)}]}
\label{2ciapa}\ee
This determines the value of the quantum measure introduced in (\ref{2fp}).
The factor (\ref{2ciapa}) counts the extra degree of freedom introduced by the $b$ field in the theory.
Indeed the three components of $b^{(0,2)}$ are subject to the gauge freedom by the shift 
of an exact $\partial^{(0,1)}\gamma^{(0,1)}$ term up to the ghost-for-ghost shifting
$\gamma^{(0,1)}$ by $\partial^{(0,1)}m$. Therefore the overall counting is $3-3+1=1$
complex modes.

\section{String field theory as generating function of open and closed HAEs}

Our claim of having found the effective space-time theory for the open B-model should be checked explicitly. 
Because of tadpole cancellation (see also \cite{Wal3} ) we know that the open theory is completely well defined only 
in its unoriented version ( as in the case of usual string theories ), so the most general case to consider is for open 
( and closed ) unoriented strings. Closed moduli are known to be unobstructed and so expansions of the amplitudes in their 
value is always possible. We will proceed similarly for open moduli.
An important result of \cite{BCOV} is that the partition function of Kodaira-Spencer theory encodes the recurrence relations of HAE via 
its Feynman diagram expansion.

The generating function of the full HAE of \cite{BT} generalized to the unoriented case should be: 
\begin{equation}\label{2t}
e^{W(x,u;t,\bar{t})}  \sim \exp\left(\sum_{g,h,c,n,m} 
\frac{\lambda^{2g - 2 + h + c}}{2^{\frac{\chi}{2} + 1}\;n!m!}
{\cal F}^{(g,h,c)}_{i_{1}\dots i_{n}\alpha_{1}\dots \alpha_{m}}x^{i_{1}}\dots x^{i_{n}}u^{\alpha_{1}}\dots u^{\alpha_{m}}\right)
\end{equation}
up to an overall $\lambda$ dependent prefactor which encodes the contact terms in one loop calculations and will be discussed later.
This prefactor $\lambda^{\dots}$  is encoded, in the field theory side, in the measure of the path integral, namely as the 
multiplicative term weighting the regularized determinants with omitted zero modes. 
From now on, in any case, we will focus on the perturbative expansion in $\lambda$.

The notation is as follows: ${\cal F}^{(g,h,c)}_{i_{1}\dots i_{n}\alpha_{1}\dots \alpha_{m}}$ is 
the string amplitude with genus $g$, $h$ boundaries, 
$c$ crosscaps, $n$ marginal operator insertions in the bulk and on the boundary and $m$ purely on the boundary.
The $x^{i}$'s are the expansion coefficients of $x$ in a base of Beltrami differentials, $x = x^{i}\mu_{i}$ 
and the $u^{\alpha}$'s are the expansion coefficients for $B_{0}$ in a basis $T_\alpha(x)$ of the open moduli 
$H^{(0,1)}(X,Ad_E)$, namely $B_{0} = u^{\alpha}T_{\alpha}$.
Thus the fields appearing as backgrounds in the field theory are the open and closed moduli themselves. 
Also the coordinate we have used to parametrize the moduli space are the canonical ones, already discussed in the first chapter. This means that operator insertions will be obtained as ordinary derivatives on the partition function.

The factor $\frac{1}{2^{\frac{\chi}{2} + 1}}$ is explained in \cite{Wal3} and obviously $\chi = 2g -2 + h + c$. 
If what we are doing is consistent it should be true that 
\begin{equation}\label{2g}
\int {\cal D}A{\cal D}B{\cal D}b\dots e^{-S_{tot}(x,B_{0}(x);t,\bar{t};A,B,b,\dots)} = e^{W(x,u;t,\bar{t})}.
\end{equation}
 
We want to compare this at tree level, that is at $g = 0, h = 0,1, c = 0$ and $g = 0, h = 0, c = 1$, and obtain in this way some explicit 
expressions for all the basic objects entering the extended HAE of (\ref{0trallalla}) computed at a generic background point. 
These amplitudes are already known and computed by worldsheet methods and the two results should of course match. 
To this end we will differentiate, at each order in $\lambda$, both members with respect to the moduli parameters $x^{i}$ and $u^{\alpha}$
and identify the corresponding coefficients.

A comment is in order. We should remember that the expression (\ref{2t}) is the partition function for the unoriented theory. 
As explained in \cite{Wal3} this differs from the oriented one simply projecting the space of operators in the theory 
to the unoriented sector that is the ones with eigenvalue $+1$ under the parity operator $\cal{P}$. 
Being these operators nothing else than deformations of the moduli space of the theory,  
we have to consider only its invariant part under $\cal{P}$ and then parametrise with $x^{i}$ and $u^{\alpha}$ its tangent space. 
This means that the $x^{i}$'s and the $u^{\alpha}$'s appearing in ( \ref{2t} ) are really a subset of the ones in the oriented case. 
Specifically it implies a restriction on the space of complex structures for what matters $x$ and a reduction to $Sp(N)/SO(N)$ groups for $u$. 
Still some amplitudes, as the sphere with three insertions, are perfectly meaningful also in the oriented case. 
This is why we will generically not specify to which space the $x^{i}$'s and the $u^{\alpha}$'s belongs: 
it is possible to restrict their value depending on the case. 

\subsection{${\bf g = 0}$, ${\bf h = c = 0}$}

Here we start the comparison between the string theory partition function and the space-time path integral (\ref{2g}).
We begin from the coefficients at lowest order in $\lambda$. 
From the point of view of (\ref{2t}) this is the amplitude at $g = h = c = 0$ with weight $\frac{1}{\lambda^{2}}$; 
on the field-theory side the contribution should come only from the Kodaira-Spencer action, also at weight $\frac{1}{\lambda^{2}}$. 
We know that the right-hand side of equation (\ref{2g}) at this order in $\lambda$ has no dependence on open moduli 
(because without boundaries, $h = 0$, there is no space for open operator insertions) and the building block amplitude being $C_{ijk}(x)$:
\[
C_{ijk}(x) = {\cal F}^{(0,0,0)}_{ijk}(x) = \sum_{n} \frac{1}{n!}{\cal F}^{(0,0,0)}_{ijki_{1}\dots i_{n}}x^{i_{1}}\dots x^{i_{n}} 
= \frac{\partial}{\partial x^{i}}\frac{\partial}{\partial x^{j}}\frac{\partial}{\partial x^{k}}W\mid_{(order \lambda^{-2})} 
\]
Being at tree level and given (\ref{2g}), the same result can be obtained ( see \cite{BCOV} ) deriving the Kodaira-Spencer action on shell ( $A = A(x)$ ) 
with respect to three $x^{i}$. The three derivative term gives\footnote{The factor $-2$ depends on our 
conventions which are slightly different from \cite{BCOV}.}
\[
-2\int_{M}\left[\left(\mu_{i} + \frac{\partial A(x)}{\partial x^{i}}\right)\wedge \left(\mu_{j} 
+ \frac{\partial A(x)}{\partial x^{j}}\right)\right]' \left(\mu_{k} + \frac{\partial A(x)}{\partial x^{k}}\right)' = C_{ijk}(x)
\]
The only point of possible confusion for the BCOV educated reader both here and in the subsequent computations, comes from the novel cross 
dependence of open and closed field on shell by each other by means of the field equations which are now modified with respect to the ones 
obtained with the open and closed actions separated. This might seem to carry on additional induced derivatives and contributions as, 
in this case, an induced open moduli dependence carried by the on shell closed field which would lead to the paradox of a non vanishing 
amplitude corresponding to a sphere with boundary insertions! 
Fortunately, integrating out the field $b$ does the job of enforcing the closed field solutions that would be obtained from the 
Kodaira-Spencer action alone! 
It will be true instead that the on shell open fields will carry some closed field dependence as the $B$-field equation is: 
$F_{B_{0}}^{\tilde{0},\tilde{2}} \equiv (d_{B_{0}}B + B^{2} + F_{0})\mid^{\tilde{0},\tilde{2}} = 0$ which is both $u$ and $x$ dependent.

This is a good place to stop and discuss the connections between our result for the coupling between the open theory and the closed one, 
and the comments made by Witten in \cite{W1} about this point. In his paper Witten uses an argument from the fatgraph description of 
a string tree level amplitude to infer that, if one considers a diagram with $n$ bulk and $m$ boundary insertions, 
in general the bulk operators will reduce to exact (with respect to the topological charge) objects and so will decouple. 
This goes through even in the case $m = 0$ as long as some boundaries are present. 
The direct consequence is that the on-shell couplings between closed and open strings are zero.
How can then one justify the non vanishing of the 
$\Delta_{ij}$ amplitudes of \cite{Wal1}, \cite{Wal2} and \cite{Wal3}?
Our answer is in a sense a weakened realization of Witten's idea, still allowing non zero amplitudes with bulk operators and boundaries. 
The key role is played by the field $b$, generated in the action to maintain the gauge symmetries in the Chern-Simons term. 
This field, once it is integrated over, fixes the closed field $A$ to be on shell with respect to the original Kodaira-Spencer equations 
and so defining a shift of an integrable complex structure. 
This translates to the fact that the original genuine coupling between open and closed fields in the action reduces to a coupling 
between an open, integrated field and an on-shell closed field. That is it represent a new Chern-Simons expansion around a new shifted 
and fixed complex structure. So the path integration of the closed field $A$ reduces to a single contribution coming from the unique 
deformation of the original complex structure with respect to which the Kodaira-Spencer action is written, 
this contribution being weighted by the corresponding Kodaira-Spencer on shell action.
If closed strings are substantially decoupled by the open theory, what is then their role? This is the next point discussed by Witten 
in \cite{W1} where their crucial role in anomaly cancellation is pointed out.
For example, in the A-model, whose effective theory is the real Chern-Simons, a well known topological anomaly is present. 
It comes from the $\eta$-invariant of \cite{W2}, whose dependence by the metric is compensated by the addition of a gravitational Chern-Simons. 
Then an additional anomaly connected to the framing of the target space is well known. 
In the case of the B-model however, the $\eta$-invariant is simply zero because the spectrum of eigenvalues of the determinant 
whose phase is $\eta$, is symmetric around zero \cite{Tho}. 
Instead we have one loop anomalies corresponding to a dependence by the wrong moduli \cite{new} ( K\"ahler moduli in this case) 
which is cured by tadpole cancellation involving unoriented contributions in the closed strings sector (Klein bottle).

\subsection{${\bf g = 0}$, ${\bf h + c = 1}$ }

In this subsection we want to compare the world-sheet and the target space perspective at order $1/\lambda$.
From the string theory side the relevant amplitudes of weight $\frac{1}{\lambda}$ ( $g = 0$, $h + c = 1$ ) entering the HAE
were discussed in (\ref{0trallalla}). From the field theory perspective all of them should be reproduced by the holomorphic Chern-Simons action.

Let us start with purely closed moduli dependence. This can come either from both the explicit dependence by $x$ in $\Omega$ and 
by the induced dependence in the $A(x)$ and $B(x,u)$ fields on shell, or implicitly through the background $B_{0}(x)$. 
We will find that the dependence w.r.t. closed moduli explicit and in the on shell fields, both closed and open, 
correspond to bulk insertion in the string amplitude, while the dependence w.r.t. closed moduli in the background open field
corresponds to induced boundary insertions\footnote{An additional closed moduli dependence in the worldsheet action would come 
also from the Warner term \cite{War}. For the B-model this additional boundary term, needed to make the action invariant, vanishes 
under the usual boundary conditions, see \ref{0trallalla}, \ref{1borg}  and \cite{mirror}.}.  

The two operators will be indicated as $ O_{i}$ and $\Psi_{i}$ (so for example $C_{ijk} = \langle  O_{i} O_{j} O_{k}\rangle_{0,0,0}$ 
where the subscript denotes the triple $g,h,c$). 

The first amplitude we want to derive is $\Delta_{ij} = \langle  O_{i} O_{j}^{[1]}\rangle_{0,1,0 \; + \; 0,0,1}$ which 
was computed in \cite{Wal1} and \cite{Wal3} as additional building block for the extended HAE.
This is the disk plus the crosscap with two bulk insertions. 
In particular $ O_i$ is a local insertion while $ O_j^{[1]}$ is an integrated one being the second step of the descent equation.
So, from (\ref{2f})
we get
\begin{equation}
\frac{1}{\sqrt{2}}\Delta_{ij}(x)  = \int_{X} d_{i}d_{j}\Omega L_{CS} + \int_{X} d_{i}\Omega d_{j}L_{CS} + \int_{X} d_{j}\Omega d_{i}L_{CS} + 
\int_{X} \Omega d_{i}d_{j}L_{CS} 
\label{2paletta}
\end{equation}
where all the fields are on shell; $d_{i}$ is the derivative with respect to the closed modulus $x^{i}$, both explicitly and 
through the dependence induced by $A(x)$ and $B(x,u)$; the factor $\frac{1}{\sqrt{2}}$ comes from the normalization in (\ref{2t}). 
Using the field equations for $B$ we obtain the identity
\[
0 = d_{j}\left(\int_{X} \frac{\delta S_{HCS}}{\delta B}\mid_{B=B(u,x)}d_iB(u,x)\right)= 
d_{j}\left(\int_{X} \Omega d_{i}L_{CS} \right) = \int_{X} d_{j}\Omega d_{i}L_{CS} + \int_{X} \Omega d_{i}d_{j}L_{CS} 
\]
that is, the last two terms in (\ref{2paletta}) cancel. This is nothing but Griffith's transversality condition for the normal function as stated in
\cite{Wal1}. So we get
\begin{equation}\label{2l}
\frac{1}{\sqrt{2}}\Delta_{ij}(x)  = \langle  O_{i} O_{j}\rangle_{0,1,0 \; + \; 0,0,1} = 
\int_{X} d_{i}d_{j}\Omega L_{CS} + \int_{X} d_{i}\Omega d_{j}L_{CS} 
\end{equation}
This differs from the expression derived in \cite{Wal1,Wal3} by the first term. However notice that (\ref{2l}) is valid 
at a generic value $x$ for closed string moduli, 
while the ones of \cite{Wal1,Wal3} are evaluated at $x = 0$, where the double derivative of $\Omega$ is vanishing. 
This comes from expression (\ref{2x}) and from the fact that 
$A(x) = O(x^{2})$ as follows by solving the Kodaira-Spencer equations iteratively.
  
Let us now consider the amplitudes with one bulk and one boundary insertion. The latter, as already stated, is obtained 
from the derivative with respect to 
the background open field $B_0$ which depends on $x$:
\[
\frac{1}{\sqrt{2}}\Delta'_{ij} = \langle  O_{i}\Psi_{j}^{[1]}\rangle_{0,1,0} = 
\left(d_j B_{0}(x) \frac{\delta}{\delta B_{0}(x)}\right)d_{i}S_{HCS}
\]
To compute this term from the space-time point of view it is easier to start from the action written in terms of 
$\hat{B}$ and $B_{0}$ (\ref{2m}). The result follows immediately as 
\begin{equation}
\frac{1}{\sqrt{2}}\Delta'_{ij} = \langle  O_{i}\Psi_{j}^{[1]}\rangle_{0,1,0} = -\int_{X} d_{i}\Omega Tr(d_{j}B_{0}(x)F_{0}) 
\end{equation}
once the e.o.m. of the open field are imposed.

Now we pass to the purely open moduli derivatives. The only term is the one derived three times or, 
equivalently, the one with three boundary operator insertions: 
$\Delta_{\alpha\beta\gamma}$. Again using the form (\ref{2m}) we need only explicit derivatives with respect to $u^{\alpha}$ 
(remind that $B_{0} = u^{\alpha}T_{\alpha}$ ). The result is 
\begin{equation}\label{2n}
\frac{1}{\sqrt{2}}\Delta_{\alpha\beta\gamma} = \langle \Theta_{\alpha}\Theta_{\beta}\Theta_{\gamma}\rangle_{0,1,0} 
= -\int_{X}\Omega Tr(T_{\alpha}T_{\beta}T_{\gamma}) 
\end{equation}
which is the same that would be derived with worldsheet methods in analogy to $C_{ijk}$.

Finally we have mixed terms. These are similarly obtained giving
\begin{equation}\label{2o}
\frac{1}{\sqrt{2}}\Pi_{\alpha i} = \langle \Theta_{\alpha} O_{i}\rangle_{0,1,0} = -\int_{X} d_{i}\Omega Tr(T_{\alpha}F_{0}) 
\end{equation}
and
\begin{equation}\label{2p}
\frac{1}{\sqrt{2}}\Delta'_{\beta i \alpha } = \langle \Theta_{\beta}\Psi_{i}^{[1]}\Theta_{\alpha}\rangle_{0,1,0} = -\int_{X} 
\Omega Tr(T_{\beta}d_{i}B_{0}T_{\alpha}) 
\end{equation}

    \chapter{Open-closed duality}
\label{sec:cap3}
\noindent

Open/closed duality is commonly believed \cite{OV1}
to be the effect of integrating out open strings
in the complete string field theory, leaving then a purely closed string theory
on a suitably modified background.
This program is very hard to be realized in the full string theory, 
but it becomes tractable
in its truncation to its BPS protected sectors, namely in topological string theories
\cite{Witten,BCOV}.
This issue has been investigated by several authors in a first quantized or {\it on shell}
framework.
Actually, the first examples were discussed in terms of geometric transitions \cite{GV}  
which have been extended to the brane sector in \cite{OV1}.
Then, this picture has been refined in terms of a proper world-sheet analysis in \cite{OV2}.
More advances on-shell computations has been prompt by \cite{remodeling} and then further
by \cite{tso} and \cite{emanuel}.

The formulation of holomorphic anomaly equations and the
target space interpretation of its structure functions are
very important tools to obtain a well defined computational framework
for open topological strings. D-branes sources for closed strings are actually
represented in the HAE by the Walcher's term \cite{Wal1} whose target space interpretation
has been given in terms of the Griffith's normal function (see also \cite{MoW}).
For the B-model this boils down to the on shell holomorphic Chern-Simons action.
A remarkable observation \cite{Oog} consists in the proof that the Walcher's term can be reabsorbed
by a shift in the string coupling constant and the closed moduli. This indeed realizes
an on shell proof of the open/closed duality, although at frozen open moduli.

In the following we will study this problem first from a worldsheet perspective and then from a second quantized point of view, which turns out to be
the most appropriate to study open/closed duality in particular for the B-model. Having worked out the BV formulation of the holomorphic
Chern-Simons theory by leaving the gravitational background (Kodaira-Spencer gravity field) off shell allows us to reformulate open-closed duality as a process of partial functional integration over the open string fields. From the BV viewpoint this procedure follows by partial integration of a proper subset of fields and anti-fields of a solution of the BV master equation by which one gets another solution
depending on a reduced set of fields. This is known as Losev trick \cite{Losev}.
In particular, at frozen open string moduli, we will show that this partial integration
exactly reproduces the shift formulas proposed in \cite{Oog}\cite{Wal2}.
More in general, our BV formulation proves the existence of definite shift formulas
also in presence of open moduli providing a computational set-up to determine them.

\section{Open-Closed string duality as a Losev trick}
\label{3zio}

Let us explain a basic argument about open-closed string duality in second quantization.
This is referred to the topological string theory at hand (B-model), but in principle should hold 
in a more general setting.

The Losev trick, as explained in \cite{Losev}, consists in a procedure to obtain solutions of the 
quantum Master Equation in Batalin-Vilkovisky quantization by partial gauge fixing.
In its generality it reads as follows.
Let $S(\Phi,\Phi^*)$ be a solution of the quantum Master equation 
\be
\Delta\left(e^{-S/\hbar}\right)=0
\label{3qme}\ee
where $\Delta=\partial_\Phi\partial_{\Phi^*}$ is the nilpotent BV Laplacian.
Suppose that the fields/anti-fields space $\calF$ is in the form of a 
fibration
$$
\begin{matrix}
\calF_2 & \hookrightarrow & \calF \\
\,      & \,              & \downarrow \\
\,      & \,              & \calF_1
\end{matrix}
$$
so that one can choose a split coordinate system $(\Phi,\Phi^*)=(\Phi_1,\Phi^*_1,
\Phi_2,\Phi^*_2)$ such that the BV Laplacian splits consistently as
$\Delta=\Delta_1 + \Delta_2$ with $\Delta_1^2=0$.
Then, assuming the existence of a non singular gauge fermion $\Psi$, 
one can consider the partially gauge fixed BV effective action
\be
e^{-\frac{1}{\hbar} S_{eff}(\Phi_1,\Phi^*_1)}
=
\int_{\calF_2}\cald\left[\Phi_2,\Phi^*_2\right] e^{-\frac{1}{\hbar} S(\Phi,\Phi^*)}
\delta\left(\Phi_2^*-\partial_{\Phi_2}\Psi\right).
\ee
which can be readily seen to satisfy the reduced BV Master equation
\be
\Delta_1 e^{-\frac{1}{\hbar} S_{eff}(\Phi_1,\Phi^*_1)}=0
\label{3qme1}\ee
Actually -- the proof is two lines -- one consider (\ref{3qme}) partially
gauge fixed on the fibers and integrated along the fiber $\calF_2$
$$
0=\int_{\calF_2}\cald\left[\Phi_2,\Phi^*_2\right] \left\{\Delta_1+\Delta_2\right\}
e^{-\frac{1}{\hbar} S(\Phi,\Phi^*)}
\delta\left(\Phi_2^*-\partial_{\Phi_2}\Psi\right)
= 
\Delta_1 e^{-\frac{1}{\hbar} S_{eff}(\Phi_1,\Phi^*_1)} + $$ $$+
\int_{\calF_2}\cald\left[\Phi_2\right] 
\left\{
\frac{d}{d\Phi_2}
           \left( 
           \left[\partial_{\Phi_2^*}
            e^{-\frac{1}{\hbar} S(\Phi_1,\Phi_2,\Phi^*_1,\Phi_2^*)}
           \right]_{\Phi_2^*=\partial_{\Phi_2}\Psi}
	   \right)
           - 
           \partial^2_{\Phi_2}\Psi \cdot \left(\partial^2_{\Phi_2^*}
e^{-\frac{1}{\hbar} S(\Phi,\Phi^*)}\right)\mid_{\Phi_2^*=\partial_{\Phi_2}\Psi}
\right\}$$
Now, the last line vanishes because of translation invariance of the path-integral along the fiber
and field/anti-field opposite statistics, so that we recover (\ref{3qme1}).
Let us notice that the resulting BV effective action depends on the particular gauge fixing 
chosen to integrate the fiber degrees of freedom. This dependence is BV trivial in the effective action.

Let us now specify the above setup to open/closed string theory, namely we identify
$\calF$ with the open and closed string theory, $\calF_2$ with the open strings
and $\calF_1$ with closed strings.
The complete theory is given by the BV action 
\be
S_{c+o}(A,B;x,u,\lambda)=S_c(A;x,\lambda)+S_o(A,B;x,u,\lambda)
\label{3c+o}\ee
where $S_c(A;x,\lambda)$ is the closed string BV action, and 
$S_o(A,B;x,u,\lambda)$ completes the open and closed BV action.
The BV Laplacian takes the form $\Delta_{c+o}=\Delta_c + \Delta_o$.
We assume that both closed and open plus closed strings have been BV formulated, so that
the corresponding quantum Master equations hold. 
Moreover, the uniqueness of closed string field theory is taken to mean that all solutions of 
the quantum Master equation, with proper boundary conditions in the string coupling dependence -- namely
the background independence of the kinetic term, 
are given by $S_c(A;x,\lambda)$ for some background $x$ and the choice of the string coupling constant $\lambda$.
For the B-model, this is explicitly proved in \cite{BCOV}.

Therefore, by specifying the Losev trick to our case, we obtain that
the effective action obtained from (\ref{3c+o}) by partial gauge fixing and integration over the open string 
field, satisfies the quantum Master equation (\ref{3qme1}) that is the quantum master equation for the {\it closed} 
string field theory.
Notice that, by definition,
\be
e^{-S_{eff}(A,x,\lambda,u)}=
e^{-S_c(A;x,\lambda)}
\int_{\tiny
\begin{matrix} gauge \\ fixed \end{matrix}
}
\cald[B]e^{-S_o(A,B;x,u,\lambda)}
\label{3yyy}\ee
approaches the required boundary condition in the string coupling constant dependence.
The actions entering (\ref{3yyy}) are required to have a canonically normalized kinetic term.
Therefore, we conclude that the effective action (\ref{3yyy}) has to be
the closed string field action (in some gauge determined by the gauge fixing in the open string sector)
for a shifted set of background moduli and a redefined string coupling constant, that is,
\be
\caln\,\, e^{-S_{c}(A;x^\star,\lambda^\star)}=
e^{-S_c(A;x,\lambda)}
\int_{\tiny
\begin{matrix} gauge \\ fixed \end{matrix}
}
\cald[B]e^{-S_o(A,B;x,u,\lambda)}
\label{3yygen}\ee
up to a field independent normalization $\caln$.

The particular case we have in mind is therefore the topological B-model, where
$S_c$ is the Kodaira-Spencer gravity action 
and $S_o$ the holomorphic Chern-Simons action 
suitably coupled to the Kodaira-Spencer field as discussed in the previous sections.
After passing to flat coordinates, (\ref{3yygen}) then specifies to
\be
\caln(u,x,\lambda^{-1}\Omega_0)\,\, e^{-\frac{1}{{\lambda^\star}^2}S_{KS}(A^*,x^\star)}=
e^{-\frac{1}{\lambda^2}S_{KS}(A,x)}
\int_{\tiny
\begin{matrix} gauge \\ fixed \end{matrix}
}
\cald[B]e^{-\frac{1}{\lambda}S_{HCS}(A,B,x,u)}
\label{3yy}\ee
where 
the closed string field gets renormalized as $A^\star/\lambda^\star=A/\lambda$.
In (\ref{3yy}) $\caln$ is a normalization factor\footnote{The particular dependence on the ratio $\Omega_0/\lambda$ is due to the fact that 
we have chosen flat coordinates $u,x$ for the moduli. See next section for a specific discussion on the relevance of the normalization factor 
in comparing with \cite{Wal2}.} and
\be
\frac{1}{\lambda^\star}=\frac{1}{\lambda} + \delta(u,x,\lambda)
\quad {\rm and} \quad
(x^\star)^i=x^i +\delta^i(u,x,\lambda)
\label{3shift}\ee
are some shifted background and string coupling. All these are {\it to be determined} 
and can be perturbatively computed from (\ref{3yy})
by Feynman diagrams expansion or with non perturbative techniques when available.
The redefinition (\ref{3shift}) is a generalization (with tunable open moduli) of the moduli shift in \cite{Oog}.
The aim of the next subsection is to show that, at frozen open moduli, the above formulas reproduce the shift of \cite{Oog}.

\subsection{Open-closed duality at frozen open moduli}\label{3z}

In this subsection we want to apply the general arguments just explained in Section \ref{3zio} to the oriented string theory
with frozen open moduli \cite{Oog}. Indeed, since the computations will be done at tree level, we do not have to deal with unoriented amplitudes.
The effect of freezing the open moduli is easily obtained by replacing the non abelian field $B$ with $N$ identical copies of an abelian one, 
reducing the trace simply to a Chan-Paton factor $\beta$, which takes into account the number of boundaries. Accordingly, we
consider a slightly modified version of (\ref{2t}) which better fits our purposes:
\begin{equation}\label{3y}
e^{W(x,\lambda^{-1})} = \lambda^{\frac{\chi}{24} - 1 - \beta^{2}\frac{N}{2}}\exp\left(\sum_{g,h,n} 
\frac{\lambda^{2g - 2 + h + n}}{n!}\beta^{h}{\cal F}^{(g,h)}_{i_{1}\dots i_{n}}x^{i_{1}}\dots x^{i_{n}}\right)
\end{equation}
(\ref{3y}) is obtained from (\ref{2t}) suppressing all the open moduli parameters $u^{\alpha}$, rescaling $\frac{\lambda}{\sqrt{2}} \rightarrow \lambda$, $x^{i}\rightarrow \lambda x^{i}$, absorbing a common factor of one half and considering the additional $\beta$-parameter dependence. 
The HAE for open strings of \cite{Wal1} are obtained expanding in powers of $x^{i}$, $\lambda$ and  $\beta$ in the following equation 
\begin{equation}\label{3z1}
\left(-\bar{\partial}_{\bar{i}} + \frac{1}{2}C^{jk}_{\bar{i}}\frac{\partial^{2}}{\partial x^{j}\partial x^{j}} 
+ G_{j\bar{i}}x^{j}\frac{\partial}{\partial\lambda^{-1}} + \beta\bar{\Delta}_{\bar{i}}^{j}\frac{\partial}{\partial x^{j}} \right)
e^{W(x,\lambda^{-1})} = 0 .
\end{equation}

In \cite{Oog} it was shown that the above HAE (\ref{3z1}) can be derived from the HAE of the closed theory
by means of a suitable change of variables 
\begin{eqnarray}\label{3q}
x^{i} \rightarrow x^{i} - \beta\bar{\Delta}^{i} \nonumber \\
\lambda^{-1} \rightarrow \lambda^{-1} + \beta\bar{\Delta} 
\end{eqnarray}

Here we spend a few lines to define $\bar{\Delta}^{i}$. We know that $\bar{\Delta}_{\bar{i}\bar{j}}$ has been defined as the disk two points amplitude computed in the antitopological theory. It can be written as

\[
\bar{\Delta}_{\bar{i}\bar{j}} = D_{\bar{i}}D_{\bar{j}}\tilde{\Delta}
\]

and we can easily interpret $\tilde{\Delta}$ as the empty antitopological disk and raise one index with the usual topological-antitopological metric, $\bar{\Delta}^{i}_{\bar{j}}=g^{i\bar{i}}\bar{\Delta}_{\bar{i}\bar{j}}$. The goal is to look for the primitive with respect to the usual $\bar{\partial}_{\bar{i}}$ derivative; the reason will become clear in a second. Defining $\bar{\Delta}=\tilde{\Delta}g^{0\bar{0}} =\tilde{\Delta}e^{K} $, it is easy to show that:

\[
\bar{\Delta}^{i}_{\bar{j}} = g^{i\bar{i}}\bar{\Delta}_{\bar{i}\bar{j}} = \bar{\partial}_{\bar{j}}\left(e^{-K}g^{\bar{k}i}\bar{\partial}_{\bar{k}}\bar{\Delta}\right)
\]

Where we have used the rule that the covariant derivative acting over $\tilde\Delta$ is $D_{\bar{j}}\tilde{\Delta} = \bar{\partial}_{\bar{j}}\tilde{\Delta} + \bar{\partial}_{\bar{j}}K\tilde{\Delta}$, because $\tilde{\Delta}$ is a section of $\bar{\cal L}^{-1}$,   while over $D_{\bar{j}}\tilde\Delta$ it has also the additional piece from the contraction with the index $\bar{j}$, $D_{\bar{i}}\left(D_{\bar{j}}\tilde\Delta\right) = \ldots -\Gamma^{\bar{k}}_{\bar{i}\bar{j}}D_{\bar{k}}\tilde\Delta$. Than the Leibniz rule is enough to get the result. So we have that 

\[
\bar{\Delta}^{i} = G^{\bar{k}i}\bar{\partial}_{\bar{k}}\bar{\Delta}
\]

is the primitive of $\bar{\Delta}^{i}_{\bar{j}}$ ( note that the index is raised from $\bar{\partial}_{\bar{k}}\bar{\Delta}$ using the $G$ metric ). The physical interpretation of $\bar{\Delta}$ is clearly the empty antitopological disk with the ``$\bar{0}$'' implicit index raised by the $g$ metric, while $\bar{\Delta}^{i}$ is the antitopological disk one point amplitude with the barred index raised by $g_{i\bar{j}}$, as we can check:

\[
g^{i\bar{j}}\left(D_{\bar{j}}\tilde{\Delta}\right) = g^{i\bar{j}}\left(\bar{\partial}_{\bar{j}}\tilde{\Delta} + \bar{\partial}_{\bar{j}}K\tilde{\Delta}\right)  = G^{i\bar{j}}e^{K}\left(\bar{\partial}_{\bar{j}}\left[e^{-K}\bar{\Delta}\right] + e^{-K}\bar{\Delta}\bar{\partial}_{\bar{j}}K \right) = G^{i\bar{j}}\bar{\partial}_{\bar{j}}\bar{\Delta}
\]

The shift (\ref{3q}) allows to rewrite (\ref{3z1}) in the same form as the master equation for purely closed strings
\begin{equation}\label{3z2}
\left(-\bar{d}_{\bar{i}} + \frac{1}{2}C^{jk}_{\bar{i}}\frac{\partial^{2}}{\partial x^{j}\partial x^{j}} 
+ G_{j\bar{i}}x^{j}\frac{\partial}{\partial\lambda^{-1}}  \right)e^{W(x - \beta\bar{\Delta}^{i},\lambda^{-1} + \beta\bar{\Delta})} = 0
\end{equation}

as follows from an easy application of the chain rule to the total derivative $\bar{d}_{\bar{i}} = \frac{d}{d\bar{t}^{\bar{i}}}$, and the fact that $\bar{\Delta}$ and $\bar{\Delta}^{i}$ are both $\lambda^{-1}$ and $x^{i}$ independent; the last statement following from the fact that they are not expanded in the parameter $x^{i}$, as it is instead for the amplitudes ${\cal F}^{(g,h)}_{i_{1}\dots i_{n}}x^{i_{1}}\dots x^{i_{n}}$. So that $\frac{\partial \bar{\Delta}}{\partial x^{i}} = 0$ ( and similarly for $\bar{\Delta}^{i}$ ).\footnote{A subtlety is here worth to be mentioned. We wrote $\frac{\partial \bar{\Delta}}{\partial x^{i}}$ and not $\frac{\partial \bar{\Delta}}{\partial t^{i}}$ which instead would have not been zero. It is true in fact that $\bar{\Delta}_{\bar{i}\bar{j}}$ has an antiholomorphic anomaly, that is $\frac{\partial \bar{\Delta}_{\bar{i}\bar{j}}}{\partial t^{k}} = - \bar{C}_{\bar{i}\bar{k}\bar{l}}\bar{\Delta}_{\bar{k}}^{l} \neq 0$ and this anomaly is inherited from its antiholomorphic primitives.} 
So the idea is that open and closed partition functions are in some way related from a change of variables; the shift (\ref{3q}) acts ``removing'' open string data while with opposite signs we can add them. In particular let us reverse the argument and suppose we start with the closed partition function

\begin{equation}\label{3aaa}
e^{W_{c}(x,\lambda^{-1})} = \lambda^{\frac{\chi}{24} - 1}\exp\left(\sum_{g,n} 
\frac{\lambda^{2g - 2 + n}}{n!}{\cal F}^{g}_{i_{1}\dots i_{n}}x^{i_{1}}\dots x^{i_{n}}\right)
\end{equation}

obeying the equation

\begin{equation}\label{3z5}
\left(-\bar{\partial}_{\bar{i}} + \frac{1}{2}C^{jk}_{\bar{i}}\frac{\partial^{2}}{\partial x^{j}\partial x^{j}} 
+ G_{j\bar{i}}x^{j}\frac{\partial}{\partial\lambda^{-1}}  \right)e^{W_{c}(x ,\lambda^{-1})} = 0
\end{equation}
 
Consider the same partition function with shifted variables, with opposite signs with respect to (\ref{3q}). It clearly obeys (\ref{3z5}) in the shifted variables, that is 

\[
\left(-\bar{\partial}_{\bar{i}} + \frac{1}{2}C^{jk}_{\bar{i}}\frac{\partial^{2}}{\partial x^{j}\partial x^{j}} 
+ G_{j\bar{i}}\left[x^{j} + \beta\bar{\Delta}^{i}\right]\frac{\partial}{\partial\lambda^{-1}}  \right)e^{W_{c}(x + \beta\bar{\Delta}^{i},\lambda^{-1} - \beta\bar{\Delta})} = 0
\]

Instead of the partial derivative $\bar{\partial}_{\bar{i}}$ let us apply the full $\bar{d}_{\bar{i}}$. We obtain

\begin{eqnarray}\label{3z6}
\bar{d}_{\bar{i}}e^{W_{c}(x + \beta\bar{\Delta}^{i},\lambda^{-1} - \beta\bar{\Delta})} = \left(\bar{\partial}_{\bar{i}} + \beta\frac{\partial\bar{\Delta}^{j}}{\partial\bar{t}^{\bar{i}}}\partial_{j} - \beta\frac{\partial\bar{\Delta}}{\partial\bar{t}^{\bar{i}}}\partial_{\lambda^{-1}} \right) e^{W_{c}(x + \beta\bar{\Delta}^{i},\lambda^{-1} - \beta\bar{\Delta})}= \nonumber \\
= \left(\frac{1}{2}C^{jk}_{\bar{i}}\frac{\partial^{2}}{\partial x^{j}\partial x^{k}} 
+ G_{j\bar{i}}\left[x^{j} + \beta\bar{\Delta}^{j}\right]\frac{\partial}{\partial\lambda^{-1}} + \beta\bar{\Delta}_{\bar{i}}^{j}\frac{\partial}{\partial x^{j}} - \beta\bar{\Delta}_{\bar{i}}\frac{\partial}{\partial\lambda^{-1}}\right) e^{W_{c}(x + \beta\bar{\Delta}^{i},\lambda^{-1} - \beta\bar{\Delta})} \nonumber \\
\end{eqnarray}

which, after a simple cancellation of two terms, is again (\ref{3z1}) with the partial derivative $\bar{\partial}_{\bar{i}}$ replaced by the total one $\bar{d}_{\bar{i}}$. That is, for what matters the holomorphic anomaly master equation, $\exp(W_{c}(x + \beta\bar{\Delta}^{i},\lambda^{-1} - \beta\bar{\Delta})) = \exp(W_{o}(x,\lambda^{-1}))$ ( $o$ is for open ). The substitution $\bar{\partial}_{\bar{i}} \rightarrow \bar{d}_{\bar{i}}$ has a deep physical meaning which we are going to explain. 
 
\subsection{Worldsheet explanation of the open-closed duality}

Let us discuss a little bit what can be the worldsheet reason for the interpretation of the closed partition function with shifted variables as an open one in the original variables. We want to apply the shift of variables

\begin{eqnarray}\label{34b}
x^{i} \rightarrow x^{i} + \beta\bar{\Delta}^{i} \nonumber \\
\lambda^{-1} \rightarrow \lambda^{-1} - \beta\bar{\Delta} 
\end{eqnarray}
from the easy application of $2-2g-h|_{g=0,h=1}=1$,
to (\ref{3aaa}) and to reinterpret the result as an open partition function. For the moment let us discard the multiplicative factor $\lambda^{\frac{\chi}{24} - 1}$, we will come back to it later. After the shift \ref{34b} we get

\begin{equation}\label{3c}
W(x^{i}+\beta\bar{\Delta}^{i},\lambda^{-1}-\beta\bar{\Delta} ) = \sum_{g,n=0}^{\infty} 
\left(\frac{1}{\lambda} - \beta\bar{\Delta} \right)^{-2g + 2 - n}\frac{{\cal F}^{g}_{i_{1}\dots i_{n}}}{n!}(x^{i_{1}} + \beta\bar{\Delta}^{i})\dots (x^{i_{n}} + \beta\bar{\Delta}^{i_{n}})
\end{equation}

We can expand and obtain

\[
\left(\frac{1}{\lambda} - \beta\bar{\Delta} \right)^{-2g + 2 - n} = \sum_{h_{1} = 0}^{\infty}\lambda^{2g - 2 + n + h_{1}}\frac{(\beta\bar{\Delta})}{h_{1}!}^{h_{1}}(2g -2 +n)\ldots(2g -2 + n + h_{1} -1)
\]

Analogously

\[
(x^{i_{1}} + \beta\bar{\Delta}^{i})\dots (x^{i_{n}} + \beta\bar{\Delta}^{i_{n}}) = \sum_{h_{2}=0}^{n}(\beta)^{h_{2}}\frac{n!}{(n-h_{2})!h_{2}!}x^{i_{1}}\dots x^{i_{n-h_{2}}}\bar{\Delta}^{i_{n-h_{2}+1}}\dots \bar{\Delta}^{i_{n}}
\]

Calling

\[
h=h_{1}+h_{2} 
\]
\[
\tilde{n}=n - h_{2}, 
\]

replacing the sum over $\sum_{n = 0}^{\infty}$ and $\sum_{h_{2} = 0}^{n}$ with $\sum_{\tilde{n} = 0}^{\infty}$ and $\sum_{h_{2} = 0}^{\infty}$ and finally using the fact that

\[
\sum_{h_{1}=0}^{\infty}\sum_{h_{2}=0}^{\infty}(\bar{\Delta})^{h_{1}}\bar{\Delta}^{i_{1}}\dots \bar{\Delta}^{i_{h_{2}}}\frac{1}{h_{1}!h_{2}!}= \sum_{h=0}^{\infty}\frac{1}{h!}(\bar{\Delta}+\bar{\Delta}^{i_{1}})\dots(\bar{\Delta}+\bar{\Delta}^{i_{h}})
\]

we can rewrite

\begin{eqnarray}\label{34d}
& & W(x^{i}+\beta\bar{\Delta}^{i},\lambda^{-1}-\beta\bar{\Delta} ) = \\
& & = \sum_{g,\tilde{n},h=0}^{\infty} 
\lambda^{2g - 2 + \tilde{n} + h}\beta^{h}\frac{{\cal F}^{g}_{i_{1}\dots i_{\tilde{n}}\alpha_{1}\dots\alpha_{h}}}{\tilde{n}!h!}x^{i_{1}} \dots x^{i_{\tilde{n}}}\bar{\Delta}^{\alpha_{1}}\dots\bar{\Delta}^{\alpha_{h}}(2g - 2 + \tilde{n} + h_{2})\dots(2g - 2 + \tilde{n} + h) \nonumber 
\end{eqnarray}

where  $\bar{\Delta}^{\alpha}=\left(\bar{\Delta},\bar{\Delta}^{i} \right) $ and the index $\alpha$ represent either the Identity operator or the usual marginal insertions:

\[
O_{\alpha}=Id \:\:\: for \;\;\; \bar{\Delta}^{\alpha}=\bar{\Delta} 
\]
\[
O_{\alpha}=O_{i} \:\:\: for \;\;\; \bar{\Delta}^{\alpha}=\bar{\Delta}^{i} 
\]

We want to interpret (\ref{34d}) as the exponent of the open partition function, that is 

\begin{equation}\label{34e}
{\cal F}^{g}_{i_{1}\dots i_{\tilde{n}}\alpha_{1}\dots\alpha_{h}}\bar{\Delta}^{\alpha_{1}}\dots\bar{\Delta}^{\alpha_{h}}(2g - 2 + \tilde{n} + h_{2})\dots(2g - 2 + \tilde{n} + h) = {\cal F}^{g,h}_{i_{1}\dots i_{\tilde{n}}}
\end{equation}

\begin{figure}[h] 
\centering
\psfrag{x2}[][][0.8]{$z_{1}$} \psfrag{x3}[][][0.8]{$z_{2}$}
\includegraphics[width=0.5\textwidth]{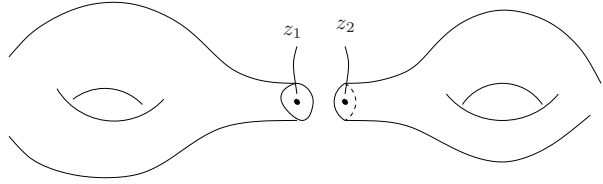}
\caption{Cut on a Riemann surface}\label{3f1}
\end{figure}

In order to do this we briefly review a way of cutting and sewing among Riemann surfaces in order to construct new ones. Consider two Riemann surfaces $\Sigma_{1}$ and $\Sigma_{2}$ and fix a point on each one, $z_{1}$ and $z_{2}$. Then chose two local coordinate systems such that the two points are in the respective origins and cut a small disk around each origin as depicted in figure (\ref{3f1}).

\begin{figure}[h!]
  \centering
\psfrag{y1}[][][0.8]{$\phi_{\alpha}$} \psfrag{y2}[][][0.8]{$\phi_{\beta}$} \psfrag{y3}[][][0.8]{$\;\;\;\;\;\;q^{h_{\alpha}}\bar{q}^{h_{\alpha}}g^{\alpha\beta}$}
  \includegraphics[width=0.5\textwidth]{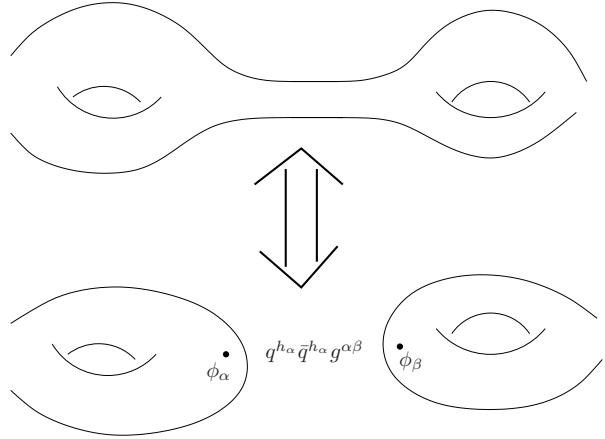}
\caption{How to replace an handle}\label{3f2}
\end{figure}

Then proceed to glue a thin annular region around the two holes through the identification

\begin{equation}
z_{1}z_{2} = q 
\end{equation}

where $q$ is some complex parameter. The result is a new Riemann surface $\Sigma_{tot}(q,z_{1},z_{2})$ with a tube with modulus $q$ joining $\Sigma_{1}$ and $\Sigma_{2}$ around the points $z_{1},z_{2}$. The point is that it is possible to replace the tube with a complete set of states, one for each hole, contracted with an appropriate inverse metric, which turns out to be exactly the inverse of the two point function on the sphere. In addition one of the two set of states should carry an additional factor of $q$ and $\bar{q}$ weighted with its conformal weight ( this comes from the conformal transformation properties of the states under rescaling of $q$ ).
In practice we can obtain the amplitude for the full Riemann surface $\Sigma_{tot}(q,z_{1},z_{2})$ considering the amplitudes $\Sigma_{1}$ and $\Sigma_{2}$ with, on each one, an additional bulk operator insertion in $z_{1}$ and $z_{2}$ running over the complete set of states, adding the proper conformal factor and contracting with the inverse metric. This is shown in figure (\ref{3f2})

The result should then be integrated over the moduli $q,z_{1}$ and $z_{2}$ with the appropriate Jacobian factors. Trying to do this in ``physical`` string theory is quite complicated mainly because there a complete set of states is given by an infinite number of objects with different conformal weights. Luckily in our case things are a little easier. What we want to do is to start with some closed topological amplitude with genus $g$, with an arbitrary number $n$ of bulk operator insertion, and to add boundaries. Add boundaries means to sew disk amplitudes to the closed surface using the procedure described. In this case out of the 6 real moduli $q,z_{1}$ and $z_{2}$ only three survives, essentially the position $z_{1}$ and the length of the tube, $|q|$ or, which is conformally equivalent, the length of the boundary. Let us note a few things. First the only bulk operator insertions allowed on the closed surface are, for anomaly reasons, only the one constructed out of the marginal operators. This is nothing but the marginal operator insertion $O_{i}$ surrounded by the two spin two supercurrents $G^{-},\bar{G}^{-}$ and integrated over the Riemann surface. Thus the disk amplitude will come with an upper index ''$i$``, that is some complete set of states contracted with the appropriate metric. If for the metric we use the topological sphere, the charge anomaly over it requires for the second set of states to have left equal to right charge equal to two ( index $a$ ). Then on the disk this state operator insertion will be dressed with some linear function of the supercurrent $G^{-}$ and $\bar{G}^{-}$ , partially integrated only along the radial direction, so also there the charge anomaly condition is satisfied. Of course marginal operators ( as well as charge two operators ) are only a finite number and, in addition, they all are scalars so that the weight $q^{h_{i}}\bar{q}^{\bar{h}_{i}}$ simply contributes to one. Also the two supercurrents encircling the operator on the closed surface correspond exactly to the Jacobian for the modulus for the position $z_{1}$ of the handle while the $G^{-}$ and $\bar{G}^{-}$ combination folded with appropriate Beltrami differentials is the Jacobian for what remains of the modulus $q$. Instead of $|O_{i}\rangle\eta^{ia}\langle O_{a}|$ for replacing the tube it is also possible to chose to twist half of the sphere in the topological way and half in the antitopological one; this brings as a result 
$|O_{i}\rangle g^{i\bar{j}}\langle O_{\bar{j}}|$. However in this case the interpretation in terms of disk amplitude is less clear. In fact the disk with only one marginal insertion is not allowed from anomaly requirement, the interpretation is in term of integrated quantities from the allowed amplitudes, that is functions of the moduli of the theory such that their derivative gives an amplitude with the corresponding insertion. This is our old friend $\bar{\Delta}^{i}$. The shift $\bar{\Delta}$ has a nice interpretation as well, that is it is exactly $\bar{\Delta}^{i}$ for $i = 0$ that is for the identity operator insertion or, more physically for the dilaton insertion. Also the factor of $2g - 2 + n +h_{2} +h_{1}$, which comes after you are inserting an additional identity operator on a surface with already $h_{2} + h_{1}$ insertions of $O_{i}$ and $Id$ operators, is perfectly justified ( as a generalization of the discussion around (\ref{0cov}) ). It is the ''weight'' the $Id$ operator carries once it is inserted on a Riemann surface with the corresponding topological data. All in all we have the picture 
 
\begin{figure}[h!]
  \centering
\psfrag{z1}{$O_{\alpha}$} \psfrag{z2}{$G^{-}$} \psfrag{z3}{$\bar{G}^{-}$} \psfrag{z4}{$= \bar{\Delta}^{\alpha}$}
    \includegraphics[width=0.5\textwidth]{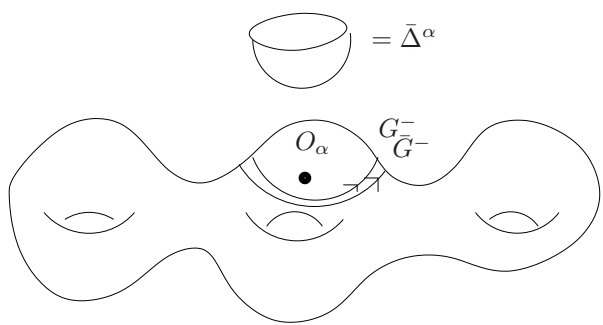}
\caption{Gluing holes}\label{3f3}
\end{figure}

Now it should be clear the issue about the replacement $\bar{\partial}_{\bar{i}} \rightarrow \bar{d}_{\bar{i}}$ in the master equation for the closed partition function in order to correspond to the one for the open partition function. In the worldsheet language a total derivative means that the corresponding operator insertion will be added not only on the closed Riemann surface but also on the disks that will be sewn, that is a partial derivative ( insertion only on the original closed surface ) plus the induced derivatives carrying factors corresponding to the derivatives of the discs ( insertions on the surrounding of the new boundaries ). 

Let us mention here that the construction we have described represents exactly the procedure of cutting and sewing Riemann surfaces only in the limit of long and narrow tubes, that is on the boundary of the moduli space.  For topological strings the contribution to the amplitudes which is not coming from the boundary is the so called holomorphic ambiguity, because not fixed by solving the HAE, and in fact the idea about the shift for the closed moduli is completly justified only in that set up. So it is resonable to suppose the presence of some additional holomorphic ambiguity for open amplitudes not inherited by the original closed one, and some computations on the resolved conifold seem to confirm this problem.

The last important issue to discuss is about the fate of the prefactor $\lambda^{\frac{\chi}{24} - 1 - \beta^{2}\frac{N}{2}}$ for the open partition function which cannot in any way be obtained from the one of the closed partition function,$\lambda^{\frac{\chi}{24} - 1}$, simply by using the shift in $\lambda$. This prefactor really originates from one loop contact terms so that we can infer that for lower genus amplitudes something different is happening. This conjecture finds confirmation in the analysis of \cite{Wal2} were it was noticed that the original shift of \cite{Oog} does not work if we want to translate the constrain 

\begin{equation}\label{3wcon}
D_{i_{n}}{\cal F}^{(g,h)}_{i_{1}\dots i_{n-1}} = {\cal F}^{(g,h)}_{i_{1}\dots i_{n}}.
\end{equation}

plus the corresponding requirement for the only nonvanishing one loop amplitudes that is

\begin{equation}\label{3wcon2}
D_{i}X = X_{i} \;\;\;\; for \;\;\;\; X = {\cal F}^{(0,2)},{\cal F}^{(1,0)} 
\end{equation}

plus

\begin{equation}\label{3wcon3}
{\cal F}^{(0,0)}_{ijk}=C_{ijk}\;\;\;\; {\cal F}^{(0,1)}_{ij}=\Delta_{ij} 
\end{equation}
 
from open amplitudes to the closed ones and vice versa. In particular there is a master equation gathering together (\ref{3wcon}), (\ref{3wcon2}) and (\ref{3wcon3}) into a single expression \cite{BCOV,Wal2}, but applying (\ref{3q}) to its open version does not kill all the open data; what remain are the terms reproducing the constrain for lower genus terms that is (\ref{3wcon2}) and (\ref{3wcon3}). The solution is to divide the shift into two parts: in particular being ${\cal T}$ the domainwall tension for the 3-brane ( see \cite{Wal1} ) we can write

\[
\bar{\Delta}_{\bar{i}\bar{j}} = D_{\bar{i}}D_{\bar{j}}\bar{{\cal T}} - \bar{C}_{\bar{i}\bar{j}\bar{k}}g^{k\bar{k}}\bar{D}_{k}{\cal T}
\]
 
and shift $x^{i}$ and $\lambda^{-1}$ only with the primitives of the firs part, $D_{\bar{i}}D_{\bar{j}}\bar{{\cal T}}$, while absorbing the remaining piece into a partition function redefinition which deletes the open string data still present in the equations. From a target space point of view it seems it is the normalization factor $\caln$ in (\ref{3yy}) to correspond to the rescaling ( eq.(3.13) of \cite{Wal2} ). The point is that for lower genus amplitudes working in this way seems unavoidable while, if we want to consider objects with higher $g$ and $h$, the formalism already considered works perfectly as (\ref{3wcon}) with high $g$ and $h$ encounters any problem in passing from the open to the purely closed amplitudes and vice versa. 

\subsection{Target space computations}

It is now possible to postulate that an analog shift for $x$ and $\lambda^{-1}$ in the path integral with the Kodaira-Spencer action 
(corresponding to the closed partition function) would allow to obtain the full path integral with the complete action ( apart from the issues involving a normalization factor ). 

In order to reproduce the power expansion 
of (\ref{3y}) from the target space field theory we have to set $x\to \lambda x$, so that any bulk operator insertion 
carries a weight $\lambda x$ as in (\ref{3y}). 
To maintain our setting we translate (\ref{3q}) into a shift for the product $\lambda x$ 
\begin{eqnarray}\label{3z3}
\lambda x^{i} &\rightarrow & \lambda x^{i} + \lambda\beta\bar{\Delta}^{i}  - \lambda^{2}\beta\bar{\Delta} x^{i} + o(\lambda^{3},\beta^{2})\nonumber \\
\lambda^{-1} &\rightarrow & \lambda^{-1} - \beta\bar{\Delta} 
\end{eqnarray}
of which we will keep only the lowest order term for the first line, discarding the $\lambda^{2}$ piece induced by the transformation of $\lambda$. 
From now on $\lambda x$ will be denoted simply  as $x$. We want to check that
\begin{equation}\label{3v}
\int {\cal D}A e^{-S_{KS}(x^{i} + \lambda\beta\bar{\Delta}^{i} + \dots,\lambda^{-1} - \beta\bar{\Delta};t,\bar{t};A)} 
\simeq \int {\cal D}A{\cal D}B{\cal D}b\dots e^{-S_{tot}(x,B_{0},\lambda^{-1};t,\bar{t};A,B,b,\dots)} 
\end{equation}
Let us consider (\ref{3v}) at the tree level. First we need the explicit expressions for $\bar{\Delta}$ and $\bar{\Delta}^{i}$; computed using the topological-antitopological metric they are

\[
\bar{\Delta} = g^{0\bar{0}}\int_{X}L_{CS}\wedge \bar{\Omega}_{0} \; \; \; \; g^{0\bar{0}} = 
\left(\int_{X}\Omega_{0} \wedge \bar{\Omega_{0}}\right)^{-1}
\]
\[
\bar{\Delta}^{i} =g^{i\bar{j}} \left(\int_{X}L_{CS}\wedge d_{\bar{j}}\bar{\Omega}\right)_{x = 0} 
\; \; \; \; g^{i\bar{j}} = \left(\int_{X}d_{i}\Omega \wedge d_{\bar{j}}\bar{\Omega}\right)^{-1}_{x = 0} 
\]

where all the fields are on shell and $x=0$. Notice also that $\bar{\Delta}$ and $\bar{\Delta}^{i}$ have been computed starting from the 
antitopological theory. 
Finally the closed field does not appear because on shell it goes as $O(x^{2})$.
Simply applying (\ref{3z3}) to the Kodaira-Spencer action gives, 
at order $\beta$ and $\lambda^{-1}$, and redefining 
$S_{KS}$ in order to have the factor $\lambda^{-2}$ explicit,
\[
\frac{1}{\lambda^{2}}S_{KS}(x^{i} + \lambda\beta\bar{\Delta}^{i} + \dots,\lambda^{-1} - \beta\bar{\Delta};t,\bar{t};A) 
= \frac{1}{\lambda^{2}}S_{KS}(x^{i},\lambda^{-1};t,\bar{t};A) - 
\]
\[
- \frac{\beta}{\lambda}\int_{M}[(A + x)(A + x)]'(\mu_{i})'\bar{\Delta}^{i} - 
\frac{2\beta\bar{\Delta}}{\lambda}S_{KS}(x^{i},\lambda^{-1};t,\bar{t};A) + O(\lambda^0,\beta^{2})
\]
Going at tree level the $O(\lambda^0,\beta^{2})$ are not taken into account; in addition the $A$ field should be 
taken on shell with respect to the Kodaira-Spencer equation in the shifted background, that is
\begin{equation}\label{3u}
A \rightarrow A(x^{i} + \lambda\beta\bar{\Delta}^{i} + \dots) = A(x) + \lambda\beta\bar{\Delta}^{i}\partial_{i}A(x) + O(\lambda^{2},\beta^{2})
\end{equation}
Then, at order $\beta$, $\frac{1}{\lambda}$, the left side of (\ref{3v}) is the exponential of
\[
\frac{1}{\lambda^{2}}S_{KS}(x^{i},\lambda^{-1};t,\bar{t};A(x)) 
- \frac{\beta}{\lambda}\int_{X}[(A(x) + x)(A(x) + x)]'(\mu_{i})'\bar{\Delta}^{i} -
\]
\[
 - \frac{2\beta\bar{\Delta}}{\lambda}S_{KS}(x^{i},\lambda^{-1};t,\bar{t};A(x)) +
\]
\begin{equation}\label{3exp}
+ \frac{\beta}{\lambda}\int_{X}\bar{\Delta}^{i}(\partial_{i}A(x))'\frac{1}{\partial}\bar{\partial}A(x) 
- [(A(x) + x)(A(x) + x)]'(\partial_{i}A(x))'\bar{\Delta}^{i}  
\end{equation}
where the last line is actually zero because of the equations obeyed by $A(x)$, and the second line reduces to 
\[
-\frac{\beta\bar{\Delta}}{3\lambda}[(A(x) + x)(A(x) + x)]'(A(x) + x)' = 
\]
\[
= -\frac{\beta\bar{\Delta}}{\lambda}[(A(x) + x)(A(x) + x)(A(x) + x)]'\Omega_{0} 
\]
Remembering the expression (\ref{2x}) we can substitute the value of $\bar{\Delta}^{i}$ in (\ref{3exp}) and get, for the second term 
in the first line,
\begin{eqnarray}\label{3s}
\frac{\beta}{\lambda}\int_{X}\Omega^{(1,2)}_{A = A(x)}\wedge (d_{i}\Omega)^{(2,1)}_{x = 0}\left(\int_{X}(d_{i}\Omega)^{(2,1)}_{x = 0} 
\wedge (d_{\bar{j}}\bar{\Omega})^{(1,2)}_{\bar{x} = 0}\right)^{-1}\cdot \nonumber \\
\cdot \int_{X}L_{CS}^{(2,1)}\mid_{B = B(u,x)}\wedge (d_{\bar{j}}\bar{\Omega})^{(1,2)}_{\bar{x} = 0}   
 = \frac{\beta}{\lambda}\int_{X}{\Omega}^{(1,2)}_{A = A(x)}\wedge L_{CS}^{(2,1)}\mid_{B = B(u,x)}
\end{eqnarray}
The last equality has been obtained using the Riemann bilinear relations:
\[ \int_{X}w\wedge\hat{w} = \sum_{a = 0}^{h_{2,1}}\int_{\delta_{a}}w\int_{\delta_{a + h_{2,1}}}\hat{w} 
- \int_{\delta_{a + h_{2,1}}}w\int_{\delta_{a}}\hat{w} 
\]
where $\delta_{a}$ is a base of 3-cycles on X.
First we express in this way the integrals containing $\Omega \wedge d_{i}\Omega$ and 
$ L_{CS}\wedge d_{\bar{j}}\bar{\Omega}$. 
Then we can define $X^{i}$ and $\bar{X}^{j}$ as three forms such that
\[
\left(\int_{X}d_{i}\Omega\wedge d_{\bar{j}}\bar{\Omega}\right)^{-1} \equiv \int_{X}X^{i}\wedge \bar{X}^{j} 
= \sum_{a = 0}^{h_{2,1}}\int_{\delta_{a}}X^{i}\int_{\delta_{a + h_{2,1}}}\bar{X}^{j} 
- \int_{\delta_{a + h_{2,1}}}X^{i}\int_{\delta_{a}}\bar{X}^{j}  
\]
respects the definition
\[
\sum_{\bar{j}}\left(\int_{X}d_{i}\Omega\wedge d_{\bar{j}}\bar{\Omega}\right)^{-1}
\int_{X}d_{k}\Omega \wedge d_{\bar{j}}\bar{\Omega} = \delta_{i,k}
\]
that is  
\[
\sum_{i}\int_{\delta_{a}}d_{i}\Omega\int_{\delta_{b}}X^{i} \equiv \delta_{a,b} \;\; \;\;\; \sum_{a = 0}^{2h_{2,1} +
 2}\int_{\delta_{a}}d_{i}\Omega\int_{\delta_{a}}X^{j}\equiv \delta_{i,j}
\]
and similarly with the barred quantities. Substituting these expressions in (\ref{3s}) we obtain the result. 

Equivalently for the term in $\bar{\Delta}$ in the second line of (\ref{3exp}) we get 
\begin{equation}
\frac{\beta}{\lambda}\int_{X}{\Omega}^{(0,3)}_{A = A(x)}\wedge L_{CS}^{(3,0)}\mid_{B = B(u,x)}
\end{equation}
In order to reconstruct the full integral $\int_{X}{\Omega}_{A = A(x)}\wedge L_{CS}\mid_{B = B(u,x)}$
from the above equation the $(0,3)$ and $(1,2)$ components of $L_{CS}$ are still missing.
Notice however that they can be recovered by requiring CPT invariance. In particular,   
we modify (\ref{3s}) as
\[
\frac{\beta}{\lambda}\int_{X}\left(\Omega^{(1,2)}_{A = A(x)} + \Omega^{(2,1)}_{A = A(x)}\right)
\wedge (d_{i}\Omega)^{(2,1)}_{x = 0}\cdot g^{i\bar{j}}\cdot
\]
\[
\cdot\int_{X}\left(L_{CS}^{(2,1)}\mid_{B = B(u,x)} + L_{CS}^{(1,2)}\mid_{B = B(u,x)}\right)
\wedge (d_{\bar{j}}\bar{\Omega})^{(1,2)}_{\bar{x} = 0} 
\]
where the extra term actually vanishes due to form degree reasons.   
This lead to an additional term
\[
 \frac{\beta}{\lambda}\int_{X}{\Omega}^{(2,1)}_{A = A(x)}\wedge L_{CS}^{(1,2)}\mid_{B = B(u,x)}
\]
An analogous modification has to be performed in order to obtain the 
$(0,3)$ component of $L_{CS}$. 

The geometrical counterpart of the above is as follows.
We know from the discussion of \cite{Wal1} that 
the coupling of the on-shell Chern-Simons action to $\Omega_0$ can be translated in mathematical terms to the pairing with the related 
normal function, $\nu$, dual to a suitable three-chain, $\Gamma$, such that
\[
\int_{X}\Omega_{0}\wedge L_{CS}\mid_{B=B(u,x)} = \int_{\Gamma}\Omega_{0} = \langle \Omega_0, \nu \rangle 
\]
and similarly for a $(2,1)$ form. Then it exists a lift of $\nu$ such that the coupling with a $(0,3)$ and $(1,2)$ forms are defined 
to be obtained by CPT invariance, that is complex conjugation of the corresponding $(0,3)$ and $(2,1)$ couplings. 

Summarizing we have shown that
\begin{eqnarray}\label{3r}
\frac{1}{\lambda^{2}}S_{KS}(x^{i} + \beta\lambda\bar{\Delta}^{i},\lambda^{-1} - \beta\bar{\Delta};t,\bar{t})\mid_{on \; shell} = \nonumber \\  
=\left(\frac{1}{\lambda^{2}}S_{KS}(x^{i},\lambda^{-1};t,\bar{t}) + \frac{\beta}{\lambda}\int_{X}\Omega \wedge L_{CS} 
- \Omega db 
\right)\mid_{on \; shell}
\end{eqnarray}
in the gauge $F_{B_{0}}^{\tilde{2},\tilde{0}}=0$.
Notice that the completion of the solution via CPT invariance obtained by adding the classical solutions of the anti-topological theory
is consistent with the fact that, in our gauge, the gauge fixing $F^{(2,0)}=0$ and the equation of motion $F^{(0,2)}=0$
of the topological theory are the same, up to a switch of role, as in the
{\it on shell} anti-topological one which is then manifestly CPT conjugate.       
    \chapter{Conclusions and open issues}
\label{sec:cap4}
\noindent

In this thesis we have discussed several issues whose common origin is the extension to open strings of results already known in the closed sector. Open strings are richer, and of course more complicated, then the purely closed ones, essentially because they introduce new degrees of freedom in the game which should be separately analyzed. 

Already in the introductory chapter (\ref{sec:cap0}) we have encountered an issue about the interpretation of the HAE fully extended to open moduli. There it was explained how the interpretation a la Witten can be generalized only partially if you restrict yourself to the case of frozen open moduli but, trying to export the same ideas to the most generic situation, encounters many problems; It seems that, even thought the worldsheet origin of the HAE relies on very similar techniques, the target space interpretation really differentiates and is far from being fully understood. Obviously we do not have a Witten-type explanation of the HAE even for the closed A-model case, but in that situation you can rely on the Calabi Yau mirror symmetry for making evident what was hidden in the formalism. In the present case, instead, the problem seems to arise from a completely different point and it would be interesting to try to shed some light on it.    

In the second chapter we discussed the issue of tadpole cancellation in the context of unoriented
topological strings, and showed from super conformal field theory arguments that this corresponds
to the decoupling of wrong moduli at all loops.
We also provided a geometrical interpretation for unoriented B-model amplitudes at one loop in terms of analytic 
torsions of vector bundles over the target space. In itself the discussion seems self consistent, as it was for the analogue problem of tadpole cancellation in superstring theory, but a few points seem worth of future work; let us for example remark that the topological open A-model free energy is expected to provide a generating
function for open Gromov-Witten invariants. However, these have not been
defined rigorously yet, except for some particular cases \cite{KL,GZ,panda}.
We observe that the inclusion of unoriented worldsheet geometries turns out
to be natural also from a purely mathematical viewpoint.
In fact the compactified moduli spaces of open Riemann and Klein surfaces
have common boundary components (see Section \ref{1comp}).
Thus string theory suggests that a proper mathematical definition
of open Gromov-Witten invariants should be obtained by including
non-orientable domains for the maps. Therefore one should consider
equivariant Gromov-Witten theory and sum over all possible involutions
of the complex double, up to equivalences.

There are other several interesting directions to be further investigated,
the most natural being the study of holomorphic anomaly equations in presence of non-trivial
open string moduli. This can be obtained by extending the holomorphic anomaly equations
studied in \cite{BT} in order to include the contribution of non-orientable worldsheets.

Actually, our method is applicable only to cases in which the D-brane/orientifold set is 
modeled on the fixed locus of a target space involution. It would be quite interesting to be able to 
generalize it to a more general framework, that is to remove the reference to a given target space involution 
to perform the orientifold projection, in order to compare with 
some of the compact examples studied in \cite{j,Grimm,kw2,w5,emanuel}.

It would be also interesting to link our B-model torsion formulae to the A-model side where open strings on orientifolds have been
understood quite recently \cite{marino} to be the dual of coloured polynomials in the Chern-Simons theory.
This should also enter a coloured extension of the conjecture stated in \cite{df}.
Notice also that interpretation of open B-model one loop amplitudes in terms of analytic torsions could be extended to 
more general target space geometries. For example one could investigate whether the notion of twisted torsions introduced in
\cite{RW,MW} could provide a definition of B-model one loop amplitudes in the presence of $H$-fluxes and more in general with a target
of generalized complex type.

The third and fourth chapters contain the results best fitted for generalizations and applications and their implementation in a clean scheme is still to be settled. There we discussed, mainly from a target space point of view, the coupling between open and closed strings for the B model, providing in this way a complete consistent effective theory. Then we integrated out the open fields and found that the path integral had reduced to the one of a closed theory with shifted background. We also gave a worldsheet explanation for this result in terms of sewing holes to closed Riemann surfaces which closely resemble the idea of \cite{OV2}, even if from a slightly different perpective. 

The first point it comes to my mind is certainly the idea of implementing our idea of open-closed duality far beyond tree level. Both the result of \cite{Oog} and the worldsheet argument we have presented in our work seem to show that it is in principle possible to obtain open amplitudes from the closed ones simply applying a shift of moduli. It would be interesting to apply this idea to explicit computations as the ones done on the  \cite{OV1,Gopakumar:1998jq}, reduced to the case of trivial Wilson lines. In those papers it is indicated the full closed partition function as well as the open one for the brane settings specified in the article. A simple expansion of the closed amplitude around its parameters, $x$ and $\lambda$, using the disk amplitudes should give, at each order, the corresponding open result. At the same time it would be interesting to push as further as possible the target space point of view, trying to go beyond tree level; clearly the two methods should agree and a dictionary between them should be obtained. Further study is certainly necessary to clarify the full picture.   

More complicated would be to try to extend our ideas to a true topological open closed duality, that is a general recipe for the topological changes in the target space that allows a duality between open and closed theories. In the topological case we have obviously in mind the conifold transition of \cite{Gopakumar:1998ki} but an attempt to re derive that result in our setup ( for the B-model case ) hits some difficulties. The most evident is that we implement open closed duality through a change of field background but the target space remains untouched. Obviously we can try to translate a shift of field background to a geometrical change, nevertheless we will never be able to accommodate a topological change. The reason behind this difference seems to rely on the presence of singularities in the Calabi Yau, whose very presence accounts for the topological shift. But a background is by definition singular on that locus so we need additional input to understand what is going on. Also here some work is important.

Last point, the picture we provided in this paper seems to allow an extension to generalized complex geometries.
This should follow by the definition of an extended Chern-Simons functional
where the 3-form $\Omega$ gets promoted to the relevant pure spinor
as in \cite{luca}. Once this is done and the $b$ field promoted to a multiform, this would extend to open strings
the proposal in \cite{pestun} to generalized complex geometry of an analog of the Kodaira-Spencer theory.
    
    
    
    \backmatter

    \bibliography{bibliography}
\label{sec:bibl}
   
\end{document}